\documentclass[nofootinbib, showkeys, unsortedaddress, superscriptaddress, notitlepage,onecolumn]{revtex4-1}
\usepackage{amssymb,amsmath,bm,graphicx}
\usepackage{array}[=2016-10-06]
\usepackage[svgnames]{xcolor}
\usepackage[colorlinks=true, allcolors=DarkBlue,linktocpage=true]{hyperref}
\usepackage{orcidlink}

\graphicspath{{./figures/}}

\def\thesection{\arabic{section}}
\def\thesubsection{\arabic{section}.\arabic{subsection}}
\def\thesubsubsection{\arabic{section}.\arabic{subsection}.\arabic{subsubsection}}

\makeatletter
\renewcommand{\p@subsection}{}
\renewcommand{\p@subsubsection}{}
\makeatother

\makeatletter
\renewcommand\@makecaption[2]{%
  \par
  \vskip\abovecaptionskip
  \begingroup
   \small\rmfamily
    \begingroup
     \samepage
     \flushing
     \let\footnote\@footnotemark@gobble
     \@make@capt@title{#1}{#2}\par
    \endgroup
  \endgroup
  \vskip\belowcaptionskip
}
\makeatother

\begin{document}

\title{Large Deviations in Switching Diffusion:\\Dynamical Condensation and Diffusivity BBP Transition}

\author{Mathis Gu{\'e}neau\,\orcidlink{0009-0001-4763-5105}}
\email{gueneau@pks.mpg.de}
\affiliation{Max-Planck-Institut f{\"u}r Physik komplexer Systeme, N{\"o}thnitzer Stra{\ss}e 38, 01187 Dresden, Germany}

\author{Timo Schorlepp\,\orcidlink{0000-0002-9143-8854}}
\email{timo.schorlepp@nyu.edu}
\affiliation{Courant Institute of Mathematical Sciences, New York University, 251 Mercer Street, New York, NY 10012, USA}

\author{Satya N. Majumdar\,\orcidlink{0000-0001-9844-6980}}
\email{satyanarayan.majumdar@cnrs.fr}
\affiliation{LPTMS, CNRS, Univ.\ Paris-Sud, Universit{\'e} Paris-Saclay - 91405 Orsay, France}

\author{Gr{\'e}gory Schehr\,\orcidlink{0000-0001-6648-5213}}
\email{schehr@lpthe.jussieu.fr}
\affiliation{Sorbonne Universit{\'e}, Laboratoire de Physique Th{\'e}orique et Hautes Energies, CNRS UMR 7589, 4 Place Jussieu, 75252 Paris Cedex 05, France}

\date{\today}

\begin{abstract}
We consider a Brownian particle with position $x(t)$ whose diffusivity changes at Poissonian
switching times, with each new value drawn independently from a bounded
distribution $W(D)$. Assuming
$W(D)\sim C(D_{\max}-D)^\nu$ near its upper edge, we study the
large-deviation regime $x(t)\simeq yt$ with $y = O(1)$ at large times, far beyond the typical diffusive
scale. In a recent work [Gu{\'e}neau et al., \textit{Phys.\ Rev.\ Lett.} {\bf 135}, 067102 (2025)], the rate function of the displacement was obtained through a renewal approach {and it was shown to display several singularities.} In this paper, we carry out a detailed study of the corresponding dynamical phase transitions and trajectory condensation. For moderate $y$, the
imposed displacement is shared among $O(t)$ displacement increments
accumulated during successive intervals between switches. Beyond a first
threshold $y_-$, one interval produces an increment of order $t$ carrying
a finite fraction of the total displacement, signaling dynamical
condensation. For $\nu>1$, a second threshold $y_+>y_-$ is reached when
this macroscopic increment carries asymptotically the entire imposed
displacement, whereas for $0<\nu\leq1$ the two thresholds merge,
$y_-=y_+$. In turn, the moment-generating function of $x(t)$ displays a complementary
transition from collective contributions of many diffusivities to
selection of states near $D_{\max}$. We show that this transition is governed by the same edge mechanism as
the Baik--Ben Arous--P{\'e}ch{\'e} (BBP) transition, indicating that diffusivity
selection and trajectory condensation are two manifestations of the same
phenomenon. We validate our predictions for the dynamical condensation against numerical results from rare event sampling.
\end{abstract}

\maketitle

\tableofcontents

\section{Introduction}

Large-deviation theory is often introduced as a
way of quantifying the probability of rare events. Yet this probability tells
only part of the story: one would also like to identify the trajectories or
configurations through which the fluctuation is realized. This question is
central in statistical physics. In macroscopic fluctuation theory~\cite{bertini-de-sole-gabrielli-etal:2015}, for instance, a rare
current or density fluctuation is described by an optimal space--time history,
whose change may produce a dynamical phase transition. In
random-matrix theory, an extreme-eigenvalue fluctuation may likewise arise
either collectively, through a deformation of the spectrum, or locally,
through a single eigenvalue detached from the bulk~\cite{dean-majumdar:2006,majumdar-vergassola:2009,majumdar-schehr:2014}.\\

This leads to a broader question: is a large fluctuation built up through
many contributions, each carrying only a vanishing fraction of the total, or
is it carried by a single exceptional event that accounts for a finite
fraction? The crossover between these two mechanisms is known as a condensation transition. It
appears in stochastic mass transport models and zero-range processes~\cite{PhysRevLett.81.3691,majumdar-evans-zia:2005, evans-majumdar-zia:2006, evans-hanney:2005, majumdar:2010, szavits-nossan-evans-majumdar:2014, corberi-sarracino:2019}, in constrained sums of random
variables~\cite{feller:1950, petrov:1975, smith:2025}, in the displacement of run-and-tumble particles~\cite{gradenigo-majumdar:2019,mori-le-doussal-majumdar-etal:2021,mori-gradenigo-majumdar:2021}, in the statistics of time-integrated observables~\cite{nickelsen-touchette:2018,den-hollander-majumdar-meylahn-etal:2019, smith-majumdar:2022, meylahn-sabhapandit-touchette:2015, yerrababu-majumdar-guiselin-etal:2025},
and many other models~\cite{berlin-kac:1952,gunton-buckingham:1968,kosterlitz-thouless-jones:1976,gradenigo-iubini-livi-etal:2021,garrahan-manai-warzel:2023, manai-warzel:2025,bernard-bouchaud-le-doussal:2026,camacho-guardian-kleebank-vewinger-etal:2026,gradenigo-lucente-salsnich:2026,PhysRevResearch.5.023103,defaveri_dynamical_2026}.
The distinction
can arise along the trajectory of a single particle: an atypical displacement
may be accumulated over many intervals of the motion or generated during one exceptional interval. The emergence of such an exceptional contribution is reminiscent of the
single-big-jump principle in the context of random walks and anomalous diffusion, according to which a rare fluctuation of a sum is
dominated by its largest term~\cite{vezzani-barkai-burioni:2019, wang-vezzani-burioni-etal:2019,vezzani-barkai-burioni:2020,bassanoni_rare_2024,bassanoni-vezzani-barkai-etal:2025,burov2026gaussian}.\\

A closely related collective-versus-localized transition occurs in
random-matrix theory. Consider a high-dimensional random matrix perturbed
along a preferred direction. When the perturbation is too weak, the largest
eigenvalue remains at the edge of the continuous spectral bulk and its
eigenvector is essentially isotropic, with a vanishing projection onto the
preferred direction: the perturbation is absorbed by the random
background. Beyond a critical strength, the largest eigenvalue detaches
from the bulk and becomes an isolated outlier, while its eigenvector
develops a finite projection onto that direction. This is known as the
Baik--Ben Arous--P{\'e}ch{\'e} (BBP) transition~\cite{baik-ben-arous-peche:2005,benaych-georges-nadakuditi:2011, Edwards_1976}.\footnote{It should not be confused with the collective-versus-local mechanism for extreme-eigenvalue fluctuations mentioned above~\cite{dean-majumdar:2006,majumdar-vergassola:2009,majumdar-schehr:2014}, which concerns atypically large deviations of the largest eigenvalue of an unperturbed ensemble and is associated with a third-order transition. The BBP transition instead concerns the typical location of the largest eigenvalue of a perturbed ensemble, as a function of the perturbation strength, see Sec.~\ref{sec:rmt} for details.} The squared projection is its natural
order parameter~\cite{benaych-georges-nadakuditi:2011,bocchi-biroli-cammarota-etal:2026} and may be viewed as the fraction carried by the selected
component. Although usually formulated as a spectral transition, BBP thus
expresses a broader change from a collective response of the background to
the emergence of one distinguished component.\\

In this paper, we show that rare displacements of a single Brownian
particle with a randomly changing diffusivity exhibit both a dynamical
condensation transition and a BBP transition. The present work thereby substantially expands upon the results recently reported in Ref.~\cite{gueneau-majumdar-schehr:2025}, which calculated the large-deviation rate function for this switching diffusion (SD) model.
We find that at fixed displacement, a
moderate fluctuation is accumulated over many intervals between switches,
whereas beyond a threshold one interval becomes macroscopic and carries a
finite fraction of the total displacement. In the moment-generating
function of the displacement, the same reorganization appears as a BBP transition: the growth
changes from a collective contribution of many diffusivity values to the
selection of values close to
the largest allowed diffusivity, $D_{\max}$. The spectral BBP transition
and the condensation of conditioned trajectories are therefore two
complementary manifestations of the same physical mechanism.

\subsection{Definition of the model}
\label{sec:sd-def}

\begin{figure}
\centering
\begin{minipage}[t]{.49\textwidth}
\vspace{0pt}
\includegraphics[width = \textwidth]{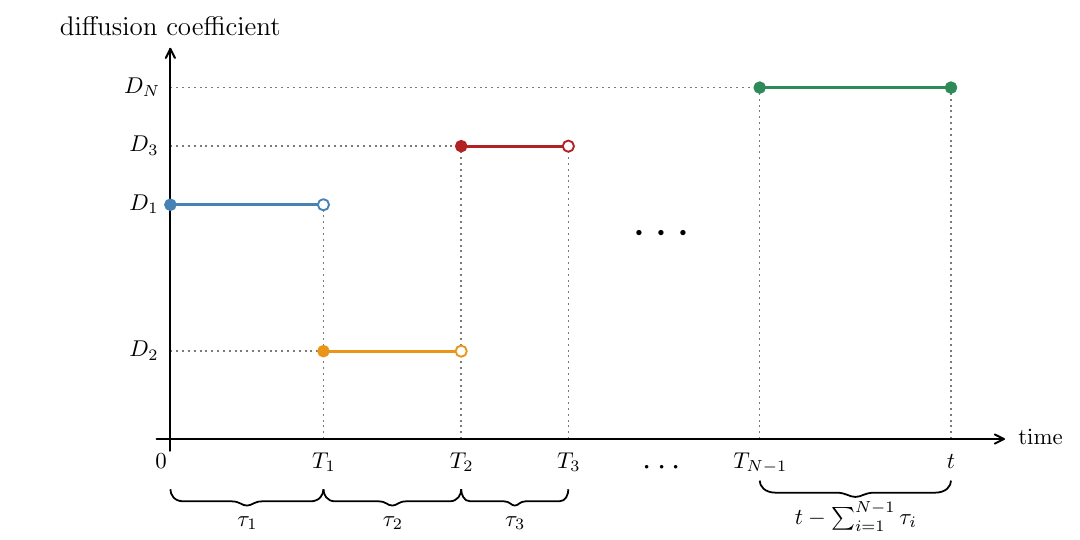}
\end{minipage}
\hfill
\begin{minipage}[t]{.49\textwidth}
\vspace{7pt}
\includegraphics[width = \textwidth]{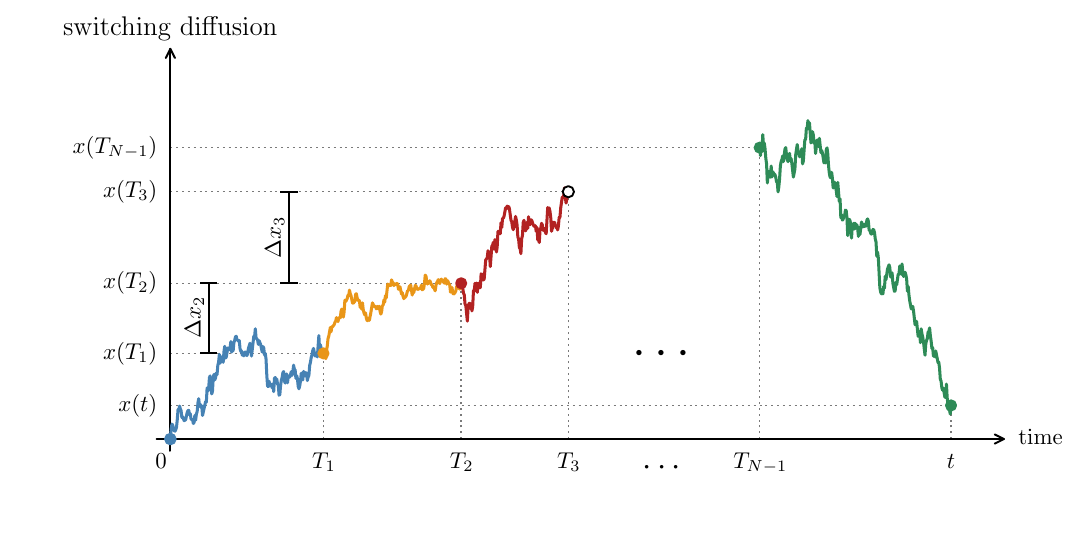}
\end{minipage}
\caption{Sketch of a sample path of the SD~\eqref{eq:sd-def}. \textbf{Left.} Diffusion coefficient sample path $D(t)$ which is piecewise constant, with values drawn independently from $W(D)$, during time intervals $T_i - T_{i -1} \sim {\rm Exp}(r)$. \textbf{Right.} Corresponding sample path of $x(t)$ with this diffusion coefficient history $D(t)$. We denote the total displacement per constant $D$ segment by $\Delta x_i = x(T_i) - x(T_{i-1})$. 
}
\label{fig:sketch-setup}
\end{figure}

Consider a one-dimensional SD process $x(t)$, $t \geq 0$, with
\begin{align}
    \dot x(t) = \sqrt{2 D(t)} \eta(t)\,, \quad x(0) = 0\,.
    \label{eq:sd-def}
\end{align}
Here, $\eta(t)$ denotes standard Gaussian white noise with $\langle \eta(t) \rangle = 0$ and $\langle \eta(t) \eta(t') \rangle = \delta(t-t')$, and the diffusion constant~$D(t)$ is an independent, pure jump and continuous time Markov process. Jumps, or ``switches'', of $D(t)$ occur at a Poissonian rate $r > 0$, such that the waiting times $\tau_i = T_i - T_{i-1}$ between jumps are independent and identically distributed as $\tau_1, \tau_2, \dots \sim \text{Exp}(r)$,
where $0 = T_0 < T_1 < T_2 < \dots$ denote the switching times (see Fig.~\ref{fig:sketch-setup} for a sketch). In other words, waiting times are exponentially distributed, with an expected waiting time between switches of $1/r$. The initial value $D(0) = D_1$, and new states after each switch, are drawn independently from a distribution $W(D)$ with bounded\footnote{For distributions with unbounded support, the Fourier-space tilted generator can nevertheless exhibit an analogous BBP transition, now controlled by the behavior of $W(D)$ near its lower edge $D=0$ -- see Sec. VI of the supplementary material of Ref.~\cite{gueneau-majumdar-schehr:2025}.} support $[0, D_{\max}]$.
We assume that, near its upper edge,
\begin{align}
    W(D) \sim C (D_{\max} - D)^\nu \quad \text{as} \quad D \uparrow D_{\max} \label{eq:edge-expo-def}
\end{align}
where $C>0$ is a constant and $\nu>-1$ ensures normalizability.
Remarkably, the qualitative structure of the phase diagram governing the long-time large-deviation behavior of the model is entirely controlled by the right-edge exponent~$\nu$ of the distribution $W(D)$.
In the following, we study trajectories that produce a rare, ballistic displacement at long times, with the endpoint fixed at
\begin{align}
    x(t) = yt\,, \quad y = O(1)\,, \quad t \gg 1/r\, ,
    \label{eq:constraint-intro}
\end{align}
where $y=x(t)/t$ is the time-averaged velocity over $[0,t]$, which we occasionally refer to simply as ``velocity''.
It is sufficient to consider $y>0$, since the process~\eqref{eq:sd-def} is symmetric in law under the transformation $x\to -x$.

\subsection{Summary of Ref.~\cite{gueneau-majumdar-schehr:2025}: rate function and dynamical phase transitions}
\label{sec:summary-prl}

In this section, we summarize the main findings of Ref.~\cite{gueneau-majumdar-schehr:2025}, which focused mainly on obtaining the rate function of the SD process at large times. These results provide the basis for the analysis developed throughout the rest of the paper. Let $p_r(x,t) = \langle \delta(x(t) - x) \rangle$ denote the position probability density function (PDF) of the process~\eqref{eq:sd-def}. We define the
moment-generating function (MGF) of the displacement as the double-sided
Laplace transform
\begin{equation}
    \hat p_r(q,t)= \left \langle e^{q x(t)} \right \rangle = \int_{-\infty}^{\infty}e^{qx}p_r(x,t)\mathrm{d} x.
    \label{eq:generating-function-summary}
\end{equation}
At long times, the end-point PDF and the
generating function have the large-deviation forms
\begin{equation}
    p_r(yt,t)\asymp e^{-tI(y)},
    \qquad
    \hat p_r(q,t)\asymp e^{t\Psi(q)},
    \label{eq:large-deviation-summary}
\end{equation}
where $\asymp$ denotes log-asymptotic equivalence, i.e., up to pre-exponential factors. By analogy with equilibrium statistical mechanics, we call the ensemble ``microcanonical'' when the end point $x=yt$ is fixed (as in $p_r(x=yt,t)$ on the left in Eq.~(\ref{eq:large-deviation-summary})), while the ``canonical'' ensemble is the one on the right where the end point $x$ is allowed to fluctuate and a weight $e^{q x(t)}$ is associated to it (cf.\ Eq.~(\ref{eq:generating-function-summary})). The end point $x(t)$ plays the role of ``energy'' in this analogy. \\

The two large-deviation functions in
Eq.~\eqref{eq:large-deviation-summary} were obtained exactly in
Ref.~\cite{gueneau-majumdar-schehr:2025}, where the emphasis was on their mathematical
structure. In particular, around $q=0$, the scaled cumulant-generating function (SCGF) $\Psi(q)$ was expressed in terms of the $R$-transform of the diffusivity distribution $W(D)$, where the $R$-transform is the generating function of the free cumulants of $W(D)$~\cite{mingo-speicher:2017}.\\ 

\begin{figure}
\centering
\includegraphics[width = \textwidth]{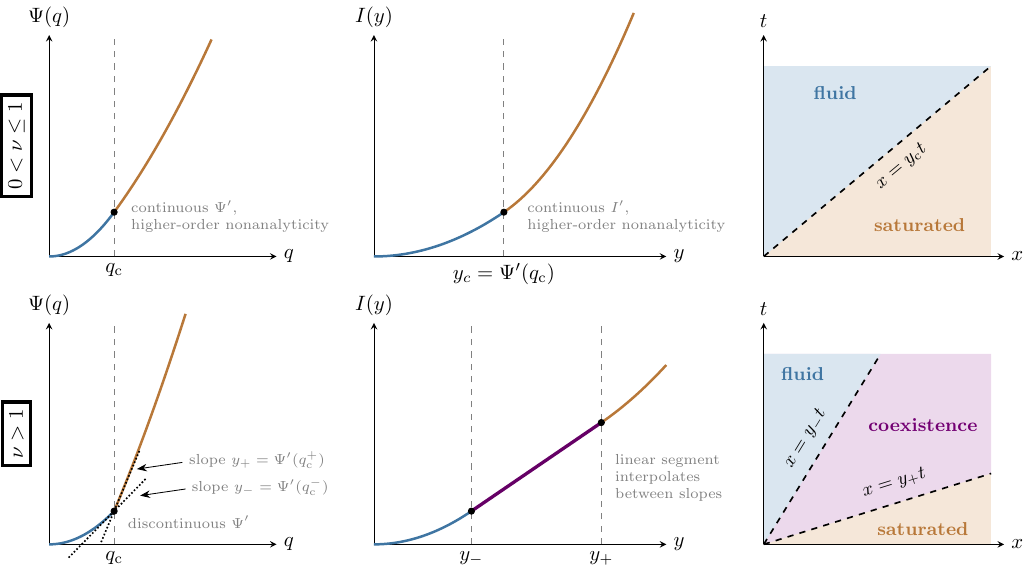}
\caption{Sketch of the SCGF $\Psi(q)$ (left column), rate function $I(y)$ (center column), and space-time diagrams of the SD displacement (right column), cf.\ Ref.~\cite{gueneau-majumdar-schehr:2025}.
\textbf{Top row.} Right edge exponent $\nu \in (0,1]$, where the SCGF dynamical phase transition is continuous. \textbf{Bottom row.} Right edge exponent $\nu > 1$, where the SCGF dynamical phase transition is first order, and an intermediate coexistence region appears for the displacement.
}
\label{fig:legendre}
\end{figure}

\begin{figure}
\centering
\includegraphics[width = .49 \textwidth]{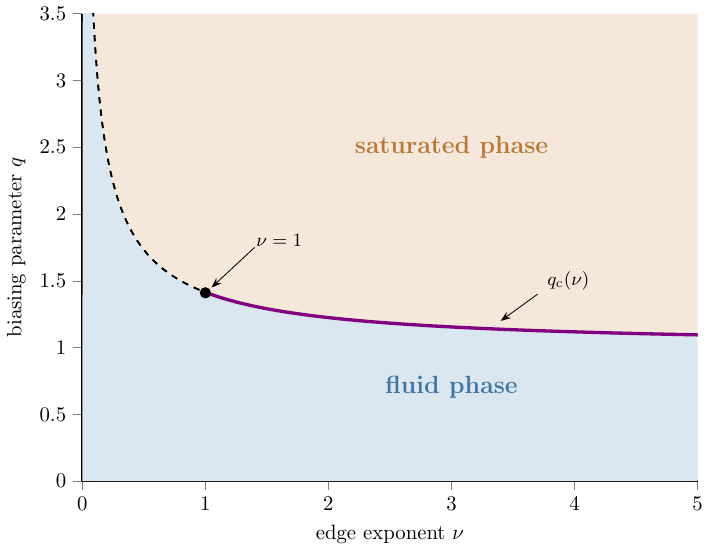}
\hfill
\includegraphics[width = .48\textwidth]{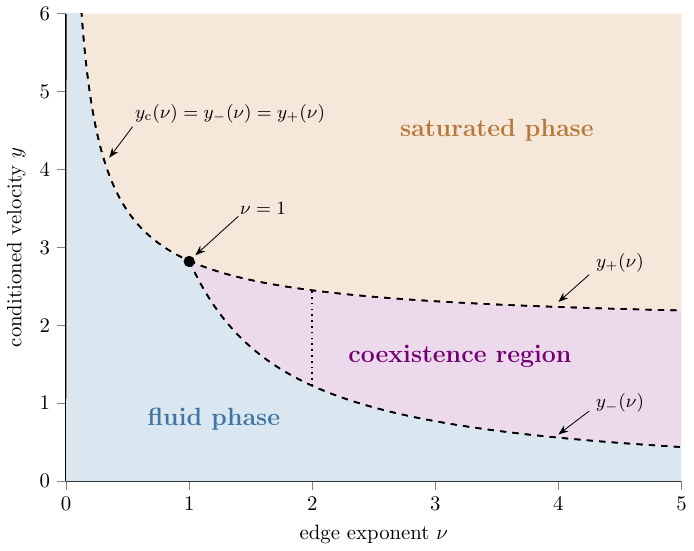}
\caption{Phase diagrams of the SCGF $\Psi(q)$ (left) and rate function $I(y)$ (right) for the example $W(D) = (\nu + 1) (1 - D)^\nu$ of the diffusion coefficient distribution (in units of $r = D_{\max} = 1$).  Dashed lines indicate continuous phase transitions, and solid lines are first-order phase transitions (consistent with the convention in Ref.~\cite{landau2013statistical}). The dotted vertical line at $\nu = 2$ in the phase diagram of $I(y)$ delimits the boundary between a normal ($\nu \geq 2$) and anomalous condensate $(1 < \nu < 2)$. The calculations to generate these phase diagrams, for the more general case of $W(D)$ a Beta distribution, are included in Appendix~\ref{app:beta-dist-phases}.}
\label{fig:phase-diag}
\end{figure}

We first recall the exact large-deviation results of
Ref.~\cite{gueneau-majumdar-schehr:2025} in more detail, which provide the framework for the physical
analysis below. In Ref.~\cite{gueneau-majumdar-schehr:2025}, these results were obtained
from a renewal approach. The natural
starting point is the SCGF $\Psi(q)$ defined in
Eq.~\eqref{eq:large-deviation-summary}: since $q$ is conjugate to the
displacement, increasing $q$ gives greater weight to trajectories with
larger positive displacement. For $q\geq0$ and
$\nu>0$, one gets~\cite{gueneau-majumdar-schehr:2025}
\begin{equation}
    \Psi(q)=
    \begin{cases}
      s^*(q), & 0\leq q\leq q_{\rm c},\\[1mm]
      D_{\max}q^2-r, & q\geq q_{\rm c}.
    \end{cases}
    \label{eq:canonical-summary}
\end{equation}
Here $s^*(q)$ denotes the regular solution, continuously connected to
$s^*(0)=0$, of
\begin{equation}
    1=r\int_0^{D_{\max}}
    \frac{W(D)\,\mathrm{d} D}{r+s^*(q)-q^2D}.
    \label{eq:fluid-branch-summary}
\end{equation}
The critical value $q_{\rm c}$ is reached when this solution touches the upper edge, $r+s^*(q_{\rm c})=q_{\rm c}^2D_{\max}$, yielding
\begin{eqnarray}\label{recapqc}
     q_{\rm c}^2=r\int_0^{D_{\max}}
    \frac{W(D)\mathrm{d} D}{D_{\max}-D}\, .
\end{eqnarray}
For $q>q_{\rm c}$, Eq.~\eqref{eq:fluid-branch-summary}
no longer has a regular solution, and the SCGF is instead given by
the edge branch $D_{\max}q^2-r$. We distinguish the following cases, based on the edge exponent $\nu$ in Eq.~\eqref{eq:edge-expo-def}:
\begin{itemize}
    \item $-1 < \nu \leq 0$: In this case, the edge integral~(\ref{recapqc}) defining $q_{\rm c}$ diverges. The regular solution branch thus persists for all $q$, and the SCGF $\Psi(q)$ remains smooth, with no transition at a finite $q$. (The same is then true for the rate function $I(y)$ below).

    \item $0 < \nu \leq 1$: In this case, there is a finite $q_{\rm c}$, and the SCGF $\Psi(q)$ in Eq.~\eqref{eq:canonical-summary} is differentiable at $q_{\rm c}$. In other words, this is a continuous phase transition. While the transition of $\Psi(q)$ is continuous, its order depends on $\nu$: Eq.~\eqref{eq:fluid-branch-summary} gives $r+s^*(q)-q^2D_{\max}\propto(q_{\rm c}-q)^{1/\nu}$ as $q\uparrow q_{\rm c}$ (with a logarithmic correction at $\nu=1$), so that $\Psi''(q_{\rm c}^-)$ diverges for $1/2<\nu\leq1$, $\Psi''$ jumps for $\nu=1/2$, and the singularity is of higher order for $0<\nu<1/2$, cf.\ Ref.~\cite{gueneau-majumdar-schehr:2025}.
    \item $\nu > 1$: In this case, there is a finite $q_{\rm c}$, and the SCGF $\Psi(q)$ in Eq.~\eqref{eq:canonical-summary} is continuous but not differentiable at $q_{\rm c}$. In other words, this is a first-order transition.
\end{itemize}
The latter two cases are sketched in Fig.~\ref{fig:legendre} (left) and Fig.~\ref{fig:phase-diag} (left). The names of the different phases in Figs.~\ref{fig:legendre} and~\ref{fig:phase-diag} refer to the structure of the corresponding trajectories, as summarized in Sec.~\ref{sec:main-results} below.\\

The two large-deviation functions in Eq.~\eqref{eq:large-deviation-summary} describe two different ensembles of rare
trajectories. In the canonical ensemble, governed by $\Psi(q)$, the endpoint remains free to fluctuate, while $q$ controls its
typical velocity $y=\Psi'(q)$ on each smooth branch of~(\ref{eq:canonical-summary}). Increasing $q$ therefore gives progressively more weight to
trajectories that display a large total displacement. The velocities selected
immediately below and above $q_{\rm c}$ are
\begin{equation}
    y_-= \Psi'(q_{\rm c}^-) = {s^*}'(q_{\rm c}^-),
    \qquad
    y_+= \Psi'(q_{\rm c}^+) =2q_{\rm c}D_{\max}.
    \label{eq:critical-velocities-summary}
\end{equation}
For $0 < \nu \leq 1$, one finds $y_- = y_+$, whereas for $\nu > 1$, $y_- < y_+$. In the microcanonical ensemble, one instead conditions directly on the fixed endpoint $x(t)=yt$, where the exponential decay of its PDF is governed by the rate
function $I(y)$ introduced in Eq.~\eqref{eq:large-deviation-summary}. The two
ensembles are related by the Legendre transform\footnote{See Appendix~\ref{app:rate-func-convex} for further discussion in view of the G{\"a}rtner--Ellis theorem and (lack of) differentiability of the SCGF $\Psi$ for $\nu > 1$. For $\nu > 1$ and  $y \in [y_-,y_+]$ the canonical ensemble at $q_{\rm c}$ cannot select $y$, which reflects the non-equivalence of path ensembles~\cite{touchette:2009,chetrite-touchette:2015}.}
\begin{equation}
    I(y)=\max_{q\geq0}\,[qy-\Psi(q)].
    \label{eq:legendre-summary}
\end{equation}
Let $q(y)$ be a value at which this maximum is reached, and let
$q_*(y)$ denote the solution on the regular branch,
${s^*}'(q_*(y))=y$. One then has 
\begin{equation}
    q(y)=
    \begin{cases}
      q_*(y), & 0\leq y\leq y_-,\\[1mm]
      q_{\rm c}, & y_-\leq y\leq y_+,\\[1mm]
      \tfrac{y}{2D_{\max}}, & y\geq y_+.
    \end{cases}
    \label{eq:q-of-y-summary}
\end{equation}
Hence the whole interval $y_-\leq y\leq y_+$ is associated with the single
conjugate variable $q_{\rm c}$. By evaluating Eq.~\eqref{eq:legendre-summary}, the rate function is\footnote{For $\nu > 1$, in the interval $y_-\leq y\leq y_+$, the position distribution then takes the traveling-front form~\cite{van2003front} $p_r(x,t)\asymp e^{-q_{\rm c}(x-vt)}$, with $v=y_+/2-2rD_{\max}/y_+$. We note that the expression for $v$ given in Ref.~\cite{gueneau-majumdar-schehr:2025} contains a typo, which is corrected here.}
\begin{equation}
    I(y)=
    \begin{cases}
      \phi_\nu(y), & 0\leq y\leq y_-,\\[1mm]
      \phi_\nu(y_-)+q_{\rm c}(y-y_-), & y_-\leq y\leq y_+,\\[1mm]
      r+\tfrac{y^2}{4D_{\max}}, & y\geq y_+,
    \end{cases}
    \label{eq:microcanonical-summary}
\end{equation}
where
$\phi_\nu(y)=q_*(y)y-s^*(q_*(y))$. 
For $0<\nu\leq1$, the slopes coincide, $y_-=y_+\equiv y_{\rm c}$, so that the
linear interval shrinks to a point and the transition at fixed displacement is direct and continuous. 
For $\nu > 1$, the rate function $I(y)$ in Eq.~\eqref{eq:microcanonical-summary} is continuously differentiable at both $y_-$ and $y_+$, i.e., displays only continuous phase transitions. See Fig.~\ref{fig:legendre} for qualitative sketches of the rate function (center), and the corresponding space-time phase diagram of the displacement (right). Furthermore, in Fig.~\ref{fig:phase-diag} (right), we show the phase diagram of the rate function as a function of the edge exponent $\nu$.

\subsection{Outline and summary of main results}
\label{sec:main-results}

\begin{figure}
\centering
\includegraphics[width = \textwidth]{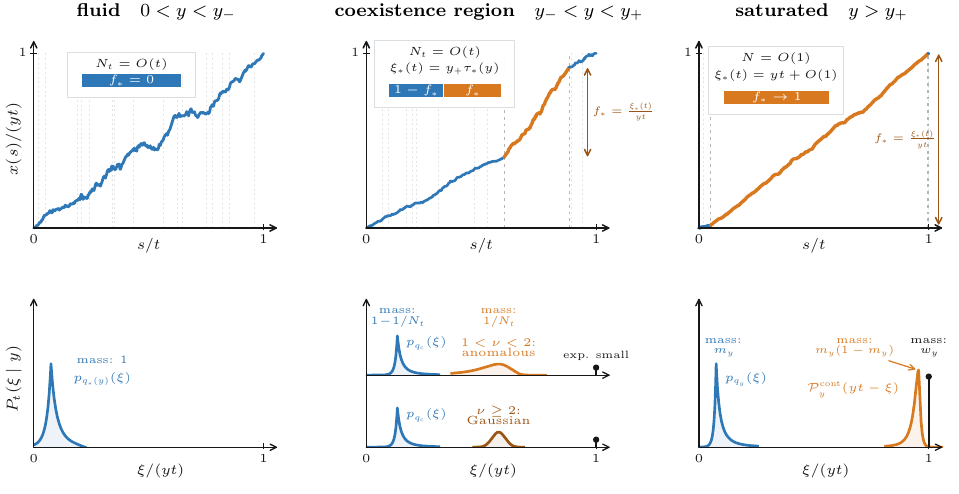}
\caption{Sketch of trajectory condensation for $\nu>1$ in the fluid, coexistence, and saturated phases, cf.\ Table~\ref{tab:phases}. \textbf{Top.} Representative conditioned trajectories $x(s)/(yt)$. The insets indicate the number of runs $N_t$ and the displacement fraction $f_*$ carried by the largest run.
\textbf{Bottom}. First-run
laws $P_t(\xi\mid y)$ on the complete axis $\xi/(yt)$. Blue, orange, and black denote the fluid sector, condensate sector, and the Dirac atom at $yt$ from trajectories without any switch. For their masses, see Sec.~\ref{sec:single-run-main-text}.
}
\label{fig:runs-dist}
\end{figure}

Our main results concern how the different regimes of the phase diagram in Fig.~\ref{fig:phase-diag} are realized at the level of trajectories:  they are summarized in Table~\ref{tab:phases} and Fig.~\ref{fig:runs-dist}. We begin our analysis by
establishing an exact mapping between the switching-diffusion problem and
the large deviations of a run-and-tumble particle (RTP) in Sec.~\ref{sec:rtp}, which will be useful for subsequent calculations and understanding the dynamical phase transitions.
Our results then are:\\

\noindent\textbf{Diffusivity BBP transition (Sec.~\ref{sec:rmt}).} In order to elucidate the connection to random-matrix theory (RMT), we replace the continuous stochastic process $D(t)$ by a Markov jump process where the diffusion coefficient takes $M$ discrete values $D_1,\ldots,D_M$ each drawn independently from $W(D)$. In this case, the evolution of $\langle e^{qx(t)}\rangle$, resolved by the current diffusivity, is generated by a random $(M\times M)$ diagonal matrix with a rank-one perturbation due to switching, whose largest eigenvalue gives $\Psi(q)$. As $M\to\infty$, this eigenvalue undergoes a BBP transition~\cite{baik-ben-arous-peche:2005,benaych-georges-nadakuditi:2011} at exactly $q=q_{\rm c}$, traversed in reverse since the strength $r/q^2$ of the rank-one perturbation decreases with $q$. For $q<q_{\rm c}$, an extensive set of diffusivities contributes collectively to the generating function, while for $q>q_{\rm c}$ only diffusivities close to $D_{\max}$ are selected (Fig.~\ref{fig_BBP_Cauchy}). We note that similar rank-one calculations have appeared in related models~\cite{garrahan-manai-warzel:2023,bernard-bouchaud-le-doussal:2026}, albeit without explicit reference to the BBP transition. The natural order parameter, the squared overlap $\chi(q)$ between the top eigenvector and the switching direction (Fig.~\ref{fig:chi-fstar}, left), jumps at $q_{\rm c}$ for $\nu>1$ and vanishes continuously for $0<\nu\leq1$. Its jump fixes the width $y_+-y_-$ of the coexistence interval of $I(y)$, which links diffusivity selection in the canonical ensemble to condensation at fixed displacement. Overall, Sec.~\ref{sec:rmt} hence clarifies the relation of the results of Ref.~\cite{gueneau-majumdar-schehr:2025} to random-matrix theory, which answers a question left open in that work.\\

\noindent\textbf{Single-run distribution and dynamical condensation (Sec.~\ref{sec:single-run-main-text}).} We condition on $x(t)=yt$ and study the law $P_t(\xi\mid y)$ of the displacement $\xi$ accumulated during the first switching interval (a ``run''). All results follow from an exact renewal equation for $P_t(\xi\mid y)$, combined with the precise (prefactor-level) asymptotics of the endpoint density. In the \emph{fluid phase}, $0<y<y_-$, the displacement is shared among $O(t)$ microscopic runs. In the \emph{coexistence phase}, $y_-<y<y_+$, which only exists for $\nu>1$, the microscopic background freezes at a critical law with algebraic tail $\xi^{-(\nu+1)}$. As in run-and-tumble particles and the single-big-jump principle~\cite{vezzani-barkai-burioni:2019,vezzani-barkai-burioni:2020,gradenigo-majumdar:2019,mori-gradenigo-majumdar:2021}, this heavy tail allows a single macroscopic run, the condensate, to absorb the excess displacement, while the background moves at the critical velocity $y_-$. The condensate fluctuates according to a completely asymmetric stable law (``anomalous condensate'') for $1<\nu<2$, and a Gaussian (``normal condensate'') for $\nu\geq2$, mirroring the condensates of mass transport models~\cite{majumdar-evans-zia:2005,evans-majumdar-zia:2006} and RTPs~\cite{mori-le-doussal-majumdar-etal:2021}. In the \emph{saturated phase}, $y>y_+$, the trajectory contains only $O(1)$ runs, and a single run carries the whole displacement up to $O(1)$. At the level of the first-run law, the transition at $y_+$ shows up as a change of the condensate weight from $O(1/t)$ to $O(1)$.\\

\noindent\textbf{Numerical validation (Sec.~\ref{sec:numerics}).} Finally, we sample exactly conditioned trajectories by reversible-jump MCMC~\cite{green:1995}, and solve the renewal equation for $P_t(\xi\mid y)$ numerically at finite $t$. Both numerical approaches agree with each other, and the asymptotic predictions converge toward the full solution in all phases, see Fig.~\ref{fig:cond-mcmc-vs-theory}. In the coexistence phase, convergence to the limiting shape of the condensate peak is slow, especially for the anomalous condensate.\\

We conclude in Sec.~\ref{sec:outlook}, and technical details are collected in the appendices. These include the convexity of $I(y)$ (Appendix~\ref{app:rate-func-convex}), phase boundaries for Beta-distributed $W(D)$ (Appendix~\ref{app:beta-dist-phases}), the RTP mapping (Appendix~\ref{app:rto-to-sd}), the asymptotic analysis of the single-run distribution (Appendix~\ref{SingleRunAppendix}), and our numerical methods (Appendices~\ref{app:numerical-algorithm} and~\ref{app:renewal-numerics}).

\begin{table}
\caption{Summary of the three dynamical phases at fixed displacement $x(t)=yt$ when $\nu>1$, with $y$ denoting the time-averaged velocity of the particle. For $0<\nu\leq1$, one has $y_-=y_+$ and the coexistence phase is absent. Here, ``runs'' are the intervals between switches, $f_*$ is the fraction of the displacement carried by the largest run, see Eq.~\eqref{eq:condensate-fraction} below, and $P_t(\xi\mid y)$ is the conditional law of the displacement of the first run, see Eq.~\eqref{eq:Pt-def} below.}
\label{tab:phases}
\renewcommand{\arraystretch}{1.25}
\begin{ruledtabular}
\begin{tabular}{>{\raggedright\arraybackslash}p{0.22\textwidth}>{\raggedright\arraybackslash}p{0.22\textwidth}>{\raggedright\arraybackslash}p{0.26\textwidth}>{\raggedright\arraybackslash}p{0.22\textwidth}}
 & \textbf{fluid} & \textbf{coexistence} & \textbf{saturated} \\
\colrule
velocity $y$ & $0<y<y_-$ & $y_-<y<y_+$ & $y>y_+$ \\
conjugate parameter $q(y)$ & $q_*(y)<q_{\rm c}$ & pinned at $q_{\rm c}$ & $y/(2D_{\max})>q_{\rm c}$\\
\colrule
number of runs & $O(t)$ & $O(t)$ plus one macroscopic & $O(1)$ \\
largest-run fraction $f_*$ & $0$ & $\in(0,1)$ & $1$ \\
microscopic runs (Sec.~\ref{sec:fluid}) & exponential tails & critical, tail $\xi^{-(\nu+1)}$ & tail $\xi^{-(\nu+1)}$, total mass $<1$ \\
condensate bump weight in $P_t(\xi\mid y)$ & exponentially small & $O(1/t)$ & $O(1)$ \\
condensate bump fluctuations (Sec.~\ref{sec:condensate}) & --- & stable, width $t^{1/\nu}$ ($1<\nu<2$); marginal, $(t \log t)^{1/2}$ ($\nu = 2$); Gaussian, width $t^{1/2}$ ($\nu>2$) & $O(1)$ \\
\end{tabular}
\end{ruledtabular}
\end{table}

\section{Mapping to a run-and-tumble particle}
\label{sec:rtp}

We start by showing that the results summarized in Sec.~\ref{sec:summary-prl} can also be obtained through an exact mapping to a one-dimensional RTP
tumbling at the same Poisson rate $r$,
with a new positive velocity $v$ drawn independently from the distribution $W(v)$ after each tumble (cf.\ Ref.~\cite{mori-le-doussal-majumdar-etal:2021}). Indeed, by conditioning the SD on a specific realization of the diffusion coefficient history $(D(s))_{s \in [0,t]}$, the conditional random variable $x(t) \mid (D(s))_{s \in [0,t]}$ is Gaussian with zero mean and variance $2 A(t) := 2 \int_0^t D(s) \mathrm{d} s$. By the law of total expectation, 
the MGF in Eq.~\eqref{eq:generating-function-summary} then becomes
\begin{equation}
   \hat p_{r, \rm SD}(q,t)= \left \langle e^{q x(t)} \right \rangle = \left \langle \left \langle e^{q x(t)} \mid (D(s))_{s \in [0,t]} \right \rangle \right \rangle_{(D(s))_{s \in [0,t]}} = \left\langle\exp\!\left(q^2\int_0^tD(s)\mathrm{d} s\right)\right\rangle,
    \label{eq:moment-identity}
\end{equation}
where $\langle \cdots\rangle_{(D(s))_{s \in [0,t]}}$ stands for 
the remaining average over the diffusion coefficient history $(D(s))_{s \in [0,t]}$. In Eq.~\eqref{eq:moment-identity}, the last step follows by a simple Gaussian integration.
Here, we write~$\hat{p}_{r,{\rm SD}}$ instead of just~$\hat{p}_r$, as in Eq.~\eqref{eq:generating-function-summary}
and elsewhere, to distinguish the SD generating function from its RTP counterpart. Note that $A(t)=\int_0^t D(s)\mathrm{d} s = D_1\tau_1 +D_2\tau_2+\cdots$ is precisely the distance traveled by an RTP until time $t$, with tumble rate $r$, run duration $\tau_i$, and velocities $D_i$ drawn from $W(D)$. 
The rare-displacement problem of the SD~\eqref{eq:sd-def} is therefore directly linked to the
displacement statistics of the RTP, up to evaluating the SCGF of the latter at squared biasing parameter $q^2$. Similarly, by conditioning, the displacement PDF of the SD is
\begin{align}
p_{r, \rm SD}(x,t) = \int_0^{D_{\max} t} p_{r,\rm RTP}(A,t)\, \frac{1}{\sqrt{4 \pi A}} \, e^{-\frac{x^2}{4 A}} \mathrm{d} A\,.
\label{eq:sd-vs-rtp-pdf-relation}
\end{align}
The integration in Eq.~\eqref{eq:sd-vs-rtp-pdf-relation} runs from $A = 0$ to $A = D_{\max}t$, as the latter is the maximum displacement that can be achieved by the RTP since its velocity is bounded by $D_{\max}$.
Now suppose the RTP displacement PDF has the long-time large deviation form $p_{r, {\rm RTP}}(A,t) \asymp \exp(-t I_{\rm RTP}(A / t))$ where $I_{{\rm RTP}}(a = A / t)$ denotes its rate function. Then, we insert this expression into Eq.~\eqref{eq:sd-vs-rtp-pdf-relation} and evaluate the SD displacement PDF at $x = yt$ for large $t$ via Laplace's method, i.e., the saddle point method:
\begin{align}
    p_{r, \rm SD}(yt,t) = \int_0^{D_{\max}} p_{r,\rm RTP}(at,t)\, \frac{t}{\sqrt{4 \pi at}} \, e^{-\frac{(yt)^2}{4 at}} \mathrm{d} a \asymp \int_0^{D_{\max}} \exp\left(-t \left(I_{\rm RTP}(a) + \frac{y^2}{4a} \right) \right) \mathrm{d} a \asymp \exp \left(-t I_{\rm SD}(y) \right)\,.
\end{align}
This hence implies the relation
\begin{align}
I_{\rm SD}(y) = \min_{0\leq a \leq D_{\max}} \left( I_{\rm RTP}(a) + \frac{y^2}{4a}\right)\,,
\label{eq:sd-rtp-rf}
\end{align}
between the rate functions of the RTP and SD displacement.
We verify in Appendix~\ref{app:rto-to-sd} that Eq.~\eqref{eq:sd-rtp-rf} indeed recovers the results for the SD rate function from Ref.~\cite{gueneau-majumdar-schehr:2025}, as summarized in Sec.~\ref{sec:summary-prl}, from those of Ref.~\cite{mori-le-doussal-majumdar-etal:2021} 
for the RTP rate function (which we briefly re-derive for completeness).
The connection between SD and RTP is particularly
suggestive since for $\nu>1$, the RTP is known to realize a sufficiently large
displacement through a condensate carried by one run~\cite{mori-le-doussal-majumdar-etal:2021},
thereby anticipating the trajectory condensation studied below. Note that in Ref.~\cite{mori-le-doussal-majumdar-etal:2021}, the shape of the condensate PDF for the RTP was only analyzed in the ``fixed-$N$'' ensemble, where one fixes the total number of switches instead of the total time $t$ and considers the limit $N \to \infty$. This is in contrast to our analysis in this paper, which proceeds directly in the ``fixed-$t$'' ensemble.

\section{Diffusivity BBP transition for the scaled cumulant-generating function}
\label{sec:rmt}

Here, we derive and interpret the dynamical phase transition
seen in the SCGF $\Psi(q)$, as shown via Eqs.~\eqref{eq:canonical-summary}, \eqref{eq:fluid-branch-summary} and Fig.~\ref{fig:phase-diag} (left), as an instance of the well-known BBP transition~\cite{baik-ben-arous-peche:2005} of random matrix theory.
We compute the SCGF of the SD in Eq.~\eqref{eq:sd-def} as
\begin{align}
    \Psi(q) = \lim_{t \to \infty} \frac{1}{t} \log \hat{p}_r(q,t) = \lim_{t \to \infty} \frac{1}{t} \log\left\langle\exp\!\left(q^2\int_0^tD(s)\mathrm{d} s\right)\right\rangle,
    \label{eq:scgf-limit}
\end{align}
where we have used Eq.~\eqref{eq:moment-identity}. The average in Eq.~\eqref{eq:scgf-limit} is over the switching history. In the following, we first replace $W(D)$ by $M$ fixed values $D_1,\ldots,D_M$ drawn independently from $W(D)$,
and consider the corresponding Markov jump process on the finite state space $\{D_1, \dots, D_M\}$ with uniform jump probabilities. As detailed below, we determine the top eigenvalue of the tilted generator of this process to compute the SCGF\footnote{Note that this corresponds to the Donsker--Varadhan approach~\cite{donsker-varadhan:1975} for long-time large deviations (see Ref.~\cite{touchette:2018} for an introduction).}, and take the limit $M \to \infty$ for the ``disorder'' in the end.

\subsection{Random-matrix formulation}

Since we know from Sec.~\ref{sec:rtp} that it suffices to study the RTP process $A(t) = \int_0^t D(s) \mathrm{d} s$, we first focus on its statistics here. We denote by  $p_{r, {\rm RTP}}(A,t)$ its PDF at time $t$. After discretizing the state space of $(D(s))_{s \in [0,t]}$ into $M$ fixed states $D_1, \dots, D_M$ drawn independently from $W(D)$,
we may decompose this PDF as
\begin{align}
    p_{r, {\rm RTP}}(A,t)\mathrm{d}A = \mathbb{P}(A(t)  \in [A, A + \mathrm{d}A] ) = \sum_{i=1}^M \mathbb{P}(A(t)  \in [A, A + \mathrm{d}A], \; D(t) = D_i) =: \sum_{i=1}^M P_{r,i}(A,t)\mathrm{d}A
\end{align}
where $P_{r,i}(A,t)$ is the
joint probability density that $A(t) \in [A, A + \mathrm{d}A]$, and the diffusivity at time $t$ is in the state $D_i$. These probabilities obey the system of evolution equations
\begin{equation}
    \partial_t P_{r,i}=-D_i\partial_A P_{r,i}-rP_{r,i}+\frac{r}{M}\sum_{j=1}^M P_{r,j}\,, \qquad P_{r,i}(A,0)= \delta(A) \frac{1}{M} \,.
    \label{eq:finite-fp}
\end{equation}
Here, the individual terms on the right-hand side (RHS) can be understood as follows: The first term $-D_i\partial_A P_{r,i}$ is a transport term due to $A(t)$ growing at rate $D_i$. The second term $-rP_{r,i}$ captures the switching at rate $r$ from the current state $D_i$ to a new one. Lastly, $+\frac{r}{M}\sum_{j=1}^M P_{r,j}$ corresponds to switching to state $D_i$ from any of the other states (including $D_i$ itself, if it is redrawn). Importantly, since the frozen states $D_1, \dots, D_M$ were drawn from $W(D)$, the switching distribution is uniform on $\{D_1, \dots, D_M\}$, and hence all terms in $+\frac{r}{M}\sum_{j=1}^M P_{r,j}$ have equal weights. We now introduce the two-sided Laplace transform of each of the $P_{r,i}$'s, evaluated at $q^2$:
\begin{equation}
    \hat p_{r,i}(q,t)=\int_{-\infty}^{\infty}e^{q^2 A}P_{r,i}(A,t)\mathrm{d} A,
    \qquad
    \hat p_r^{(M)}(q,t):=\sum_{i=1}^M\hat p_{r,i}(q,t).
    \label{eq:resolved-generating-function}
\end{equation}
Thus $\hat p_{r,i}(q,t)$ is the contribution to $\hat p_r^{(M)}(q,t)$ from
trajectories whose diffusivity at time $t$ is $D_i$.  We stress that $\hat{p}_r^{(M)}(q,t)$ as defined in Eq.~\eqref{eq:resolved-generating-function} indeed coincides with Eq.~\eqref{eq:generating-function-summary} for the SD process (in the large $M$ limit) by virtue of Eq.~\eqref{eq:moment-identity}, i.e., $\hat p_r^{(M)}(q,t) \to \hat p_r(q,t)$ as $M \to \infty$.
From Eq.~\eqref{eq:finite-fp}, we find by taking the Laplace transform that
\begin{align}
   \partial_t \hat p_{r,i}=q^2 D_i \hat p_{r,i}-r\hat p_{r,i}+\frac{r}{M}\sum_{j=1}^M \hat p_{r,j}.
   \label{eq:finite-fp-laplace-trafo}
\end{align}
We now rewrite this system of evolution equations~\eqref{eq:finite-fp-laplace-trafo} in vector form. The column vector
$\widehat{\bm p}_r=(\hat p_{r,1},\ldots,\hat p_{r,M})^\top$ obeys
$\partial_t\widehat{\bm p}_r=\bm H(q)\widehat{\bm p}_r$, where the so-called ``tilted generator''~\cite{touchette:2018} is given by
\begin{equation}
    \bm H(q)=q^2\bm D-r\bm I+r\bm e\bm e^\top,
    \qquad
    \bm e=\frac{1}{\sqrt M}\begin{pmatrix}
        1\\\vdots\\1
    \end{pmatrix} \in \mathbb{R}^M \;.
    \label{eq:auxiliary-matrix}
\end{equation}
Here, we denote $\bm D=\mathrm{diag}(D_1,\ldots,D_M)$, while $\bm I$ is the $M \times M$ identity matrix. Let $\lambda_1 > \lambda_2 \geq \dots \geq \lambda_M$ be the eigenvalues of $\bm H(q)$, with orthonormal eigenvectors $\bm v_1, \dots, \bm v_M$. Using these, we can write the solution of~\eqref{eq:finite-fp-laplace-trafo} as
\begin{align}
    \widehat{\bm p}_r(q,t) = \exp \left(\bm H(q)t \right) \widehat{\bm p}_r(q,0) = \sum_{k = 1}^M e^{\lambda_k t} [\bm v_k^\top \widehat{\bm p}_r(q,0)] \bm v_k \overset{t \to \infty}{\sim} e^{\lambda_1 t} [\bm v_1^\top \widehat{\bm p}_r(q,0)] \bm v_1
    \label{eq:widehat-p-solution}
\end{align}
as long as $\bm v_1^\top \widehat{\bm p}_r(q,0) \neq 0$. Note that from Eqs.~\eqref{eq:finite-fp} and~\eqref{eq:resolved-generating-function}, we have $\hat p_{r,i}(q,0)=\mathbb{P}(D(0)=D_i)=1/M$
for all $i$, i.e., $\widehat{\bm p}_r(q,0)  = \bm e / \sqrt{M}$. Let us denote the squared \textit{overlap} between the leading eigenvector $\bm v_1$ of $\bm H(q)$, and the direction $\bm e$, by
\begin{align}
    \chi^{(M)}(q):= |\bm v_1^\top \bm e|^2\,.
\end{align}
Here, the superscript $(M)$ again indicates the size of the involved vectors and matrices (and we write the absolute value $|\cdot|$ for analogy with quantum mechanics). With that, we find from Eqs.~\eqref{eq:finite-fp-laplace-trafo} and~\eqref{eq:widehat-p-solution}
\begin{align}
\hat p_r^{(M)}(q,t)=\sum_{i=1}^M\hat p_{r,i}(q,t) = \sqrt{M}\, \bm e^\top \widehat{\bm p}_r(q,t) \overset{t \to \infty}{\sim} \chi^{(M)}(q) e^{\lambda_1 t}\,.
\end{align}
Consequently, the SCGF of the discretized process is
\begin{align}
    \Psi^{(M)}(q) = \lim_{t \to \infty} \frac{1}{t} \log \hat{p}_r^{(M)}(q,t) = \lambda_1\,.
\end{align}
In other words, the largest eigenvalue $\lambda_1(\bm H(q))$ of the
evolution operator~\eqref{eq:auxiliary-matrix} gives the SCGF, while its associated eigenvector
$\bm v_1$ resolves the contributions of the different diffusivity
states. More precisely, since $\bm H(q)$ is symmetric, its left and right leading eigenvectors coincide, so that $v_{1,i}^2$ is the long-time fraction of time that the tilted (Doob) process~\cite{chetrite-touchette:2015} spends in state $D_i$, whereas $v_{1,i}$ itself is, by Eq.~\eqref{eq:eigenvector-finite} below, proportional to the probability $r/[M(r+\Psi^{(M)}(q)-q^2D_i)]$ that a single run of the tilted process has diffusivity $D_i$, cf.\ Sec.~\ref{sec:fluid-duration}.
We emphasize that the spectral language used below refers only
to the evolution of these diffusivity-resolved generating functions, not
to the original particle dynamics.\\

We now re-organize Eq.~\eqref{eq:auxiliary-matrix} to make the connection with the BBP transition transparent and rewrite
\begin{equation}
   \tilde{\bm H}(q) := \frac{\bm H(q)+r\bm I}{q^2}
    =\bm D+\kappa(q)\bm e\bm e^\top,
    \qquad \kappa(q)=\frac{r}{q^2}.
    \label{eq:bbp-form}
\end{equation}
Here, ${\bm D}$ is a random (diagonal) $M \times M$ matrix whose limiting spectral density has compact support $[0, D_{\max}]$, and the rank-one perturbation on the RHS is due to the switching; in the finite-$M$ representation, a
new diffusivity is chosen uniformly among the $M$ states. In general, the eigenvalues of the unperturbed matrix ${\bm D}$ and the perturbed matrix $\tilde{\bm H}(q)$ are interlaced, with the largest eigenvalue of $\tilde{\bm H}(q)$ lying above the largest eigenvalue of ${\bm D}$ for any finite matrix size~$M$. The question is whether this largest eigenvalue of $\tilde{\bm H}(q)$ tends to $D_{\max}$ as $M \to \infty$, i.e., is absorbed by the ``bulk'', or whether it stays isolated, i.e., produces an ``outlier''. In the usual BBP
scenario~\cite{baik-ben-arous-peche:2005}, increasing the strength of the rank-$1$ perturbation above a critical threshold creates an
isolated eigenvalue whose eigenvector acquires a finite projection onto the perturbation direction.
In Eq.~\eqref{eq:bbp-form}, the effective ``spike strength'' $\kappa(q)=r/q^2$ in fact decreases when $q$ increases. The isolated collective branch is therefore lost, rather than created, and
the transition is traversed in the reverse direction. Note that, from Eq.~(\ref{eq:moment-identity}), we have
$\hat p_{r,{\rm SD}}(q,t)=\hat p_{r,{\rm RTP}}(q^2,t)$, and hence, at the level of the SCGF,
$\Psi_{\rm SD}(q)=\Psi_{\rm RTP}(q^2)$. Thus, the same BBP mechanism is already present in the RTP model in the fixed-$t$ ensemble, upon identifying the RTP Laplace parameter with $q^2$. We stress that, as $M\to\infty$, only the limiting spectral density $W(D)$ of $\bm D$ enters, so that the randomness of $\bm D$ plays no essential role. We also note that structurally, the mechanism is analogous to Bose--Einstein condensation~\cite{gunton-buckingham:1968,camacho-guardian-kleebank-vewinger-etal:2026}, with $(r+\Psi(q))/q^2$ playing the role of the chemical potential.

\subsection{Outlier criterion and critical point}

\begin{figure}
    \centering
\includegraphics[width=0.9\linewidth]{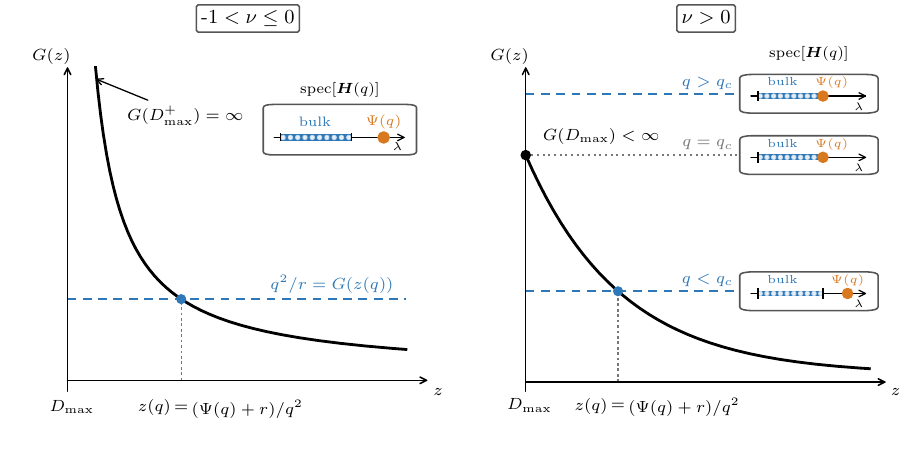}
    \caption{Graphical interpretation of the diffusivity BBP transition. The secular equation
$q^{2}/r=G(z(q))$, with $z(q)=(\Psi(q)+r)/q^{2}$, determines whether the leading eigenvalue is detached from or pinned to the upper edge of the bulk. \textbf{Left.} For
$-1<\nu\leq0$, we have $G(D_{\max}^{+})=\infty$, so the horizontal level
$q^{2}/r$ always intersects $G(z)$ at $z>D_{\max}$ and the leading
eigenvalue $\Psi(q)$ remains detached from the bulk. \textbf{Right.} For $\nu>0$,
$G(D_{\max})<\infty$ defines a critical value through
$q_{c}^{2}/r=G(D_{\max})$.
The insets illustrate the spectrum of the matrix $\bm H(q)$ in Eq.~\eqref{eq:auxiliary-matrix}.
}
    \label{fig_BBP_Cauchy}
\end{figure}

Let $\bm v_1(q)$
be the normalized eigenvector of $\bm H(q)$ associated with the largest
eigenvalue $\Psi^{(M)}(q)$, such that $\bm H(q) \bm v_1(q) = \Psi^{(M)}(q) \bm v_1(q)$, and $\tilde{\bm H}(q) \bm v_1(q) = \frac{\Psi^{(M)}(q) + r}{q^2} \bm v_1(q)$. On the branch where the top eigenvalue $\Psi^{(M)}(q)$ is separated
from the continuous band, we can use the rank-$1$ structure of Eq.~\eqref{eq:bbp-form} to simplify its eigenvalue equation, which is known as the secular equation for rank-$1$ perturbed matrices. We have
\begin{align}
    \tilde{\bm H}(q) \bm v_1(q) = \frac{\Psi^{(M)}(q) + r}{q^2} \bm v_1(q) = {\bm D}  \bm v_1(q) + \kappa(q)  [\bm e^\top  \bm v_1(q)] \bm e \quad \Leftrightarrow \quad \left(\frac{\Psi^{(M)}(q) + r}{q^2} \bm I - \bm D\right) \bm v_1(q) =  \kappa(q)  [\bm e^\top  \bm v_1(q)] \bm e\,.
\end{align}
From the last expression, we obtain the implicit expression
\begin{align}
    \bm v_1(q) =  \kappa(q)  [\bm e^\top  \bm v_1(q)]  \left(\frac{\Psi^{(M)}(q) + r}{q^2} \bm I - \bm D\right)^{-1} \bm e \quad \text{or} \quad v_{1,i}(q)=\frac{r\,[\bm e^\top \bm v_1(q)]}
    {\sqrt M\,[r+\Psi^{(M)}(q)-q^2D_i]}
    \label{eq:eigenvector-finite}
\end{align}
for the eigenvector. Taking the inner product of Eq.~\eqref{eq:eigenvector-finite} with $\bm e$ and canceling out $\bm e^\top  \bm v_1(q)$ on both sides finally results in the secular equation
\begin{align}
    1 = \kappa(q) \bm e^\top \left(\frac{\Psi^{(M)}(q) + r}{q^2} \bm I - \bm D\right)^{-1} \bm e =\frac{r}{M}\sum_{i=1}^M
    \frac{1}{r+\Psi^{(M)}(q)-q^2D_i}.
    \label{eq:secular-finite}
\end{align}
Taking $M\to\infty$ for Eq.~\eqref{eq:secular-finite} and supposing $\Psi(q) = \lim_{M \to \infty} \Psi^{(M)}(q)$ yields (cf.~Eq.~\eqref{eq:fluid-branch-summary})
\begin{equation}
    1=r\int_0^{D_{\max}}
    \frac{W(D)\,\mathrm{d} D}{r+\Psi(q)-q^2D}.
    \label{eq:secular}
\end{equation}
In terms of the Cauchy transform of $W$, $G(z) = \int \tfrac{W(D)}{z-D} \mathrm{d} D$, this can be written in the familiar form
\begin{align}
    \frac{1}{\kappa(q)} = \frac{q^2}{r} = G \left(\frac{\Psi(q) + r}{q^2} \right)\,.
    \label{eq:cauchy}
\end{align}
A graphical interpretation of Eq.~\eqref{eq:cauchy} and the resulting
diffusivity BBP transition are illustrated in Fig.~\ref{fig_BBP_Cauchy}. For the top eigenvalue to be isolated, the argument $z = (\Psi(q) + r)/q^2$ in the Cauchy transform must strictly exceed~$D_{\max}$. Because $G(z)$ decreases monotonically for $z > D_{\max}$, the maximum possible value the RHS of Eq.~\eqref{eq:cauchy} can take is at the bulk edge: $G(D_{\max})$. Hence, for any $W(D)$ with right edge exponent $\nu \in (-1,0]$, since $G(D_{\max}) \to \infty$ diverges, the top eigenvalue of the perturbed matrix will always be isolated for any $\kappa > 0$ (corresponding to $q_{\rm c} = \infty$ below), whereas for edge exponents $\nu > 0$, the integral converges, yielding a finite $q_{\rm c}$ at which the top eigenvalue starts to stick to the bulk.\\

Expanding further on this, for $q<q_{\rm c}$, the largest eigenvalue of $\bm H(q)$ lies above the continuous band in the limit $M \to \infty$ and its
eigenvector retains a finite projection onto the uniform direction
$\bm e$. It is hence extended over an extensive set of diffusivity states, which
therefore contribute collectively to the generating function. Interestingly, localization onto the spike in the random-matrix theory sense hence corresponds to \textit{delocalization} over the diffusivities in the physical sense in our model. The isolated top eigenvalue reaches the upper edge of the band at
\begin{equation}
    \frac{1}{\kappa(q_{\rm c})} = G(D_{\max}) \quad \Leftrightarrow \quad q_{\rm c}^2=r\int_0^{D_{\max}}
    \frac{W(D)\mathrm{d} D}{D_{\max}-D},
    \label{eq:critical-q}
\end{equation}
which is finite for $\nu>0$ as argued above. For $q>q_{\rm c}$, the largest eigenvalue sits at the
edge of the bulk of $\bm H(q)$ (cf.~Eq.~\eqref{eq:canonical-summary}), so that
$
    \Psi(q)=D_{\max}q^2-r.
$
At finite $M$, its eigenvector is then concentrated on the components
associated with the largest sampled diffusivities; as $M\to\infty$, these
diffusivities approach $D_{\max}$. Thus the leading eigenvector of the
generating-function evolution changes from an extended structure involving
many diffusivities to one concentrated near the upper edge $D_{\max}$, and in this reversed BBP transition, the collective
contribution of many diffusivities gives way to diffusivity selection as $q$ increases past $q_{\rm c}$.

\subsection{Diffusivity selection as spike overlap}\label{overlapsection}

\begin{figure}
    \centering
    \includegraphics[width=0.66\linewidth]{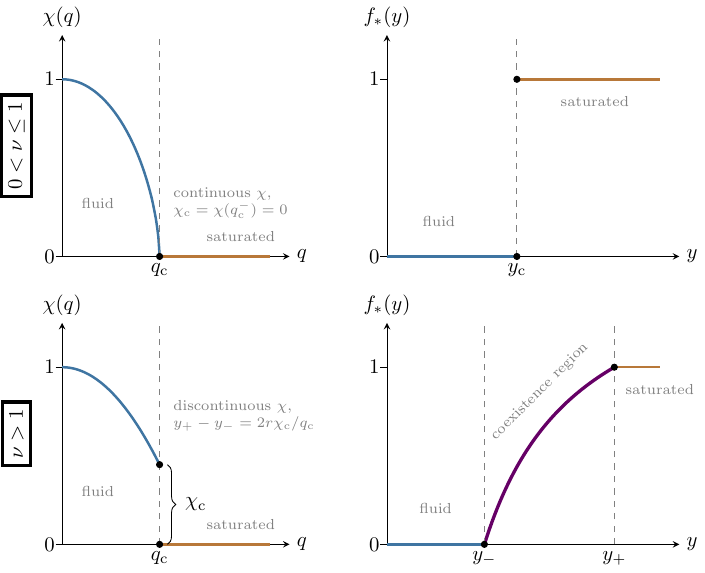}
    \caption{\textbf{Left.} Squared overlap $\chi(q) = \lim_{M \to \infty} |\bm e^\top \bm v_1(q)|^2$ for different edge exponents $\nu$. This serves as an order parameter for the BBP transition, cf.\ Refs.~\cite{benaych-georges-nadakuditi:2011,bocchi-biroli-cammarota-etal:2026}. \textbf{Right.} Fraction $f_*(y)$ of the total displacement carried by the single macroscopic run, cf.\ Eq.~\eqref{eq:condensate-fraction}.}
    \label{fig:chi-fstar}
\end{figure}

Because the spike vector $\bm e$ gives equal weight to all diffusivity states, the squared overlap
$\chi^{(M)}(q)=|\bm e^\top\bm v_1(q)|^2$ between the spike and the top eigenvector $\bm v_1(q)$ measures the collective contribution associated
with switching among many diffusivities. Normalizing
Eq.~\eqref{eq:eigenvector-finite} gives
\begin{equation}
    1= r^2\chi^{(M)}(q)\frac{1}{M}\sum_{i=1}^M
    \frac{1}{\left(r+\Psi^{(M)}(q)-q^2D_i \right)^2}.
    \label{eq:overlap-normalization}
\end{equation}
Therefore, on the separated branch, the overlap in the continuum limit $M \to \infty$ is
\begin{equation}
\begin{aligned}
    \chi(q)=\lim_{M\to\infty}\chi^{(M)}(q)=\left[r^2\int_0^{D_{\max}}
    \frac{W(D)\mathrm{d} D}{\left(r+\Psi(q)-q^2D \right)^2}\right]^{-1} = \frac{1}{\kappa(q)^2 |G'((\Psi(q) + r) / q^2)|}.
\end{aligned}
    \label{eq:projection}
\end{equation}
As $q \uparrow q_{\rm c}$, we have
\begin{equation}
    \chi_{\rm c}:= \chi(q_{\rm c}^-)=\frac{q_{\rm c}^4}{r^2}
    \left[\int_0^{D_{\max}}
    \frac{W(D) \mathrm{d} D}{\left(D_{\max}-D \right)^2}\right]^{-1} = \frac{1}{\kappa(q_{\rm c})^2 |G'(D_{\max})|} .
    \label{eq:critical-projection}
\end{equation}
Thus $\chi_{\rm c}>0$ for edge exponents $\nu>1$, whereas $\chi_{\rm c}=0$ for $0<\nu\leq1$, see Fig.~\ref{fig:chi-fstar} (left). Taking the squared overlap $\chi(q)$ as an order parameter of the dynamical phase transition for the SCGF, the BBP transition is therefore discontinuous for $\nu > 1$, as the collective contribution disappears abruptly at $q_{\rm c}$, and continuous for $0 < \nu \leq 1$ (cf.~Fig.~\ref{fig:phase-diag}, left). This is in accordance with Ref.~\cite[Proposition 2.4]{benaych-georges-nadakuditi:2011}, and has also been recently pointed out in Ref.~\cite{bocchi-biroli-cammarota-etal:2026}. The connection between the overlap and the slope of the SCGF becomes apparent by differentiating the implicit Eq.~\eqref{eq:cauchy} in~$q$, which yields
\begin{align}
    \Psi'(q) = \frac{2}{q} \Psi(q) + \frac{2r}{q} \left( 1 - \chi(q) \right)\,,
    \label{eq:slope-gap}
\end{align}
so that a jump in $\chi$ results in a jump of $\Psi'$. Since the SCGF is continuous at $q_{\rm c}$, with $\Psi(q_{\rm c})=D_{\max}q_{\rm c}^2-r$, Eq.~(\ref{eq:slope-gap}) yields $y_- =\Psi'(q_{\rm c}^-)=2D_{\max}q_{\rm c}-2r\chi_{\rm c}/q_{\rm c}$ and $y_+ =\Psi'(q_{\rm c}^+)=2D_{\max}q_{\rm c}$, such that
\begin{eqnarray}
    y_+ - y_- = \frac{2 r  \chi_{\rm c}}{q_{\rm c}}\, .
\end{eqnarray}
Via the Legendre transform~\eqref{eq:legendre-summary}, the jump of the overlap thus produces the finite coexistence interval
$[y_-,y_+]$ of the rate function for $\nu>1$, while for $0<\nu\leq1$ the interval collapses -- see Fig.~\ref{fig:chi-fstar}. This is the central link between diffusivity selection and the conditioned
trajectories studied in Sec.~\ref{sec:single-run-main-text}\footnote{See also Eq.~\eqref{eq:overlap-vs-run-duration} below, where we further establish an interpretation of $\chi$ in terms of the switching rate of the conditioned process, in accordance with the framework of Ref.~\cite{chetrite-touchette:2015} on tilted (Doob) processes.}.

\section{Single-run distribution and dynamical condensation}
\label{sec:single-run-main-text}

In this section, we characterize the trajectory picture summarized in Table~\ref{tab:phases} and Fig.~\ref{fig:runs-dist}, by studying the typical behavior of the rare trajectories conditioned on the ballistic endpoint $x(t)=yt$. As in Sec.~\ref{sec:sd-def}, let $0=T_0<T_1<T_2<\cdots$ denote the switching times. The increment of a
completed run is the displacement between two successive switching times,
$\Delta x_i=x(T_i)-x(T_{i-1})$. When the first switching has not yet occurred
at the observation time $t$, $\Delta x_1$ instead denotes the displacement
accumulated from time $0$ to time $t$.  Our main observable in this section is defined by
\begin{equation}
\mathbb{P}\!\left(\Delta x_1\in[\xi,\xi+\mathrm{d}\xi]\mid x(t)=yt\right)
=P_t(\xi\mid y)\,\mathrm{d}\xi,
        \label{eq:Pt-def}
\end{equation}
with a possible Dirac contribution, in case there are no switches at all, treated explicitly below.\footnote{Since $D(0)\sim W$, the diffusivity process is stationary and reversible, and, given a diffusivity history, the Brownian path measure is invariant under $D(s)\mapsto D(t-s)$, $x(s)\mapsto x(t)-x(t-s)$. This preserves the constraint $x(t)=yt$ and exchanges the first and the last (unfinished) run, so that both have the same conditional law $P_t(\xi\mid y)$. Runs with a fixed index $k=O(1)$ show the same scaling regimes, but the $O(1)$ masses found below in the saturated phase are specific to the first (or last) run.} Since every conditioned trajectory in the coexistence phase contains $O(t)$ microscopic runs and one condensate, the condensate shows up in $P_t(\xi\mid y)$ as a peak at $\xi=O(t)$ of weight $O(1/t)$, whereas in the saturated phase it carries a weight of order one. We mainly focus on
$\nu>1$, for which all three phases exist. For $0<\nu\leq1$, where $y_-=y_+$, the coexistence phase is absent, while our results for the fluid phase $y<y_-$ and the saturated phase $y>y_+$ remain valid, since they only require $\nu>0$.\\

\begin{figure}
    \centering
    \includegraphics[width=1\linewidth]{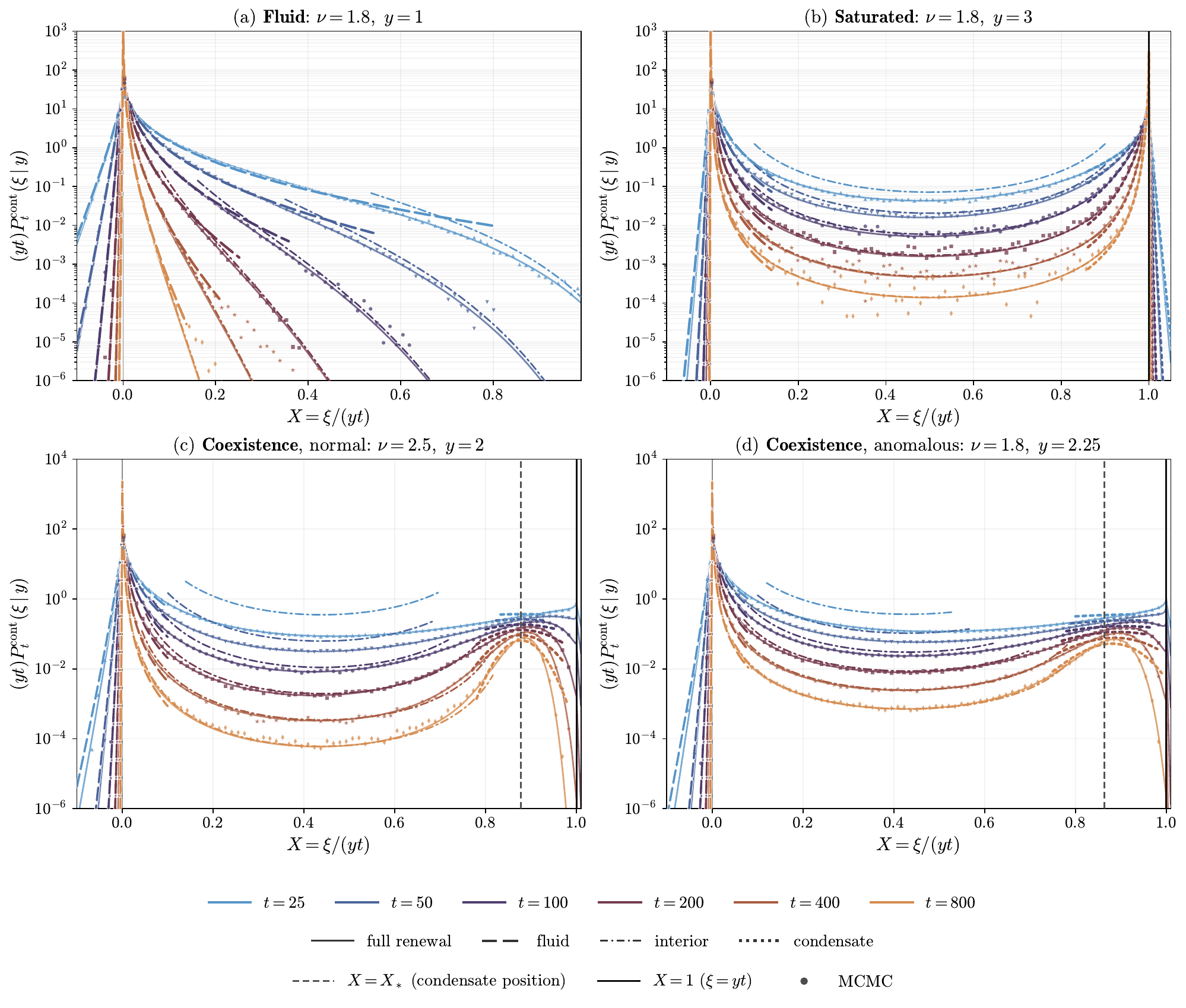}
    \caption{
    Single-run marginal PDF from MCMC (markers) for different times $t$ compared to theoretical predictions (lines). On the horizontal axis, we show the single-run increment size $\xi$, normalized by the total displacement $yt$, and on the vertical axis, its PDF (rescaled accordingly by $yt$).
    Solid lines show the numerical solution of the exact renewal equation for the single-run marginal PDF, while other line styles show asymptotic theoretical predictions in the large $t$ limit for different segments of the PDF. \textbf{Top left.} fluid phase $ y < y_-$.\textbf{ Top right.} saturated phase $y > y_+$. \textbf{Bottom
    left.} normal condensate for $\nu = 2.5 > 2$ in coexistence phase $y_- < y < y_+$.
    \textbf{Bottom right.} anomalous condensate for $1 < \nu = 1.8 < 2$  in coexistence phase $y_- < y < y_+$. We use the diffusion coefficient distribution $W(D) = (\nu + 1)(1 - D)^{\nu}$ with $r = D_{\max} = 1$, consistent with Fig.~\ref{fig:phase-diag} (right). For the theoretical derivations of the PDF asymptotics, see Sec.~\ref{sec:single-run-main-text}, and for a description of the numerics and sampling, see Sec.~\ref{sec:numerics}.
    }
    \label{fig:cond-mcmc-vs-theory}
\end{figure}

All results below follow from a single exact renewal equation for $P_t(\xi\mid y)$, which we derive in Sec.~\ref{sec:renewal}, together with the precise long-time asymptotics of the endpoint PDF needed to analyze it. The remaining subsections consider separately the different scaling regimes of the displacement $\xi$ and the duration $\tau$ of the first run; lengthy technical calculations are deferred to Appendix~\ref{SingleRunAppendix}.
Our resulting asymptotic predictions are compared to numerical results in Fig.~\ref{fig:cond-mcmc-vs-theory}; we describe this figure in more detail below in Sec.~\ref{sec:numerics}.\\

Before delving into the details of this calculation, we remark here that the different branches of the rate function defined in Eq.~(\ref{eq:microcanonical-summary}) already hint at a condensation mechanism using the following heuristics. First, the small-$y$ behavior $I(y) \sim  y^2 / (4\langle D\rangle)$ (which follows by Taylor expanding Eq.~\eqref{eq:microcanonical-summary}) matches the Gaussian distribution of the typical fluctuations
\begin{eqnarray}
    p_r(x,t) \overset{t\to \infty}{\sim}  \frac{1}{\sqrt{4\pi \langle D\rangle t}} e^{-\frac{x^2}{4\langle D \rangle t}}\, , \qquad x= O(\sqrt t)\, .
\end{eqnarray}
Indeed, over a long observation time, the particle undergoes many switches, and its time-averaged diffusivity converges to $\langle D \rangle$. This collective mechanism persists throughout the first branch: the imposed displacement is shared among many microscopic runs, as sketched in Fig.~\ref{fig:runs-dist} (left). We hence refer to this regime as the fluid phase.\\

On the other hand, the last branch of Eq.~(\ref{eq:microcanonical-summary}), up to pre-exponential factors, gives 
\begin{eqnarray}\label{rareLD}
    p_r(x,t) \asymp e^{-rt} e^{-\frac{x^2}{4D_{\max}t}}\, ,
\end{eqnarray}
where $e^{-rt}$ is the probability that no switch occurs during the observation time, while the Gaussian factor corresponds to free diffusion with diffusivity fixed at its maximal value $D_{\max}$.
This mechanism is reminiscent of dynamical phase transitions encountered in stochastic resetting processes~\cite{majumdar-sabhapandit-schehr:2015} and mortal Brownian walkers~\cite{yerrababu-majumdar-guiselin-etal:2025}, where rare fluctuations may be dominated by trajectories containing an exceptionally long interval without resetting or death.
Here, it anticipates the saturated phase in Fig.~\ref{fig:runs-dist} (right), where a single run occupies asymptotically the whole observation time and carries the entire imposed displacement, although additional microscopic runs may still occur.\\

Finally, the intermediate linear branch suggests coexistence between a fluid at its critical velocity $y_-$, and a macroscopic run with diffusivity close to $D_{\max}$, carrying displacement at velocity $y_+$. Indeed, introducing
\begin{eqnarray}
    \theta=\frac{y-y_-}{y_+-y_-},
\end{eqnarray}
one can write
\begin{eqnarray}
    y=(1-\theta)y_-+\theta y_+, \qquad
I(y_-<y<y_+)=(1-\theta)I(y_-)+\theta I(y_+)\, .
\end{eqnarray}
The physical interpretation is the following: a fraction $1-\theta$ of the observation time is occupied by the critical fluid, while a fraction $\theta$ is occupied by the macroscopic run, as shown in Fig.~\ref{fig:runs-dist} (center). The latter carries the displacement $y_+\theta t$ with the remaining displacement $y_-(1-\theta)t$ accumulated by the fluid. Increasing $y$ therefore increases the condensate fraction until it covers asymptotically the whole trajectory in the saturated phase when $y \to y_+$.

\subsection{Exact renewal equation and normalization}
\label{sec:renewal}

Between two successive switching events $T_i$ and $T_{i+1}$, a run has duration $\tau=T_{i+1}-T_i$ and a fixed diffusion constant $D$ drawn from $W(D)$. The displacement accumulated during such a run under the unconditioned process~\eqref{eq:sd-def} has the Gaussian PDF
\begin{equation}
        G_{D,\tau}(\xi)
        =
        \frac{1}{\sqrt{4\pi D\tau}}
        \exp\!\left(-\frac{\xi^2}{4D\tau}\right).
        \label{eq:G-def}
\end{equation}
We first consider the joint density of the displacement $\Delta x_1=\xi$ accumulated during the first run and the final position $x(t)=yt$, which we write as $\mathcal P_t(\xi,yt)$. An exact renewal decomposition with respect to the first switching time gives
\begin{align}
        \mathcal P_t(\xi,yt)
        ={}&
        e^{-rt}
        \int_0^{D_{\max}}\mathrm{d}D\,
        W(D)G_{D,t}(yt)\,
        \delta(\xi-yt)
        +
        \int_0^t \mathrm{d}\tau
        \int_0^{D_{\max}}\mathrm{d}D\,
        r e^{-r\tau}W(D)G_{D,\tau}(\xi)
        p_r(yt-\xi,t-\tau).
        \label{eq:jointfirst}
\end{align}
The first term in Eq.~\eqref{eq:jointfirst} corresponds to trajectories for which no switching event occurs up to time $t$, with probability $e^{-rt}$. In this case, the first run spans the entire observation time, so that $\Delta x_1=x(t)=yt$, which gives the Dirac delta $\delta(\xi-yt)$. The diffusion constant remains equal to its initial value throughout the trajectory and is averaged over its distribution~$W(D)$. The second term in Eq.~\eqref{eq:jointfirst} corresponds to trajectories whose first switching event occurs at time $\tau<t$. The factor $r e^{-r\tau}\mathrm{d}\tau$ is the probability that the first run has duration in $[\tau,\tau+\mathrm{d}\tau]$, while $W(D)G_{D,\tau}(\xi)$ gives the probability density that this run has diffusivity $D$ and accumulates a displacement $\xi$. After the first switch, the process renews independently and must accumulate the remaining displacement $yt-\xi$ during the remaining time $t-\tau$, giving the factor $p_r(yt-\xi,t-\tau)$. The conditional law is obtained by dividing by the endpoint density
\begin{equation}
        Z(x,t):=p_r(x,t)=\langle \delta(x(t)-x) \rangle,
        \qquad
        Z_t(y):=Z(yt,t)\, ,
        \label{eq:Z-def}
\end{equation}
where the quantity $Z_t(y)$ acts as the normalization factor of the conditional single-run PDF $P_t(\xi\mid y)$, motivating the partition-function-like notation $Z$. Hence,
\begin{align}
        P_t(\xi\mid y)
        ={}&
        a_t(y)\,\delta(\xi-yt)
        +
        \frac{1}{Z_t(y)}
        \int_0^t \mathrm{d}\tau
        \int_0^{D_{\max}}\mathrm{d}D\,
        r e^{-r\tau}W(D)G_{D,\tau}(\xi)
        Z(yt-\xi,t-\tau),
        \label{eq:exactfirst}
\end{align}
where
\begin{align}
        a_t(y)
        =
        \frac{e^{-rt}}{Z_t(y)}
        \int_0^{D_{\max}}
        W(D)G_{D,t}(yt)\,\mathrm{d}D
        \label{eq:at-def}
\end{align}
is the conditional weight of the no-switching Dirac delta peak. It will be convenient below to denote the weight of a completed first run with displacement $\xi$ and duration $\tau$ by
\begin{equation}
        Q(\xi,\tau)
        :=r\int_0^{D_{\max}}\mathrm{d}D\,
        e^{-r\tau}W(D)G_{D,\tau}(\xi),
        \label{eq:C-Qdef}
\end{equation}
so that Eq.~\eqref{eq:exactfirst} becomes 
\begin{align}
P_t(\xi\mid y)
        = a_t(y)\,\delta(\xi-yt)
        + Z_t(y)^{-1}\int_0^t \mathrm{d}\tau\, Q(\xi,\tau) Z(yt-\xi,t-\tau)\,.
        \label{eq:renewal-with-q}
\end{align}
The renewal equation~\eqref{eq:renewal-with-q} is exact for any $t$, and it is the central object of the analysis in this section. Its structure is straightforward: the first run, with a priori weight $Q(\xi,\tau)$, is followed by an independent copy of the process, which must produce the residual displacement $yt-\xi$ in the residual time $t-\tau$. Conditioning on the endpoint therefore reweights the a priori law of the first run by the ratio of endpoint densities $Z(yt-\xi,t-\tau)/Z_t(y)$.\\

To extract the long-time behavior of $P_t(\xi\mid y)$ from Eq.~\eqref{eq:exactfirst}, the log-asymptotic form $Z_t(y)\asymp e^{-tI(y)}$ of Eq.~\eqref{eq:large-deviation-summary} is not sufficient. Since $P_t(\xi\mid y)$ is normalized, the contributions to the numerator of Eq.~\eqref{eq:exactfirst} which survive as $t\to\infty$ have the same exponential cost $e^{-tI(y)}$ as the denominator $Z_t(y)$, and these exponential factors cancel. Whether a given class of first runs then carries a probability of order one, a probability of order $1/t$, or a vanishing weight, is decided by the pre-exponential factors of $Z$, both at the conditioning point $(yt,t)$ and at the shifted arguments $(yt-\xi,t-\tau)$. We therefore need the precise large-deviation asymptotics of the endpoint PDF, including its algebraic prefactor, in each of the three regimes. The derivations for the fluid and saturated phases are given in Appendix~\ref{sec:endpoint-ldt-precise}, and we only state the main results here.\\

\noindent \textbf{Fluid phase: $0<y<y_-$}. The endpoint PDF is governed by the regular saddle $q_*(y)<q_{\rm c}$, yielding
\begin{equation}
        Z_t(y) \overset{t \to \infty}{\sim} B_{\rm reg}(y)t^{-1/2}e^{-tI(y)},
        \qquad
        I(y)= \phi_\nu(y) = q_*(y)y-\Psi(q_*(y)),
        \qquad y=\Psi'(q_*(y)).
        \label{eq:Z-regular-intro}
\end{equation}
The prefactor in Eq.~\eqref{eq:Z-regular-intro} is (see also the supplementary material of Ref.~\cite{gueneau-majumdar-schehr:2025})
\begin{equation}
        B_{\rm reg}(y)
        ={\chi(q_*(y))\over\sqrt{2\pi\Psi''(q_*(y))}},
        \qquad
        \chi(q)={1\over r^2J_2(q)},
        \qquad
        J_2(q)=\int_0^{D_{\max}}
        {W(D) \,\mathrm{d} D\over (r+\Psi(q)-q^2D)^2} .
        \label{eq:Breg-intro}
\end{equation}
Here, $\chi(q)$ is precisely the squared spike overlap of Eq.~\eqref{eq:projection}. Thus, the residue of the leading pole of the generating function, which controls the amplitude of $Z_t(y)$, is the overlap order parameter of the BBP transition of Sec.~\ref{sec:rmt}.\\

\noindent \textbf{Coexistence phase: $y_-<y<y_+$}. Throughout the coexistence interval, the conjugate parameter is pinned at $q_{\rm c}$, and the endpoint PDF takes the form
\begin{equation}
        Z_t(y) \overset{t \to \infty}{\sim} B_{\rm int}(y)t^{-\nu}e^{-tI(y)},
        \qquad
        I(y)=q_{\rm c}y-\Psi(q_{\rm c}).
        \label{eq:I-linear-intro}
\end{equation}
Since $q_{\rm c}$ is a non-analytic point of $\Psi(q)$, the algebraic factor $t^{-\nu}$ and the amplitude $B_{\rm int}(y)$ do not follow from a simple saddle point in $q$: they are derived in Sec.~\ref{sec:cond-intermediate}, see Eq.~\eqref{eq:Bint-intro} below. Until then, we only use that $B_{\rm int}(y)$ is a smooth function of $y$ in the interior of the coexistence interval.\\

\noindent \textbf{Saturated phase: $y>y_+$}. Beyond $y_+$, the endpoint asymptotics are controlled by the upper edge $D=D_{\max}$, giving
\begin{equation}
        Z_t(y)
         \overset{t \to \infty}{\sim}
        {\cal A}(y)t^{-(\nu+3/2)}e^{-tJ(y)},
        \qquad
        J(y)=r+{y^2\over4D_{\max}},
        \label{eq:Z-sat-intro}
\end{equation}
with
\begin{equation}
        {\cal A}(y)=
        {C\Gamma(\nu+1)\over
        \sqrt{4\pi D_{\max}}\,
        q_y^{2\nu+2}
        \left(1-{q_{\rm c}^2 /  q_y^2}\right)^2},
        \qquad
        q_y={y\over2D_{\max}} .
        \label{eq:A-y-intro}
\end{equation}

Recall that $C>0$ in Eq.~\eqref{eq:A-y-intro} and below denotes the constant in the density $W(D)$ as defined in Eq.~\eqref{eq:edge-expo-def}.\\

With Eq.~\eqref{eq:exactfirst} and the asymptotics~\eqref{eq:Z-regular-intro}--\eqref{eq:A-y-intro} at hand, the long-time behavior of $P_t(\xi\mid y)$ follows by identifying the scaling regimes of the first-run variables $(\xi,\tau)$ for which the numerator in Eq.~\eqref{eq:exactfirst} has the same exponential cost as $Z_t(y)$, and by then comparing the prefactors. We consider in turn:
(i) microscopic first runs, $\xi=O(1)$ and $\tau=O(1)$, which constitute the fluid (Sec.~\ref{sec:fluid});
(ii) trajectories with no switch before time $t$, i.e., the Dirac contribution in Eq.~\eqref{eq:exactfirst} (Sec.~\ref{sec:no-switch});
(iii) macroscopic first runs, $\xi=O(t)$ and $\tau=O(t)$, near the optimal values selected by the endpoint constraint, which constitute the condensate (Sec.~\ref{sec:condensate});
(iv) macroscopic first runs with $\xi$ between the fluid and the condensate, which describe the interior of the distribution (Sec.~\ref{sec:interior}).
Since $P_t(\xi\mid y)$ is normalized, the masses carried by these contributions must add up to one. As we shall see, the mass carried by the fluid alone thereby already indicates whether macroscopic contributions with a finite weight must be present.

\subsection{Microscopic runs: the fluid component}
\label{sec:fluid}

We first analyze the microscopic contribution to the first-run marginal, obtained
by taking $\tau=O(1)$ and $\xi=O(1)$ in Eq.~\eqref{eq:exactfirst}. In this scaling regime, the Dirac contribution at $\xi=yt$ plays no role, the first run
carries only a microscopic displacement, and the large endpoint $yt$ is
realized by the independent trajectory following the first switch. For the endpoint density $Z(x,t)$, we use the refined large-deviation form of Sec.~\ref{sec:renewal}, i.e., including the pre-exponential factor, which we write generically as
\begin{equation}
        Z(yt,t)\sim \mathcal{B}(y,t)e^{-tI(y)},
        \qquad
        Z(x',t')\sim \mathcal{B}(x'/t',t')
        \exp(-t'I(x'/t')) \;.
        \label{eq:fluid-refined-LD}
\end{equation}
We want to evaluate the ratio
\begin{eqnarray}
    {Z(yt-\xi,t-\tau)\over Z(yt,t)}
    \sim
    {\mathcal{B}(y',t-\tau)\over \mathcal{B}(y,t)}
    \exp\left\{-(t-\tau)I(y')+tI(y)\right\},
    \label{eq:ratio-pre-expansion}
\end{eqnarray}
present in the second term of Eq.~\eqref{eq:exactfirst}, where for $\tau=O(1)$ and $\xi=O(1)$:
\begin{eqnarray}
    y'
    ={yt-\xi\over t-\tau}
    =y+{y\tau-\xi\over t-\tau}
    =y-{\xi-y\tau\over t}+O\left(t^{-2}\right).
\end{eqnarray}
Let $q=q(y)$ denote the conjugate parameter selected by the endpoint
conditioning, given in Eq.~\eqref{eq:q-of-y-summary}; on the edge branch, we write $q(y)=q_y:=y/(2D_{\max})$.
Away from the crossover regions at $y_-$ and $y_+$, the rate function is
differentiable within each phase and satisfies
\begin{equation}\label{eq:legendre-local}
        I'(y)=q(y),
        \qquad
        I(y)=q(y)y-\Psi(q(y)).
\end{equation}
Thus, for fixed $\xi$ and $\tau$, the exponent in Eq.~\eqref{eq:ratio-pre-expansion} becomes
\begin{equation}
        -(t-\tau)I(y')+tI(y)
        \longrightarrow
        q(y)\xi-\tau\Psi(q(y)).
        \label{eq:fluid-exponent-ratio}
\end{equation}
Since in each phase the pre-exponential factor has the form $\mathcal{B}(y,t)\sim \mathcal{B}_y\,t^{-\alpha}$, with an amplitude $\mathcal{B}_y$ that is smooth in $y$ (namely $B_{\rm reg}(y)$, $B_{\rm int}(y)$, or ${\cal A}(y)$, with $\alpha=1/2$, $\nu$, or $\nu+3/2$, respectively; cf.\ Sec.~\ref{sec:renewal}), we have
\begin{equation}
        {\mathcal{B}(y',t-\tau)\over \mathcal{B}(y,t)}\longrightarrow 1 .
        \label{eq:fluid-prefactor-ratio}
\end{equation}
Combining Eqs.~\eqref{eq:fluid-exponent-ratio} and
\eqref{eq:fluid-prefactor-ratio} hence shows that Eq.~\eqref{eq:ratio-pre-expansion} leads to
\begin{equation}
        {Z(yt-\xi,t-\tau)\over Z(yt,t)}
        \xrightarrow{t \to \infty}
        \exp\left(q(y)\xi-\tau\Psi(q(y))\right) .
        \label{eq:ratiofluid}
\end{equation}
The resulting density therefore describes the shape of the microscopic fluid component, but its integral over $\xi$ need not exhaust the total probability: as we show in Sec.~\ref{sec:normalization}, part of the probability may reside at macroscopic $\xi=O(t)$, where Eq.~\eqref{eq:ratiofluid} no longer applies. Now, substituting Eq.~\eqref{eq:ratiofluid} into the continuous part of
Eq.~\eqref{eq:exactfirst}, and writing $q=q(y)$ for brevity, yields
\begin{equation}
        P_t(\xi\mid y)
        \simeq
        e^{q\xi}\int_0^\infty\mathrm{d}\tau\int_0^{D_{\max}}\mathrm{d}D\,
        rW(D)e^{-[r+\Psi(q)]\tau}
        {e^{-\xi^2/(4D\tau)}\over\sqrt{4\pi D\tau}} .
        \label{eq:fluid-microscopic-integral}
\end{equation}
Here the upper limit of the $\tau$-integral has been extended from $t$ to
$\infty$, since the contribution considered here is localized on the
microscopic scale $\tau=O(1)$.
Using the identity
\begin{equation}
        \int_0^\infty \tau^{-1/2}
        \exp\left(-(r+\Psi(q))\tau-\frac{\xi^2}{4D\tau}\right)
        \,\mathrm{d}\tau
        =
        \sqrt{\frac{\pi}{r+\Psi(q)}}
        \exp\left(-|\xi|
        \sqrt{\frac{r+\Psi(q)}{D}}\right),
        \label{eq:Laplace-Bessel-identity}
\end{equation}
we obtain the effective one-run density of the microscopic fluid component,
\begin{equation}
        P_t(\xi\mid y)
        \simeq p_q(\xi)
        :={re^{q\xi}\over 2\sqrt{r+\Psi(q)}}
        \int_0^{D_{\max}}{W(D)\over\sqrt D}
        \exp\!\left(-|\xi|\sqrt{r+\Psi(q)\over D}\right)\mathrm{d} D .
        \label{eq:pq}
\end{equation}
The variable $q$ in $p_q(\xi)$ is the conjugate variable to $y$ in Eq.~\eqref{eq:q-of-y-summary}, and $p_q(\xi)$ is the law of a single run of the tilted process: compared to its a priori law, each run is reweighted by $e^{q\xi-\Psi(q)\tau}$. We now determine the tails, the total mass, and the mean duration of the runs described by $p_q(\xi)$, and then specialize to the three phases, i.e., fluid, coexistence and saturated phase.

\subsubsection{Tails of the fluid law}\label{sec:fluid-tails}
We first note that Eq.~\eqref{eq:pq} implies the exact identity
\begin{equation}
    p_q(-\xi)=e^{-2q\xi}p_q(\xi),
    \qquad \xi>0.
    \label{eq:pq-asymmetry}
\end{equation}
Consequently, the asymptotic behavior of the left tail can be deduced
directly from that of the right tail, and vice versa.\\

We now consider the right tail $\xi\to+\infty$. Under the upper-edge behavior in Eq.~\eqref{eq:edge-expo-def},
the integral over $D$ in Eq.~\eqref{eq:pq} is dominated by the region close to $D_{\max}$. We write
$
    D=D_{\max}-\varepsilon
$,
$\varepsilon\ll 1.
$
Then
\begin{equation}
    \sqrt{\frac{r+\Psi(q)}{D}}
    =
    \sqrt{\frac{r+\Psi(q)}{D_{\max}}}
    +
    \frac{\sqrt{r+\Psi(q)}}{2D_{\max}^{3/2}}
    \varepsilon
    +
    O(\varepsilon^2).
\end{equation}
Rescaling $\varepsilon$ by $1/\xi$ and using $W(D_{\max}-\varepsilon)\sim C\varepsilon^\nu$ then gives the right tail, and using Eq.~\eqref{eq:pq-asymmetry} the left tail follows. Hence,
\begin{equation}\label{fluidtailsgeneral}
    p_q(\xi)
    \sim
    \begin{cases}
        B(q)\,\xi^{-(\nu+1)}e^{-\lambda_+(q)\xi},
        & \xi\to+\infty,\\[1mm]
        B(q)\,|\xi|^{-(\nu+1)}e^{-\lambda_-(q)|\xi|},
        & \xi\to-\infty,
    \end{cases}
\end{equation}
where
\begin{equation}
    B(q)
    =
    rC\Gamma(\nu+1)
    \frac{2^\nu D_{\max}^{(3\nu+2)/2}}
    {[r+\Psi(q)]^{(\nu+2)/2}},
    \qquad
    \lambda_\pm(q)
    =
    \sqrt{\frac{r+\Psi(q)}{D_{\max}}}\mp q.
    \label{eq:Bq-lambda}
\end{equation}
For $q<q_{\rm c}$, the regular solution satisfies
\begin{equation}\label{eq:gap-positive}
    r+\Psi(q)>q^2D_{\max},
\end{equation}
so that $\lambda_+(q)>0$ and both tails are exponentially cut off. At $q=q_{\rm c}$, the regular branch reaches the upper edge,
$
    r+\Psi(q_{\rm c})=q_{\rm c}^2D_{\max},
$
and hence $\lambda_+(q_{\rm c})=0$. For $q>q_{\rm c}$, the SCGF follows the upper-edge branch $\Psi(q)=q^2D_{\max}-r$, so that $\lambda_+(q)=0$ as well. Thus, at and beyond $q_{\rm c}$, the right tail becomes algebraic, while the left tail remains exponentially suppressed, with $\lambda_-(q)=2q$, as follows from Eq.~(\ref{eq:pq-asymmetry}):
\begin{equation}
        p_q(\xi)\sim
        \begin{cases}
        B(q)\,\xi^{-(\nu+1)}, & \xi\to+\infty,\\[1mm]
        B(q)\,|\xi|^{-(\nu+1)}e^{-2q|\xi|}, & \xi\to-\infty,
        \end{cases}
        \qquad q\geq q_{\rm c}.
        \label{eq:edge-tails}
\end{equation}
The tail amplitudes at $q=q_{\rm c}$ and at $q=q_y>q_{\rm c}$, which will be needed below, follow from Eq.~\eqref{eq:Bq-lambda} as
\begin{equation}
        A_{\rm c}:=B(q_{\rm c})=r \, C\, \Gamma(\nu+1){2^\nu D_{\max}^\nu\over q_{\rm c}^{\nu+2}},
        \qquad
        A_y^{(+)}:=B(q_y)=A_{\rm c}\left({y_+\over y}\right)^{\nu+2} .
        \label{eq:Ac-intro}
\end{equation}

\subsubsection{Normalization and mass of the fluid component}
\label{sec:normalization}

The microscopic calculation above determines the local shape of
$p_q(\xi)$.  We now determine its total mass.  Importantly, while for $q\leq q_{\rm c}$, its mass is $1$, on the upper-edge branch $q>q_{\rm c}$, we find that the microscopic
component carries only part of the probability.
For convenience, we set
$\alpha_D(q)=\sqrt{(r+\Psi(q)) / D}$. Integrating $p_q(\xi)$ from Eq.~\eqref{eq:pq} over $\xi$ leads to the following integral
\begin{align}
        \int_{-\infty}^{\infty}e^{q\xi-\alpha_D(q)|\xi|}\,\mathrm{d}\xi
        &={1\over \alpha_D(q)-q}+{1\over \alpha_D(q)+q}
        ={2\alpha_D(q)\over \alpha_D(q)^2-q^2},
        \label{eq:fluid-mass-xi-integral}
\end{align}
and hence the fluid mass is
\begin{equation}
        M_{\rm fluid}(q):=\int_{-\infty}^\infty p_q(\xi)\,\mathrm{d}\xi
        =r\int_0^{D_{\max}}{W(D) \,\mathrm{d}D\over r+\Psi(q)-q^2D} .
        \label{eq:mass-general}
\end{equation}
For $q<q_{\rm c}$, the regular pole solution exists and satisfies Eq.~\eqref{eq:secular}, while at $q=q_{\rm c}$ it reaches the upper edge. Hence, for $q\leq q_{\rm c}$,
\begin{equation}
        1=r\int_0^{D_{\max}}{W(D) \,\mathrm{d}D\over r+\Psi(q)-q^2D} .
        \label{eq:renewalinterior}
\end{equation}
Comparing Eqs.~\eqref{eq:mass-general} and \eqref{eq:renewalinterior}, we find
\begin{equation}
        M_{\rm fluid}(q)=1,
        \qquad q\le q_{\rm c} .
        \label{eq:mass-one}
\end{equation}
Exactly at the edge $q=q_{\rm c}$ this reduces to Eq.~\eqref{eq:critical-q}, i.e.,
\begin{equation}
        r+\Psi(q_{\rm c})=q_{\rm c}^2D_{\max},
        \qquad
        1={r\over q_{\rm c}^2}\int_0^{D_{\max}}{W(D) \,\mathrm{d}D\over D_{\max}-D} .
        \label{eq:criticalcondition}
\end{equation}
For $q>q_{\rm c}$, that solution no longer exists and the growth rate follows the
upper-edge branch,
$
        \Psi(q)=q^2D_{\max}-r
$.
In that case, Eq.~\eqref{eq:mass-general} gives
\begin{equation}
        M_{\rm fluid}(q)
        ={r\over q^2}\int_0^{D_{\max}}{W(D) \,\mathrm{d}D
        \over D_{\max}-D}
        ={q_{\rm c}^2\over q^2} .
        \label{eq:mass-edge}
\end{equation}
Thus, in the saturated phase,
\begin{equation}
        M_{\rm fluid}(q_y)=\left({q_{\rm c}\over q_y}\right)^2
        =\left({y_+\over y}\right)^2=:m_y .
        \label{eq:mass-my}
\end{equation}
Note in particular that $M_{\rm fluid} < 1$ for $q>q_c$.

\subsubsection{Mean run duration in the fluid background}\label{sec:fluid-duration}

The microscopic representation above also gives the distribution of the run duration in the fluid background. Integrating the
single-run weight Eq.~(\ref{eq:fluid-microscopic-integral}) over the displacement yields
\begin{equation}
        \pi_q(D,\tau)
        =
        rW(D)
        \exp\!\left(
        -\left[r+\Psi(q)-q^2D\right]\tau
        \right),
        \label{eq:tilted-duration}
\end{equation}
where $\pi_q(D,\tau)$ denotes the joint probability density of the
diffusivity and duration of a microscopic run in the fluid background. Note that
for $q\leq q_{\rm c}$, the pole equation ensures correct normalization
\begin{equation}
        \int_0^\infty \mathrm{d}\tau
        \int_0^{D_{\max}}\mathrm{d}D\,
        \pi_q(D,\tau)
        =1.
\end{equation}
Its mean duration is
\begin{equation}
        \mu(q)
        :=
        \langle \tau\rangle_q
        =
        r\int_0^{D_{\max}}
        \frac{W(D)\,\mathrm{d}D}
        {\left[r+\Psi(q)-q^2D\right]^2}.
        \label{eq:mean-duration-q}
\end{equation}
At the critical point, using
$r+\Psi(q_{\rm c})=q_{\rm c}^2D_{\max}$, this becomes
\begin{equation}
        \mu_{\rm c}
        :=
        \mu(q_{\rm c})
        =
        \frac{r}{q_{\rm c}^4}
        \int_0^{D_{\max}}
        \frac{W(D)\,\mathrm{d}D}
        {(D_{\max}-D)^2}.
        \label{eq:muc}
\end{equation}
Thus, $\mu_{\rm c}$, which will reappear in the endpoint amplitude $B_{\rm int}(y)$ of the coexistence phase, Eq.~\eqref{eq:Bint-intro}, is the mean duration of a microscopic run in the
critical fluid background. Comparing with the overlap definition (see Sec.~\ref{overlapsection}), we see the simple relation
\begin{equation}
    \mu(q) = \frac{1}{r\chi(q)}.
    \label{eq:overlap-vs-run-duration}
\end{equation}
Hence, the overlap $\chi(q)=(1/r)/\mu(q)$ can also be interpreted as the ratio between the bare mean time $1/r$ between switches and the mean run duration in the fluid background.

\subsubsection{The fluid component in the three phases}

For a prescribed mean velocity $y$, the \textit{fluid component} of the first-run law is thus given by
\begin{equation}
        {\cal F}_y(\xi)=
        \begin{cases}
        p_{q_*(y)}(\xi), & 0<y<y_-,\\[1mm]
        p_{q_{\rm c}}(\xi), & y_-<y<y_+,\\[1mm]
        p_{q_y}(\xi), & y>y_+ ,
        \end{cases}
        \label{eq:fluid-density-main}
\end{equation}
with $p_q(\xi)$ from Eq.~\eqref{eq:pq}, i.e., with the conjugate parameter $q(y)$ of Eq.~\eqref{eq:q-of-y-summary}. The function ${\cal F}_y(\xi)$ describes the contribution of microscopic runs in the different dynamical phases to the PDF $P_t(\xi \mid y)$. As we will see below, in addition to this contribution, others may be present. We now collect the consequences of the results for ${\cal F}_y(\xi)$ above in each phase.\\

\noindent\textbf{Fluid phase: $0<y<y_-$.} By Eq.~(\ref{eq:mass-one}), the microscopic component exhausts the probability mass, and the conditional first-run law has a normalized limit on the $O(1)$ scale,
$P_t(\xi\mid y)={\cal F}_y(\xi)+o(1)$. Since $\lambda_+(q_*(y))>0$, both tails of ${\cal F}_y(\xi)$ are exponentially cut off, cf.~Eq.~(\ref{fluidtailsgeneral}), so that fluid runs have a typical $O(1)$ length scale. This regime is sketched in Fig.~\ref{fig:runs-dist} (left column).\\

\noindent\textbf{Coexistence phase: $y_-<y<y_+$.} The microscopic background is described by the fluid law at $q_{\rm c}$, which again has unit mass by Eq.~\eqref{eq:mass-one}, but now has the algebraic right tail $A_{\rm c}\xi^{-(\nu+1)}$ of Eq.~\eqref{eq:edge-tails}, while its left tail remains exponentially suppressed. This asymmetry reflects the large, \textit{positive} displacement imposed at final time. Due to its scale-free right tail, we call this fluid ``critical''. Since it still carries unit mass, any macroscopic contribution to $P_t(\xi\mid y)$ can only carry a vanishing weight as $t\to\infty$. This regime is sketched in Fig.~\ref{fig:runs-dist} (center column).\\

\noindent\textbf{Saturated phase: $y>y_+$.} Here $q_y>q_{\rm c}$ lies on the edge branch, $\Psi(q_y)=q_y^2D_{\max}-r$, so that the fluid component~is
\begin{equation}
        \mathcal{F}_y(\xi)=p_{q_y}(\xi)
        ={re^{q_y\xi}\over 2q_y\sqrt{D_{\max}}}
        \int_0^{D_{\max}}{W(D)\over\sqrt D}
        \exp\left(-q_y|\xi|\sqrt{D_{\max}\over D}\right)\mathrm{d}D ,
        \label{eq:pyfluid}
\end{equation}
with tails given by Eq.~\eqref{eq:edge-tails}. In contrast to the other two phases, this microscopic law is defective: by Eq.~(\ref{eq:mass-my}), it carries only the mass $m_y=(y_+/y)^2<1$. The complementary mass $1-m_y$ must be carried by first runs  of duration $\tau$ in Eq.~(\ref{eq:jointfirst}) which are not microscopic, as we show in Secs.~\ref{sec:no-switch} and~\ref{sec:cond-saturated}. This regime is sketched in Fig.~\ref{fig:runs-dist} (right column).

\subsection{No-switching contribution}
\label{sec:no-switch}

We next consider the first term of Eq.~\eqref{eq:exactfirst}, i.e., trajectories without any switch up to time $t$, for which the first run spans the whole observation time and carries the full displacement, $\Delta x_1=yt$. Its weight $a_t(y)$, given in Eq.~\eqref{eq:at-def}, is the ratio of the no-switching endpoint density to $Z_t(y)$. To evaluate its numerator for fixed $y>0$, note that the Gaussian factor $G_{D,t}(yt)$ strongly favors the largest diffusion constants, so that the $D$-integral is dominated by a boundary layer close to $D_{\max}$. We therefore set $D=D_{\max}-\zeta/t$ with $\zeta=O(1)$. Then
\begin{equation}
        e^{-rt}G_{D,t}(yt)
        \sim
        {e^{-tJ(y)}\over\sqrt{4\pi D_{\max}t}}e^{-q_y^2\zeta},
        \qquad
        J(y)=r+{y^2\over4D_{\max}} .
        \label{eq:no-switching-local-edge}
\end{equation}
The edge law $W(D_{\max}-\zeta/t)\sim C(\zeta/t)^\nu$ from Eq.~\eqref{eq:edge-expo-def} and the Jacobian
$\mathrm{d}D=\mathrm{d}\zeta/t$ then give
\begin{equation}
        e^{-rt}\int_0^{D_{\max}}W(D)G_{D,t}(yt)\,\mathrm{d}D
        \sim
        {C\Gamma(\nu+1)\over
        \sqrt{4\pi D_{\max}}\,q_y^{2\nu+2}}
        t^{-(\nu+3/2)}e^{-tJ(y)} .
        \label{eq:no-switching-numerator-asympt}
\end{equation}
The exponential cost $J(y)$ is the cost of a ballistic displacement at velocity $y$ during the whole time $t$ with the maximal diffusion constant $D_{\max}$, and no switch, while the algebraic prefactor reflects the width $O(1/t)$ of the relevant boundary layer of diffusion constants. Dividing Eq.~\eqref{eq:no-switching-numerator-asympt} by the asymptotic forms of $Z_t(y)$ from Sec.~\ref{sec:renewal} gives the no-switching weights in the three phases reported below.\\

\noindent\textbf{Fluid and coexistence phases: $0<y<y_+$.} Here, the no-switching cost exceeds the rate function, $J(y)>I(y)$. In the fluid phase, this follows from $I(y)=q_*y-\Psi(q_*)<q_*y-(q_*^2D_{\max}-r)\leq J(y)$, with $q_* = q_*(y)$, using Eq.~\eqref{eq:gap-positive}. In the coexistence phase, using $I(y)=q_{\rm c}y-\Psi(q_{\rm c})$ and $r+\Psi(q_{\rm c})=q_{\rm c}^2D_{\max}$, one finds explicitly
\begin{equation}
        a_t(y)=\exp\left(-t[J(y)-I(y)]+o(t)\right),
        \qquad
        J(y)-I(y)={ (y_+-y)^2\over 4D_{\max}},
        \qquad y_-<y<y_+ .
        \label{eq:delta-exp-small-main}
\end{equation}
The no-switching weight is therefore exponentially small for $y<y_+$, and the Dirac contribution in Eq.~\eqref{eq:exactfirst} plays no role at long times.\\

\noindent\textbf{Saturated phase: $y>y_+$.} Here $I(y)=J(y)$, and $Z_t(y)$ in Eq.~\eqref{eq:Z-sat-intro} has the same power $t^{-(\nu+3/2)}$ as the no-switching numerator~\eqref{eq:no-switching-numerator-asympt}. Comparing with the prefactor~\eqref{eq:A-y-intro}, in which $(1-q_{\rm c}^2/q_y^2)^2=(1-m_y)^2$ with $m_y=(y_+/y)^2$ the mass of the fluid component, Eq.~\eqref{eq:mass-my}, we obtain a finite weight of the no-switching Dirac delta peak:
\begin{equation}
        a_t(y)\longrightarrow w_y:=(1-m_y)^2,
        \qquad y>y_+ .
        \label{eq:my-w-intro}
\end{equation}
Thus, in the saturated phase, the event that no switch at all occurs before time $t$ has a finite probability under the endpoint conditioning. It accounts for part of the mass $1-m_y$ missing from the fluid component; the remainder is derived in Sec.~\ref{sec:cond-saturated}.

\subsection{Macroscopic runs: the condensate}
\label{sec:condensate}

We now turn to first runs of macroscopic duration and displacement, $\tau=O(t)$ and $\xi=O(t)$, in Eq.~(\ref{eq:jointfirst}). As for the no-switching contribution, such a first run can only be realized with diffusion constants in a boundary layer close to $D_{\max}$. Setting $\xi=\rho t$ and $\tau=\theta t$, one finds (see Appendix~\ref{app:Q-edge} for details)
\begin{equation}
\begin{split}
        Q(\rho t,\theta t)
        &\sim
        {rC\Gamma(\nu+1)\over
        \sqrt{4\pi D_{\max}}\,q_{\rho,\theta}^{2\nu+2}}
        (\theta t)^{-(\nu+3/2)}
        \exp\left\{-t\left[r\theta+{\rho^2\over4D_{\max}\theta}\right]\right\},
        \qquad
        q_{\rho,\theta}={\rho\over2D_{\max}\theta} \;,
\end{split}
        \label{eq:C-Q-edge-global}
\end{equation}
for $Q$ as defined in Eq.~\eqref{eq:C-Qdef}.
This asymptotics of the first-run factor, together with the precise asymptotics of the endpoint density of the remaining trajectory in Eqs.~\eqref{eq:I-linear-intro} and~\eqref{eq:Z-sat-intro}, determines all macroscopic contributions below.
We first discuss the coexistence phase, where the condensate carries a finite fraction of the displacement, and then the saturated phase, where it carries all of it.

\subsubsection{Coexistence phase: the condensate bump}
\label{sec:cond-intermediate}

In the coexistence phase, the microscopic background is the critical fluid of Sec.~\ref{sec:fluid}, whose large positive increments are only algebraically suppressed, cf.~Eq.~\eqref{eq:edge-tails}. Under the endpoint constraint $x(t)=yt$ with $y>y_-$, this heavy-tailed background favors a single-big-jump mechanism, in which the extensive excess displacement is carried by one macroscopic run. We now determine the position, the width, the weight, and the shape of the corresponding peak of $P_t(\xi\mid y)$. The lengthy technical details are given in Appendix~\ref{app:critical-profile}.\\

\noindent\textbf{Position and duration of the condensate.} For a first run with
$\xi=\rho t$ and $\tau=\theta t$, with $0 \leq \rho \leq y$ and $0\leq \theta \leq 1$, the remaining trajectory has duration
$(1-\theta)t$, displacement $(y-\rho)t$, and therefore velocity
\begin{equation}
        z={y-\rho\over1-\theta}.
        \label{eq:C-zdef}
\end{equation}
We are interested in the sector where the first run is the unique macroscopic
run.  The remaining trajectory must then stay on the fluid branch, or at its
endpoint, so that
\begin{equation}
        z\leq y_-.
        \label{eq:C-one-condensate-domain}
\end{equation}
On this branch, recall from Eq.~\eqref{eq:microcanonical-summary} that the rate function is given by
\begin{equation}
        \phi_\nu(z)=q(z)z-\Psi(q(z)),
        \qquad
        z=\Psi'(q(z)),
        \qquad
        q(z)\leq q_{\rm c} \;.
        \label{eq:C-Ifluid}
\end{equation}
By Eqs.~\eqref{eq:Z-regular-intro} and~\eqref{eq:C-Q-edge-global}, the integrand of Eq.~\eqref{eq:exactfirst} is therefore equivalent, to leading (exponential) order for large $t$, to $e^{-t s_y(\theta,\rho)}$ with
\begin{equation}
        s_y(\theta,\rho)
        =r\theta+{\rho^2\over4D_{\max}\theta}
        +(1-\theta)\phi_\nu\left({y-\rho\over1-\theta}\right),
        \label{eq:C-effective-action}
\end{equation}
which has to be minimized over both $\theta$ and $\rho$, subject to Eq.~\eqref{eq:C-one-condensate-domain}. An interior stationary point with $z<y_-$ would require $r+\Psi(q)=q^2D_{\max}$ at $q=\phi_\nu'(z)<q_{\rm c}$, which is impossible by Eq.~\eqref{eq:gap-positive}. The minimum is therefore attained on the boundary $z=y_-$, where one finds the exact identity
\begin{equation}
\begin{split}
        s_y(\theta,\rho)-I(y)
        ={(\rho-y_+\theta)^2\over4D_{\max}\theta},
        \qquad
        y_+=2D_{\max} q_{\rm c}.
\end{split}
        \label{eq:C-action-square}
\end{equation}
The minimum is hence attained at $\rho_*=y_+\theta_*$, with zero excess cost. Together with the boundary condition $z=y_-$, this fixes the duration and the displacement of the condensate,
\begin{equation}
        \tau_*(y)=t{y-y_-\over y_+-y_-},
        \qquad
        \xi_*(t)=y_+\tau_*(y)=\rho_*(y)t,
        \qquad
        \rho_*(y)=y_+{y-y_-\over y_+-y_-}<y.
        \label{eq:taustar}
\end{equation}
Equivalently, the endpoint constraint can be written as
\begin{equation}
        yt-\xi_*(t)=y_-\bigl(t-\tau_*(y)\bigr).
        \label{eq:coexistenceidentity}
\end{equation}
Thus, the macroscopic run carries displacement at the velocity $y_+$, while the remaining microscopic background stays at the critical velocity~$y_-$: once the condensate run is removed, the remaining trajectory lies precisely at the endpoint of the fluid branch. The fraction $\tau_*(y)/t$ of time spent in the condensate is given by the lever rule for coexisting phases with ``densities'' $y_-$ and $y_+$, familiar from equilibrium statistical mechanics (see, e.g., Ref.~\cite{peliti2024statistical}), except that the phase coexistence here occurs in \textit{time}. Including the fluid phase and the saturated phase (Sec.~\ref{sec:cond-saturated}), the fraction of the endpoint displacement carried by the largest run is then (cf.~Fig.~\ref{fig:chi-fstar}, right)
\begin{equation}
        f_*(y):=\lim_{t\to\infty}{\xi_*(t)\over yt}
        =
        \begin{cases}
        0, & 0<y<y_-,\\[1mm]
        \left(1 - \frac{y_-}{y} \right) / \left(1 - \frac{y_-}{y_+} \right),
        & y_-<y<y_+,\\[3mm]
        1, & y>y_+ .
        \end{cases}
        \label{eq:condensate-fraction}
\end{equation}
Since the excess cost~\eqref{eq:C-action-square} vanishes at the saddle, the numerator has the same exponential factor $e^{-tI(y)}$ as the denominator $Z_t(y)$, and the condensate carries a subexponential weight. Note that the restriction~\eqref{eq:C-one-condensate-domain} is essential here: values $z>y_-$ correspond to remaining trajectories which are themselves condensed, i.e., to configurations with a second macroscopic run. These have the same exponential cost but a smaller algebraic prefactor, see the weight of the condensate below and Sec.~\ref{sec:interior}.\\

\noindent\textbf{Width of the condensate.} The saddle point fixes the condensate only to leading order. To resolve its shape, we expand the exact integral in Eq.~\eqref{eq:exactfirst} in a subextensive neighborhood of the saddle by writing
\begin{equation}
        \xi=\xi_*(t)+bw_t,
        \qquad
        \tau=\tau_*(y)+{aw_t\over y_+},
        \label{eq:C-plan-local}
\end{equation}
with $a,b=O(1)$. The scale $w_t=o(t)$ is not assumed beforehand, but has to be determined. Two sources of fluctuations contribute to the bump, which can be seen by analyzing Eq.~(\ref{eq:jointfirst}). First, the displacement of the condensate run fluctuates around $y_+\tau$, with Gaussian fluctuations of variance $2D_{\max}\tau_*(y)=O(t)$ inherited from $G_{D,\tau}$ as defined in Eq.~\eqref{eq:G-def}.
Second, the endpoint of the remaining critical background, of duration $T_*=t-\tau_*(y)$, fluctuates around $y_-T_*$. Because the exponential factors of the two contributions cancel exactly against $e^{-tI(y)}$ in the neighborhood of the saddle, the profile of the bump is the convolution of these two distributions (see Appendix~\ref{app:critical-profile} for details). The background fluctuations are described by the law of the rescaled endpoint displacement $u=(x-y_-T)/w_t$ of a background trajectory of duration $T$, tilted by $e^{q_{\rm c}x}$,
\begin{equation}
        F_T(u):=r\mu_{\rm c} w_t\,
        e^{q_{\rm c}(y_-T+uw_t)-T\Psi(q_{\rm c})}
        Z(y_-T+uw_t,T),
        \label{eq:C-FT-real-def}
\end{equation}
where the factor $r\mu_{\rm c}$ ensures the normalization $\int_{\mathbb R}F_T(u)\,\mathrm{d}u\to1$ shown below. Its characteristic function $\widehat F_T(k)=\int_{\mathbb R}e^{iku}F_T(u)\,\mathrm{d}u$ is obtained from the two-sided Laplace transform of the endpoint density, evaluated at the complex argument $q=q_{\rm c}+ik/w_t$. On the fluid side of $q_{\rm c}$, this transform behaves for large $T$ as $\chi(q)e^{T\Psi(q)}$, with the residue $\chi(q)$ of Eq.~\eqref{eq:Breg-intro}, which tends to $\chi_{\rm c}=1/(r\mu_{\rm c})$ as $q\uparrow q_{\rm c}$ (cf.\ Sec.~\ref{sec:fluid-duration}). This gives
\begin{equation}
\begin{split}
        \widehat F_T(k)
        \sim &\;r\mu_{\rm c}\,
        \chi\left(q_{\rm c}+{ik\over w_t}\right)
        \exp\left\{
        T\left[
        \Psi\left(q_{\rm c}+{ik\over w_t}\right)
        -\Psi(q_{\rm c})-{ik y_-\over w_t}
        \right]\right\},
\end{split}
        \label{eq:C-FT-master}
\end{equation}
where $r\mu_{\rm c}\,\chi(q_{\rm c}+ik/w_t)\to1$ for $w_t\to\infty$. The scale $w_t$ is fixed by requiring the exponent in
Eq.~\eqref{eq:C-FT-master} to have a finite and non-zero limit, so that the background fluctuations have a non-degenerate limiting law. Since the linear term $-y_-h$ of $\Psi(q_{\rm c}-h)$ is canceled by the centering at $y_-T$, this limit is controlled entirely by the leading correction beyond the linear term in the expansion of the fluid branch $\Psi(q)$ as $q\uparrow q_{\rm c}$: the non-analytic term $h^\nu$ for $1<\nu<2$, the marginal term $h^2\log(1/h)$ for $\nu=2$, and the regular quadratic term for $\nu>2$, where the non-analytic $h^\nu$ term is subleading. Which case applies is determined by the edge exponent $\nu$, which also sets the power-law tail $\xi^{-(\nu+1)}$ of the critical fluid. All three expansions follow from the pole equation~\eqref{eq:secular}, see Appendix~\ref{sec:crit-background-analysis}. In the following, we use the finite integral
\begin{equation}
        J_{2,c}:=
        \int_0^{D_{\max}}{W(D) \,\mathrm{d}D\over(D_{\max}-D)^2},
        \label{eq:C-J2c-def}
\end{equation}
which is related to the quantities of Sec.~\ref{sec:fluid} by $J_{2,c}=q_{\rm c}^4J_2(q_{\rm c})$ and $\mu_{\rm c}=rJ_{2,c}/q_{\rm c}^4$. We also write $y_-=\Psi'(q_{\rm c}^-)=:2q_{\rm c}D_{\rm eff}$, which defines an effective diffusion constant $D_{\rm eff}$ of the critical fluid. There are three cases:
\begin{itemize}
\item For $1<\nu<2$, the critical fluid has infinite variance, and the fluid branch has the expansion
\begin{equation}
        \Psi(q_{\rm c}-h)
        =\Psi(q_{\rm c})-y_-h+K_\nu h^\nu+o(h^\nu),
        \qquad h\downarrow0,
        \qquad
        K_\nu:=
        {\pi C q_{\rm c}^2\over |\sin(\pi\nu)|J_{2,c}}
        \left({2q_{\rm c}\over rJ_{2,c}}\right)^\nu>0.
        \label{eq:C-Psi-stable}
\end{equation}
The exponent in Eq.~\eqref{eq:C-FT-master} then has a finite limit for
\begin{equation}
        w_t=
        \left[K_\nu\left|\cos{\pi\nu\over2}\right|T_*\right]^{1/\nu},
        \label{eq:C-wt-stable}
\end{equation}
and $\widehat F_T(k)$ converges to the characteristic function of a completely asymmetric stable law. This corresponds to a L{\'e}vy central limit theorem for the sum of the $O(t)$ heavy-tailed critical increments~\cite{evans-majumdar-zia:2006,nolan:2020}, since the condensate displacement is given by
\begin{align}
    \Delta x_{\rm condensate} = y t - \sum_{i \in {\rm fluid}} \Delta x_i\,.
    \label{eq:condensate-fluid-balance}
\end{align}
\item For $\nu=2$, the expansion contains a marginal logarithm,
\begin{equation}
        \Psi(q_{\rm c}-h)
        =\Psi(q_{\rm c})-y_-h
        +B_{\log}h^2\log{1\over h}+A_2h^2+o(h^2),
        \qquad
        B_{\log}
        ={4C(D_{\max}-D_{\rm eff})^2\over J_{2,c}},
        \label{eq:C-Psi-marginal}
\end{equation}
where $A_2$ is a regular, non-universal coefficient. This requires, to leading order,
\begin{equation}
        w_t=\left(B_{\log}T_*\log T_*\right)^{1/2},
        \label{eq:C-wt-marginal}
\end{equation}
and $\widehat F_T(k)\to e^{-k^2/2}$ becomes Gaussian.
\item For $\nu>2$, the critical fluid has finite variance and $\Psi(q)$ is twice differentiable on the fluid side of $q_{\rm c}$,
\begin{equation}
        \Psi(q_{\rm c}-h)
        =\Psi(q_{\rm c})-y_-h+{\sigma_{\rm c}^2\over2}h^2+o(h^2),
        \qquad
        \sigma_{\rm c}^2
        ={2\over J_{2,c}}
        \left[
        4\int_0^{D_{\max}}{W(D)(D_{\rm eff}-D)^2\over(D_{\max}-D)^3}\,\mathrm{d}D
        +\int_0^{D_{\max}}{D W(D)\,\mathrm{d}D\over(D_{\max}-D)^2}
        \right],
        \label{eq:C-Psi-Gaussian}
\end{equation}
such that the usual central limit theorem applies in Eq.~\eqref{eq:condensate-fluid-balance}, with
\begin{equation}
        w_t=\sqrt{\sigma_{\rm c}^2T_*},
        \label{eq:C-wt-Gaussian}
\end{equation}
and $\widehat F_T(k)\to e^{-k^2/2}$.
\end{itemize}
In summary, the width of the condensate peak is
\begin{equation}
        w_t\simeq
        \begin{cases}
        t^{1/\nu}, & 1<\nu<2 \;,\\[1mm]
        (t\log t)^{1/2}, & \nu=2\;,\\[1mm]
        t^{1/2}, & \nu>2 \;.
        \end{cases}
        \label{eq:wt-intro}
\end{equation}
up to multiplicative $O(1)$ constants given in Eqs.~\eqref{eq:C-wt-stable}, \eqref{eq:C-wt-marginal}, and~\eqref{eq:C-wt-Gaussian}, with $T_*=t-\tau_*(y)=t(y_+-y)/(y_+-y_-)$. For $1<\nu\leq2$, $w_t\gg t^{1/2}$, so that the Gaussian fluctuations of the condensate run itself are negligible, and the shape of the bump is entirely determined by the critical background. For $\nu>2$, both contributions are of the same order $t^{1/2}$.
The case $\nu=2$ is marginal.  Indeed, Eq.~\eqref{eq:edge-tails}
gives
$p_{q_{\rm c}}(\xi)\sim A_{\rm c}\xi^{-3}$ as $\xi\to+\infty$ for the critical fluid when $\nu = 2$, so the second moment of a
critical increment diverges only logarithmically.  The sum of $O(t)$ critical
increments therefore fluctuates on the scale $(t\log t)^{1/2}$, rather than
$t^{1/2}$, signaling the crossover between the stable scale $t^{1/\nu}$ for
$1<\nu<2$ and the Gaussian scale $t^{1/2}$ for $\nu>2$. The latter is directly related to the $h^2\log(1/h)$ singularity of $\Psi(q)$ in Eq.~\eqref{eq:C-Psi-marginal}. The non-Gaussian $k^2\log|k|$ structure survives in the first, slowly decaying $O(1/\log t)$ correction to scaling, which is discussed in Appendix~\ref{sec:crit-background-analysis}.\\

\noindent\textbf{Weight of the condensate.} It remains to compare the pre-exponential factor of the numerator in Eq.~\eqref{eq:exactfirst}, integrated over the bump, with the prefactor of the denominator $Z_t(y)$. The numerator amplitude follows directly by integrating the local form of the bump, which is normalized. Its ratio to the denominator is obtained in Appendix~\ref{app:critical-amplitudes} from the double Laplace transform of the endpoint density near $(q_{\rm c},\Psi(q_{\rm c}))$, where a single contribution of the edge layer $D\uparrow D_{\max}$ selects configurations with a single condensate.
The numerator is of order $t^{-(\nu+1)}e^{-tI(y)}$, while the denominator is of order $t^{-\nu}e^{-tI(y)}$. Configurations with $k\geq2$ macroscopic runs correspond to higher powers of the edge contribution in the same Laplace-space expansion; a direct power counting (Appendix~\ref{app:critical-amplitudes}) shows that they are suppressed by a relative factor $O(t^{(k-1)(1-\nu)})$, which justifies the restriction to the single-condensate sector for $\nu>1$. This is confirmed by the explicit split-condensate asymptotics of Sec.~\ref{sec:interior}. In particular, this computation completes the endpoint asymptotics~\eqref{eq:I-linear-intro} announced in Sec.~\ref{sec:renewal}, with the amplitude
\begin{equation}
        B_{\rm int}(y)=
        {C\Gamma(\nu+1)(y_+-y)(y_+-y_-)^{\nu-1}
        \over
        \mu_{\rm c}^2q_{\rm c}^{2\nu+2}(y-y_-)^{\nu+1}},
        \label{eq:Bint-intro}
\end{equation}
where $\mu_{\rm c}$ is the mean duration~\eqref{eq:muc} of a critical fluid run. This expression is valid away from the crossover regions near $y_-$ and $y_+$. Consequently, the condensate peak has mass $1/N_t$, with
\begin{equation}
        N_t\sim {t-\tau_*(y)\over \mu_{\rm c}} .
        \label{eq:Nt}
\end{equation}
This result has a simple interpretation. Since the
background has duration $t-\tau_*(y)$ and mean run duration $\mu_{\rm c}$,
cf.\ Eq.~\eqref{eq:muc}, $N_t$ is the number of microscopic runs in the background. A typical conditioned trajectory contains $N_t$ microscopic background runs
and one condensate run. Hence, a specified run is the condensate with
probability asymptotic to $1/N_t$, while it belongs to the microscopic
background with the complementary probability $1-1/N_t$.\\

\noindent\textbf{Results.} Altogether, in the coexistence phase $y_-<y<y_+$, the fluid part freezes at its critical form and a second peak, the ``condensate bump'',
appears around $\xi_*(t)$ to reach the required total displacement $x(t) = y t$:
\begin{equation}
        \left\{
        \begin{array}{ll}
        P_t(\xi\mid y)\simeq {\cal F}_y(\xi),
        & \xi=O(1),\\[2mm]
        \displaystyle
        P_t(\xi\mid y)\sim {1\over N_tw_t}
        \Phi_{\nu,y}\!\left({\xi-\xi_*(t)\over w_t}\right),
        & \xi=\xi_*(t)+O(w_t),
        \end{array}
        \right.
        \label{eq:inter-main-result}
\end{equation}
with ${\cal F}_y(\xi)=p_{q_{\rm c}}(\xi)$, $\xi_*(t)$ from Eq.~\eqref{eq:taustar}, $N_t$ from Eq.~\eqref{eq:Nt}, and $w_t$ from Eqs.~\eqref{eq:C-wt-stable}--\eqref{eq:C-wt-Gaussian}, while the no-switching weight~\eqref{eq:delta-exp-small-main} is exponentially small. On the microscopic scale, the first-run law thus converges to the critical fluid law; since the condensate takes the mass $1/N_t$ and all other contributions are $o(1/t)$, the total mass of the fluid sector is $1-1/N_t+o(1/t)$. The normalized profile of the condensate bump is
\begin{equation}
        \Phi_{\nu,y}(b)=
        \begin{cases}
        \gamma_y V_\nu(-\gamma_y b), & 1<\nu<2,\\[1mm]
        {\cal G}_{\Sigma_{2,y}}(b), & \nu=2,\\[1mm]
        {\cal G}_{\Sigma_{>,y}}(b), & \nu>2,
        \end{cases}
        \qquad
        \gamma_y={y_+-y_-\over y_+},
        \qquad
        \int_{-\infty}^\infty\Phi_{\nu,y}(b)\,\mathrm{d} b=1 .
        \label{eq:Phi-profile-main}
\end{equation}
For $1<\nu<2$, the function $V_\nu$ appearing in
Eq.~\eqref{eq:Phi-profile-main} is the completely asymmetric stable
scaling function
\begin{equation}
        V_\nu(u)={1\over 2\pi}
        \int_{-\infty}^\infty \exp\left(-iku-|k|^\nu\left(1-i\mathrm{sgn}(k)\tan\left(\pi\nu\over2\right)\right)\right)\,\mathrm{d}k ,
        \label{eq:stablelaw}
\end{equation}
which is the same as in Eq.~(95) of Ref.~\cite{mori-le-doussal-majumdar-etal:2021}.
The centered Gaussian of variance $\Sigma$ in
Eq.~\eqref{eq:Phi-profile-main} is
\begin{equation}
        {\cal G}_\Sigma(b)={1\over\sqrt{2\pi\Sigma}}
        \exp\left(-{b^2\over2\Sigma}\right),
        \label{eq:Gaussian-profile}
\end{equation}
and the variances are
\begin{equation}
        \Sigma_{2,y}={1\over\gamma_y^2},
        \qquad
        \Sigma_{>,y}
        ={1\over\gamma_y^2}
        \left[1+{2D_{\max}\beta^2\tau_*(y)\over
        \sigma_{\rm c}^2T_*}\right],
        \qquad
        \beta={y_-\over y_+}.
        \label{eq:C-Sigma-Gaussian}
\end{equation}
Note that the notation defined by Eq.~\eqref{eq:Gaussian-profile} is slightly different from the Gaussian introduced earlier in Eq.~(\ref{eq:G-def}).\\

For $\nu>2$, the second term in $\Sigma_{>,y}$ is the contribution of the Gaussian fluctuations of the condensate run itself. All three cases are derived in
 Appendix~\ref{app:critical-profile}. Because of Eq.~\eqref{eq:Phi-profile-main}, we refer to the condensate as ``anomalous'' for $1 < \nu < 2$ and ``normal'' for $\nu \geq 2$, with $\nu=2$ marginal. As in mass transport models~\cite{evans-majumdar-zia:2006}, the distinction reflects whether the single condensate run, which balances the sum of $O(t)$ critical fluid increments under the constraint $x(t)=yt$ via Eq.~\eqref{eq:condensate-fluid-balance}, inherits Gaussian or stable fluctuations from this sum. For $\nu\geq2$, the profile is centered and its maximum lies at $\xi_*(t)+o(w_t)$, whereas for $1<\nu<2$ the profile is skewed, and its maximum is shifted to $\xi_*(t)-m_\nu w_t/\gamma_y+o(w_t)$, where $m_\nu$ denotes the mode (i.e., the maximum) of $V_\nu$. For $1<\nu<2$, the profile of the condensate peak has the asymptotic behavior
\begin{equation}
        \Phi_{\nu,y}(b)\sim
        \begin{cases}
        A_\nu\gamma_y^{-\nu}|b|^{-(1+\nu)},
        & b\to-\infty,\\[1mm]
        B_\nu\gamma_y^{1-\eta_\nu}b^{-\eta_\nu}
        \exp\left(-C_\nu\gamma_y^{\nu/(\nu-1)}b^{\nu/(\nu-1)}\right),
        & b\to+\infty,
        \end{cases}
        \label{eq:B-tails-intro}
\end{equation}
where the constants describing the tails of the stable law are
\begin{equation}
\begin{split}
        A_\nu&={2\Gamma(\nu+1)\over\pi}\sin{\pi\nu\over2},
        \qquad
        \eta_\nu={\nu-2\over2(\nu-1)},\\[1mm]
        B_\nu&={1\over\sqrt{2\pi(\nu-1)}}
        \left({|\cos(\pi\nu/2)|\over\nu}\right)^{1/[2(\nu-1)]},\quad 
        C_\nu=(\nu-1)\nu^{-\nu/(\nu-1)}
        |\cos(\pi\nu/2)|^{1/(\nu-1)} .
\end{split}
        \label{eq:stable-tail-constants}
\end{equation}
Observe that it is the \textit{left} tail of the anomalous condensate profile $\Phi_{\nu,y}(b)$ in Eq.~\eqref{eq:B-tails-intro} that displays a power law, although the stable law $V_\nu(u)$ itself is heavy-tailed to the right -- this is because of the reflection $b\to-\gamma_y b$ in Eq.~\eqref{eq:Phi-profile-main}, which in turn stems from the minus sign in the balance condition~\eqref{eq:condensate-fluid-balance}. As shown in Sec.~\ref{sec:interior}, this algebraic tail connects continuously to configurations in which the condensate is split between two macroscopic runs. For $\nu\geq2$, in contrast, the limiting profile $\Phi_{\nu,y}(b)$ is Gaussian, and the Gaussian form describes the bump only up to values of $|b|$ which grow logarithmically with $t$: on its left flank, $P_t(\xi\mid y)$ eventually crosses over to the algebraic decay $A_{\rm c}(\xi_*(t)-\xi)^{-(\nu+1)}$ of the split-condensate configurations, cf.\ Eq.~\eqref{eq:interior-coex-main} for $\omega\to1$. Comparing the two expressions, this crossover occurs at $b\simeq-\sqrt{(\nu-2)\Sigma_{>,y}\log t}$ for $\nu>2$, and at $|b|=O(\sqrt{\log\log t})$ for $\nu=2$.

\subsubsection{Saturated phase: the condensate carries the full displacement}
\label{sec:cond-saturated}

For $y>y_+$, Secs.~\ref{sec:fluid} and~\ref{sec:no-switch} have shown that the fluid component carries the mass $m_y$ and the no-switching Dirac peak the mass $w_y$, which leaves a mass $m_y(1-m_y)$ to be accounted for. This mass is located at
\begin{equation}
        \xi=yt-u,
        \qquad
        u=O(1),
        \label{eq:saturated-u}
\end{equation}
i.e., the macroscopic first run has displacement $yt-O(1)$, while the
remaining trajectory carries only the finite residual displacement $u$. A first run carrying the displacement $yt-u$ must occupy
almost the whole observation time, and must again have a diffusion constant close to $D_{\max}$. We therefore write
\begin{equation}
        \tau=t-s,
        \qquad
        D=D_{\max}-{\zeta\over t},
        \qquad
        s,u,\zeta=O(1).
        \label{eq:E-local-variables}
\end{equation}
The variables have a direct interpretation: $s$ is the duration of the
residual trajectory, $u$ is its total displacement, and $\zeta$ resolves the
boundary layer of diffusion constants near $D_{\max}$.  Expanding the exponential
cost of the long first run in Eq.~\eqref{eq:exactfirst} gives 
\begin{equation}
\begin{split}
        r(t-s)+{(yt-u)^2\over
        4(D_{\max}-\zeta/t)(t-s)}
        ={}&t\left(r+{y^2\over4D_{\max}}\right)
        -{y\over2D_{\max}}u+\left({y^2\over4D_{\max}}-r\right)s
        +{y^2\over4D_{\max}^2}\zeta+o(1).
\end{split}
        \label{eq:E-exponent-expansion}
\end{equation}
Each term can be read directly.  The first is the saturated
large-deviation cost $J(y)=I(y)$ of the full trajectory, which cancels against the exponential factor of $Z_t(y)$.  The second is the correction due
to the $O(1)$ displacement $u$ carried by the residual trajectory.  The third
accounts for its finite duration $s$, while the last term resolves the
boundary layer $D_{\max}-D=\zeta/t$. Since the numerator and the denominator $Z_t(y)$ have the same power $t^{-(\nu+3/2)}$ and the same edge amplitude, their ratio has a finite limit (Appendix~\ref{app:saturated}). This gives the continuous part of the single-run PDF at macroscopic run size,
\begin{equation}
        {\cal P}_y^{\rm cont}(u)=r w_y
        \exp\!\left({yu\over2D_{\max}}\right)
        \int_0^\infty
        \exp\!\left(-s\left({y^2\over4D_{\max}}-r\right)\right)
        Z(u,s)\,\mathrm{d} s .
        \label{eq:Pcont-main}
\end{equation}
Since $y^2/(4D_{\max})-r=\Psi(q_y)$ and $y/(2D_{\max})=q_y$, the residual trajectory is tilted in exactly the same way as the microscopic fluid runs in Sec.~\ref{sec:fluid}. Correspondingly, expanding its endpoint density $Z(u,s)$ in a renewal series (Appendix~\ref{app:saturated}) gives the convolution representation
\begin{equation}
        \mathcal{P}_y^{\rm cont}(u)=w_y
        \sum_{n\geq1}\mathcal{F}_y^{*n}(u),
        \label{eq:E-Pcont-renewal-series}
\end{equation}
where $\mathcal{F}_y^{*n}(u)$ denotes the $n$-fold convolution, evaluated at $u$, of the defective fluid density
$\mathcal{F}_y(\xi)$ from Eq.~\eqref{eq:pyfluid}.  The term with index $n$ represents a residual cloud containing $n$
microscopic runs. Since $\mathcal{F}_y$ has mass $m_y$, Eq.~\eqref{eq:E-Pcont-renewal-series} gives the mass $w_y\sum_{n\geq1}m_y^n=m_y(1-m_y)$, as anticipated.\\

\noindent\textbf{Results.} For the saturated phase $y>y_+$, we define the macroscopic sector ${\cal P}_y^{\rm sat}(u):={\cal P}_y^{\rm cont}(u)+w_y\delta(u)$, consisting of the continuous density~\eqref{eq:Pcont-main} and the no-switching Dirac peak. There are hence two distinct scaling regimes,
\begin{equation}
        \left\{
        \begin{array}{ll}
        P_t(\xi\mid y)\sim {\cal F}_y(\xi),
        & \xi=O(1),\\[2mm]
        P_t(\xi=yt-u\mid y)\sim {\cal P}_y^{\rm sat}(u),
        & u=yt-\xi=O(1),
        \end{array}
        \right.
        \label{eq:main-cases}
\end{equation}
where ${\cal F}_y(\xi)$ is given in Eq.~\eqref{eq:pyfluid}. The fluid component, the Dirac peak, and the continuous macroscopic density carry the weights
\begin{equation}
        \int_{-\infty}^\infty {\cal F}_y(\xi)\,\mathrm{d}\xi=m_y,
        \qquad
        w_y=(1-m_y)^2,
        \qquad
        \int_{-\infty}^\infty {\cal P}_y^{\rm cont}(u)\,\mathrm{d} u=m_y(1-m_y),
        \label{eq:saturated-masses-main}
\end{equation}
which add up to one, with $m_y=(y_+/y)^2$ from Eq.~\eqref{eq:mass-my}. Notably, these weights are all $O(1)$, which is the
most direct signature of the second transition at $y_+$: unlike in the coexistence phase, where the condensate sector of a specified
run has weight $O(t^{-1})$, both the microscopic and macroscopic sectors
carry finite probability mass in the saturated phase. Asymptotically, the renewal series~\eqref{eq:E-Pcont-renewal-series} together with the tails~\eqref{eq:edge-tails} of ${\cal F}_y$ gives (Appendix~\ref{app:saturated})
\begin{equation}
        {\cal P}_y^{\rm cont}(u)\sim
        \begin{cases}
        A_y^{(+)}\,u^{-(\nu+1)}, & u\to+\infty,\\[1mm]
        A_y^{(+)}\,|u|^{-(\nu+1)}e^{-2q_y|u|}, & u\to-\infty .
        \end{cases}
        \label{eq:Pcont-tails-intro}
\end{equation}
Notice in
particular that the variable $u=yt-\xi$ may be negative: first-run
displacements larger than $yt$ are allowed, provided that the remaining
runs compensate with a negative total displacement.

\subsection{Interior regime: split condensates}
\label{sec:interior}

Finally, we probe the macroscopic interior of the conditional first-run distribution, i.e., first-run displacements $\xi=\rho t$ of order $t$ which lie between the microscopic fluid at $\xi=O(1)$ and the condensate: $0<\rho<\rho_*$ in the coexistence phase, and $0<\rho<y$ in the fluid and saturated phases. In other words, we are investigating the form of the marginal distribution in the regime between the end of the fluid background and the start of the condensate bump in Fig.~\ref{fig:runs-dist} (bottom) and Fig.~\ref{fig:cond-mcmc-vs-theory} (bottom) and we call this the interior regime.
The analysis follows the same lines as in Sec.~\ref{sec:cond-intermediate}: the first-run factor is given by the edge asymptotics~\eqref{eq:C-Q-edge-global} with $\tau=\theta t$, and the remaining trajectory, of duration $(1-\theta)t$ and velocity $z=(y-\rho)/(1-\theta)$, contributes its endpoint density with the precise asymptotics of Sec.~\ref{sec:renewal}. In contrast to Sec.~\ref{sec:cond-intermediate}, $z$ is now not restricted to the fluid branch, and the $\theta$-integral is evaluated by Laplace's method around a saddle point $\theta_\rho$. The physical mechanism and the resulting asymptotic form depend qualitatively on the phase in which the remaining trajectory finds itself at this saddle. The calculations are given in Appendix~\ref{app:interior}.\\

\noindent\textbf{Fluid interior.}
For $0<y<y_-$, typical conditioned trajectories
consist entirely of microscopic runs, and a macroscopic
first run is an atypical fluctuation: the remaining trajectory stays on the
regular fluid branch, $z_\rho<y_-$, and no condensate is formed. The saddle $\theta_\rho$ is determined by
\begin{equation}
D_{\max}q_{\rho,\theta_\rho}^2
=
r+\Psi( q_*(z_\rho)),
\qquad
z_\rho=\frac{y-\rho}{1-\theta_\rho},
\end{equation}
and one finds
\begin{equation}
P_t^{\rm cont}(\rho t\mid y)
\sim
\mathcal{H}_y(\rho)\,
t^{-(\nu+1)}
e^{-tJ_y^{(1)}(\rho)},
\qquad
0<\rho<y\, ,
\label{eq:interior-fluid-main}
\end{equation}
with the rate
\begin{equation}
J_y^{(1)}(\rho)
=
q_*(z_\rho)(y-\rho)
-
\Psi\!\left(q_*(z_\rho)\right)
+
\rho
\sqrt{
\frac{r+\Psi\!\left(q_*(z_\rho)\right)}
{D_{\max}}
}
-
I(y)>0\, ,
\end{equation}
and the amplitude $\mathcal{H}_y(\rho)$ given in Eq.~(\ref{eq:fluid-interior-amplitude}). Hence, a macroscopic first run in the fluid phase is exponentially rare. For $\rho\to0$, one has $J_y^{(1)}(\rho)\simeq\lambda_+(q_*(y))\rho$ and $\mathcal{H}_y(\rho)\sim B(q_*(y))\rho^{-(\nu+1)}$, such that Eq.~\eqref{eq:interior-fluid-main} matches the exponentially cut-off right tail~\eqref{fluidtailsgeneral} of the fluid law for $1\ll\xi\ll t$.\\

\noindent\textbf{Coexistence interior.}
For $y_-<y<y_+$, we probe a macroscopic first-run displacement
$0<\xi<\xi_*(t)=\rho_* t$. At the saddle $\theta_\rho=\rho/y_+$, the velocity $z_\rho$ of the remaining trajectory lies inside the coexistence interval, $y_-<z_\rho<y_+$, for all $0<\rho<\rho_*$. Since the rate function is linear on this interval, the excess cost is again given by the perfect square~\eqref{eq:C-action-square}, which vanishes at the saddle, so that there is no exponential suppression at all. Writing
$\omega=\xi/\xi_*(t)\in(0,1)$, the first run carries the fraction
$\omega$ of the condensate, while the remaining trajectory is itself in
the coexistence phase and contains a second condensate carrying the
complementary fraction $1-\omega$. The critical fluid background remains
unchanged. One finds
\begin{equation}
P_t^{\rm cont}(\omega\rho_* t\mid y)
\sim
\frac{A_{\rm c}}
{(\rho_* t)^{\nu+1}[\omega(1-\omega)]^{\nu+1}},
\qquad
0<\omega<1\, ,
\label{eq:interior-coex-main}
\end{equation}
with $A_{\rm c}$ from Eq.~(\ref{eq:Ac-intro}). Thus the usual condensate of size $\rho_* t$ is split between two macroscopic runs of sizes $\omega\rho_* t$ and $(1-\omega)\rho_* t$, and the total conditioned displacement decomposes as
\begin{eqnarray}
    yt=\rho t+(\rho_*-\rho)t+(y-\rho_*)t\, ,
\end{eqnarray}
where the first two terms are the two macroscopic runs, while the last term is carried by the critical fluid background at velocity $y_-$. The two limits of Eq.~\eqref{eq:interior-coex-main} match the neighboring regimes: for $\omega\to0$, with $1\ll\xi\ll t$, one recovers the algebraic right tail $A_{\rm c}\xi^{-(\nu+1)}$ of the critical fluid law, Eq.~\eqref{eq:edge-tails}, while for $\omega\to1$, with $1\ll\xi_*(t)-\xi\ll t$, one recovers the algebraic left tail of the condensate bump~\eqref{eq:B-tails-intro} for $1<\nu<2$, including its amplitude (Appendix~\ref{app:interior}). In contrast, macroscopic first runs beyond the condensate, $\rho_*<\rho<y$, are exponentially suppressed. The remaining trajectory would then have to fall below the critical velocity $y_-$ of the fluid branch, at a strictly positive excess cost.\\

\noindent\textbf{Saturated interior.}
For $y>y_+$, we probe a macroscopic displacement
$0<\xi<yt$ and introduce $\omega=\xi/(yt)\in(0,1)$. At the saddle $\theta_\rho=\rho/y$, the remaining trajectory has velocity $z_\rho=y>y_+$ and the excess cost vanishes. The first run carries
the fraction $\omega$ of the total macroscopic displacement, while the
remaining trajectory remains in the saturated phase and carries the
complementary fraction $1-\omega$. Hence the saturated condensate is again
split between two macroscopic runs, and
\begin{equation}
P_t^{\rm cont}(\omega yt\mid y)
\sim
\frac{A_y^{(+)}}
{(yt)^{\nu+1}[\omega(1-\omega)]^{\nu+1}},
\qquad
0<\omega<1\, ,
\label{eq:interior-sat-main}
\end{equation}
where $A_y^{(+)}$ is defined in Eq.~(\ref{eq:Ac-intro}). In contrast to the coexistence phase, there is no extensive fluid background in this case, only subextensive residual contributions. Again, the limits $\omega\to0$ and $\omega\to1$ of Eq.~\eqref{eq:interior-sat-main} match, respectively, the right tail~\eqref{eq:edge-tails} of the fluid component, and the algebraic tail ${\cal P}_y^{\rm cont}(yt-\xi)\sim A_y^{(+)}(yt-\xi)^{-(\nu+1)}$ of the continuous saturated density~\eqref{eq:Pcont-tails-intro} for large positive residual displacement $yt-\xi$.

\section{Comparison to rare event sampling}
\label{sec:numerics}

To further validate our theoretical predictions for the single-run PDF and the trajectory condensation, we perform Monte Carlo simulations of the SD process conditioned on achieving a large displacement $x(t) = yt$ at the final time~$t \gg 1/r$. Because the target end point $x(t) = y t$
lies far in the tails of the unconditioned distribution, standard direct sampling is computationally prohibitive~\cite{bucklew:2013}. Instead, we sample directly from the conditional path distribution $((x(s))_{s \in [0,t]}, D_1, D_2, \dots, \tau_1, \tau_2, \dots) \mid x(t) = yt$
via Markov Chain Monte Carlo (MCMC) on the configuration space of the paths, roughly similar to Ref.~\cite{mori-le-doussal-majumdar-etal:2021} which sampled the condensation transition for RTPs at late times. Here, in contrast to Ref.~\cite{mori-le-doussal-majumdar-etal:2021}, we design the proposal moves to \textit{exactly} preserve the condition $x(t) = yt$ in all MCMC steps, and use reversible-jump MCMC~\cite{green:1995}
to correctly allow for fluctuations of the number of constant-$D$ segments within the fixed time interval $[0,t]$. This allows us to generate a large number of samples from the conditioned path ensemble\footnote{It does not allow us, however, to obtain the absolute probability, i.e., the normalization constant of the conditional path space distribution, of $x(t) \approx yt$ from these samples.}. A detailed description of the algorithm is deferred to Appendix~\ref{app:numerical-algorithm}.\\

We show a representative set of numerical results for the single run marginal PDF in the different phases in Fig.~\ref{fig:cond-mcmc-vs-theory}. This can be viewed as a quantitative version of the sketches in Fig.~\ref{fig:runs-dist} (bottom). In Fig.~\ref{fig:cond-mcmc-vs-theory}, we compare the results from MCMC, as $t$ increases, against a numerical solution of the exact renewal equation (see Appendix~\ref{app:renewal-numerics} for details) as well as the analytical approximations for large $t$ from Sec.~\ref{sec:single-run-main-text}. First of all, we see excellent agreement between the MCMC results and the renewal equation solution, which is expected since both of them should yield the true PDF without approximations (up to numerical errors) and hence serves as a consistency check. We point out that each MCMC histogram uses $1.2 \times 10^5$ independent samples, which explains why for large $t$, e.g., in the saturated phase for $t = 800$, the MCMC histograms in the deep ``valley'' between the fluid and condensate or no-switching contribution become very noisy once their probability (conditional on $x(t) = y t$) is of the order of $10^{-5}$.\\

We turn now to our analytical predictions for the various phases at large $t$ (dashed and dash-dotted lines). In the fluid and saturated phases in the top row of Fig.~\ref{fig:cond-mcmc-vs-theory}, we see that by patching together the asymptotic approximations to the PDF at $\xi = O(1)$ and $\xi = O(t)$, already at $t = 100$, we get good agreement with the full renewal solution and MCMC. For the condensate bump in the coexistence phase in the bottom row of Fig.~\ref{fig:cond-mcmc-vs-theory}, convergence as $t$ grows is much slower. Furthermore, as expected, for the anomalous condensate, convergence is visibly slower than for the normal condensate. At the largest value of $t = 800$ where we were able to obtain MCMC results, the asymptotic predictions for the condensate bump (dashed lines) already describe the location of the condensate well, while clear differences in the shape of the PDF are still visible. Using the numerical solution to the renewal equation at larger $t$ (not shown), we checked that closer quantitative agreement is only obtained for $t$ of the order of $10^5$. This slow convergence is due to finite-time corrections to the limiting profile $\Phi_{\nu,y}$ that decay only as small powers of $t$. For the anomalous condensate, the Gaussian fluctuations of the condensate run itself, which are discarded in the limit~\eqref{eq:C-Gamma-stable-vanishes}, have a variance relative to the stable scale of order $D_{\max}\tau_*(y)/w_t^2\propto t^{1-2/\nu}$, and the regular $h^2$ term in the expansion~\eqref{eq:C-Psi-stable} of $\Psi$ yields corrections of the same relative order $w_t^{-(2-\nu)}\propto t^{-(2-\nu)/\nu}$, i.e., $t^{-1/9}$ at $\nu=1.8$\footnote{Concretely, for instance for $\nu=1.8$, $y=2.25$ and $t=800$, the stable scale of the bump is $w_t/\gamma_y\approx79$, whereas the Gaussian fluctuations of the condensate run contribute a standard deviation $\beta\sqrt{2D_{\max}\tau_*(y)}/\gamma_y\approx44$ in $\xi$, cf.\ Eq.~\eqref{eq:C-Fourier-product}.}. For the normal condensate, the leading corrections stem from the non-analytic $h^\nu$ term in Eq.~\eqref{eq:C-Psi-Gaussian} and decay as $t^{-(\nu-2)/2}$, i.e., $t^{-1/4}$ at $\nu=2.5$.
Lastly, for a visualization of sample paths in each phase from the MCMC simulations, we refer to Appendix~\ref{app:further-num-results}.

\section{Conclusion and outlook}
\label{sec:outlook}

We have shown that rare displacements in SD undergo a
condensation transition onto a single run, which for right edge exponent $\nu>1$ of the diffusion coefficient distribution in Eq.~\eqref{eq:edge-expo-def} occurs in two steps, at $y_-$ and $y_+$ (Table~\ref{tab:phases}). The rate function $I(y)$ only signals the locations $y_-$ and $y_+$ of these thresholds, through its non-analyticities. The physical content of the transitions is visible only at the level of the conditioned trajectories: the condensate weight in the first-run law changes from $O(1/t)$ to $O(1)$ at $y_+$, and the condensate shape crosses over from anomalous for $1<\nu<2$ to normal for $\nu\geq2$.
These predictions are confirmed by Monte Carlo sampling of the conditioned trajectories and by numerical solutions of the exact renewal equation.\\

The origin of the two thresholds is transparent from the generating-function side, where the evolution operator of the diffusivity-resolved MGF undergoes a reversed BBP transition. Eq.~\eqref{eq:slope-gap} converts the loss of
overlap with the uniform switching direction into the width $y_+-y_-$ of
the coexistence interval. This explains why a discontinuous selection
transition for $\nu>1$ produces two transitions at fixed displacement,
whereas a continuous one for $0<\nu\leq1$ produces only one. For
$-1<\nu\leq0$, the absence of a finite-$q$ selection transition is mirrored
by the absence of condensation at any finite velocity.\\

Finally, Eq.~\eqref{eq:moment-identity} makes the relation with
run-and-tumble condensation clear. The integrated diffusivity
$\int_0^tD(s)\mathrm{d} s$ is itself the position of a renewal process whose
velocities are the successive values of $D$. SD adds a
Gaussian displacement conditioned on this integrated diffusivity. This
accounts for the common stable scaling function and illustrates that the same
mechanism should arise whenever a large dynamical fluctuation can be shared
by many intervals or carried by a single exceptional one in the presence of a heavy-tailed microscopic distribution. Establishing such a unified perspective on condensation transitions across stochastic systems in more quantitative detail hence appears as a natural direction for future work. We further mention the important direction of extending the results of this paper to systems of multiple interacting SDs, for instance with indirect interaction mediated via a shared switching protocol~\cite{majumdar-schehr:2026}.

\section*{Acknowledgments}

T.S.\ thanks Ayoub Belhadji, Tobias Grafke, and Chokri Manai for helpful discussions.
The authors thank the organizers of the DPG Spring Meeting in Dresden 2026, the KITP program ``New Trends in Non-equilibrium Dynamics'', and the ICTS program ``Two and a half decades of the Macroscopic Fluctuation Theory'' for stimulating scientific discussions and hospitality. GS and SNM acknowledge support from ANR Grant No. ANR-23-CE30-0020-01 EDIPS. This research was supported in part by grant NSF PHY-2309135 to the Kavli Institute for Theoretical Physics (KITP) and in part by the International Centre for Theoretical Sciences (ICTS) for participating in the program - Two and a half decades of the Macroscopic Fluctuation Theory 2026 (code: ICTS/MFTHEORY2026/09). The authors acknowledge use of generative AI to assist in derivations, coding, and manuscript writing. All AI output was checked by the authors, and they take full responsibility for the results presented in this work.

\appendix
\renewcommand{\thesection}{\Alph{section}}
\renewcommand{\thesubsection}{\Alph{section}.\arabic{subsection}}
\renewcommand{\thesubsubsection}{\Alph{section}.\arabic{subsection}.\arabic{subsubsection}}

\section{G{\"a}rtner--Ellis theorem and convexity of the rate function}
\label{app:rate-func-convex}

In order to guarantee that the Legendre transform~\eqref{eq:legendre-summary} of the SCGF $\Psi$ in case of a \textit{discontinuous first derivative} at~$q_{\rm c}$ for $\nu > 1$ indeed yields the correct rate function $I = I_{\rm SD}$, and not merely the convex hull of the true, non-convex rate function, we need to prove that $I_{\rm SD}$ is convex through other means, since the G{\"a}rtner--Ellis theorem does not apply (see Ref.~\cite[Section 4.4]{touchette:2009} for a detailed discussion). In principle, this has already been done in Ref.~\cite{gueneau-majumdar-schehr:2025} by numerically finding $I_{\rm SD}(y)$ from the renewal equation, which confirmed the convexity, and linear branch, of $I_{\rm SD}(y)$. Here, we provide an alternative, theoretical argument for completeness. For this, we consider the time-integrated diffusion coefficient process $A(t) = \int_0^t D(s) \mathrm{d}s$  (cf.\ Sec.~\ref{sec:rtp}). If $A = (A(s))_{s \in [0,t]}$ obeys a long-time LDP with rate function $I_{\rm RTP}$
\begin{align}
    \mathbb{P} \left(A(t) \approx t a \right) \asymp \exp \left\{-t I_{\rm RTP}(a) \right\}\,,
\end{align}
then the SD rate function $I_{\rm SD}$ is related to $I_{\rm RTP}$ via Eq.~\eqref{eq:sd-rtp-rf}.
We will proceed in two steps: show that (i) if $I_{\rm RTP}$ is convex, then $I_{\rm SD}$ is convex as well, and (ii) $I_{\rm RTP}$ is convex. The latter uses a standard subadditivity technique for Markov additive processes (cf., e.g., Ref.~\cite{iscoe-ney-nummerlin:1985}), and in particular relies on having a switching rate $r$ which is constant and independent of the diffusivity state $D$.\\

(i) Suppose that $a \mapsto I_{\rm RTP}(a)$ is convex. Then $(y,a) \mapsto I_{\rm RTP}(a)$ is convex as well, and so is $(y,a) \mapsto y^2/(4a)$. The latter follows from computing the Hessian, which is positive semi-definite (trace positive, determinant 0). As a sum of convex functions, $(y,a) \mapsto \frac{y^2}{4a} + I_{\rm RTP}(a)$ is then convex as well. Lastly, convexity is preserved under partial minimization~\cite[section 3.2.5]{boyd-vanderghe:2004}, thereby finishing the proof via Eq.~\eqref{eq:sd-rtp-rf}.\\

(ii) Fix $a_1, a_2 > 0$ and $\mu \in (0,1)$. We want to show that
    $I_{\rm RTP}(\mu a_1 + (1 - \mu) a_2) \leq \mu I_{\rm RTP}(a_1) + (1- \mu) I_{\rm RTP}(a_2)$. We write $a := \mu a_1 + (1 - \mu) a_2$. For any time $t > 0$, define $t_1 = \mu t$ and $t_2 = (1 - \mu) t$, such that $t = t_1 + t_2$. We will later consider $t \to \infty$, in which case $t_1, t_2 \to \infty$ as well. To eventually upper bound the rate function $I_{\rm RTP}(a)$, we need to lower bound the probability
    \begin{align}
        \mathbb{P}\left(A(t) \approx a t \right) \geq \mathbb{P} \left(A(t_1) \approx a_1 t_1\,, \;A(t) - A(t_1) \approx a_2 t_2 \right)
    \end{align}
    where the RHS is one particular way of achieving $A(t) \approx at$. We continue by rewriting the RHS in terms of the conditional distribution
    \begin{align}
        &\quad \mathbb{P} \left(A(t_1) \approx a_1 t_1\,, \;A(t) - A(t_1) \approx a_2 t_2 \right) \\
        &= \mathbb{P} \left(A(t_1) \approx a_1 t_1 \right) \mathbb{P} \left(A(t) - A(t_1) \approx a_2 t_2 \mid A(t_1) \approx a_1 t_1 \right)\\
        &= \mathbb{P} \left(A(t_1) \approx a_1 t_1 \right)  \int_0^{D_{\max}}  \mathbb{P} \left(A(t) - A(t_1) \approx a_2 t_2 \mid D(t_1)= D_1 \right) \mathrm{d}\mathbb{P}(D_1 \mid A(t_1) \approx a_1 t_1)\\
        &\geq \mathbb{P} \left(A(t_1) \approx a_1 t_1 \right) \inf_{D_0 \in [0, D_{\max}]} \mathbb{P}_{D(0) = D_0} \left(A(t_2) \approx a_2 t_2 \right)
    \end{align}
    where we have used in the third line that $D$ is a Markov process, so that $A(t) - A(t_1)$ only depends on $D(t_1)$, and then uniformly lower bounded the integral by the smallest probability, shifting the process back to time $0$ by time homogeneity. It remains to relate the last factor back to the original process, where $D(0) \sim W$. Fix some time $\delta > 0$ (which does not grow with $t$). The probability that the first switch happens before $\delta$ is $\mathbb{P}(\tau_1 \leq \delta) = 1 - e^{- r \delta} =: c$. Then, for any initial $D_0$ above, we can bound
    \begin{align}
         \mathbb{P}_{D(0) = D_0} \left(A(t_2) \approx a_2 t_2 \right) \geq  \mathbb{P}_{D(0) = D_0} \left(A(t_2) \approx a_2 t_2 \,, \; \tau_1 \leq \delta \right) = c \underbrace{\mathbb{P}_{D(0) = D_0} \left(A(t_2) \approx a_2 t_2 \mid \tau_1 \leq \delta \right)}_{\asymp \mathbb{P}_{D(0) \sim W} \left(A(t_2) \approx a_2 t_2  \right)}\,,
    \end{align}
    since the initial condition is forgotten as $t \to \infty$ under the condition $\tau_1 \leq \delta$, and does not contribute to the overall displacement at $O(t)$. In total, we have shown that there exists a $c > 0$, independent of $t$, such that
    \begin{align}
        \mathbb{P}\left(A(t) \approx a t \right) \geq c \mathbb{P} \left(A(t_1) \approx a_1 t_1\right)  \mathbb{P} \left(A(t_2) \approx a_2 t_2\right)\,.
    \end{align}
    Taking $-\log$ of this inequality, and dividing by $t$, yields
\begin{align}
       \underbrace{-\frac{1}{t} \log \mathbb{P}\left(A(t) \approx a t \right)}_{\to I_{\rm RTP}(a)} \leq \underbrace{-\frac{\log c}{t}}_{\to 0} + \mu \underbrace{\left(-\frac{1}{t_1} \log  \mathbb{P} \left(A(t_1) \approx a_1 t_1\right)\right)}_{\to I_{\rm RTP}(a_1)} + (1-\mu) \underbrace{\left(-\frac{1}{t_2} \log \mathbb{P} \left(A(t_2) \approx a_2 t_2\right)\right)}_{\to I_{\rm RTP}(a_2)}\,.
    \end{align}

This establishes the convexity of $I_{\rm RTP}(a)$ and, by step (i), of $I_{\rm SD}(y)$, thereby justifying the Legendre-transform construction of the SD rate function used in the main text.

\section{Phase boundaries when $W(D)$ is a Beta distribution}
\label{app:beta-dist-phases}

The general properties of the SCGF and rate function discussed in the main text hold for any $W(D)$ with right edge exponent $\nu$; however, the detailed shape of the critical lines depends on the whole spectral distribution and not just the right edge. In this appendix, we assume a global spectral density given by the Beta distribution on $[0, D_{\max} = 1]$ (and set $r = 1$ for simplicity in the following):
\begin{equation}
    W(D) = \frac{1}{B(\beta, \nu+1)} D^{\beta-1} (1-D)^\nu\,,
    \label{eq:beta}
\end{equation}
where $\beta > 0$, and $B(x, y) = \Gamma(x)\Gamma(y) / \Gamma(x+y)$ is the Beta function. For $\beta = 1$, this reduces to the example $W(D) = (\nu + 1) (1 - D)^\nu$ for which the phase diagrams of $\Psi$ and $I$ are shown in Fig.~\ref{fig:phase-diag}. Here, we derive the necessary equations that define the critical lines for illustrative purposes.\\

To find the critical conjugate parameter $q_{\rm c} = q_{\rm c}(\nu)$ for the SCGF $\Psi(q)$, using Eq.~\eqref{eq:critical-q}, we evaluate the Cauchy transform of Eq.~\eqref{eq:beta} at the right edge $z=1$ for $\nu > 0$ where it is finite:
\begin{align}
    G(1) &= \int_0^1 \frac{1}{1-D} \left[ \frac{1}{B(\beta, \nu+1)} D^{\beta-1} (1-D)^\nu \right] \mathrm{d}D = \frac{1}{B(\beta, \nu+1)} \int_0^1 D^{\beta-1} (1-D)^{\nu-1} \mathrm{d}D = \frac{B(\beta, \nu)}{B(\beta, \nu+1)}\,.
\end{align}
Expressing the result using Gamma functions and the property $\Gamma(z+1) = z\Gamma(z)$ yields $G(1) = (\nu+\beta) / \nu$.
Thus, by~\eqref{eq:critical-q}, the critical line for the SCGF is located at
\begin{equation}
    \kappa_{\rm c} = \frac{1}{G(1)} = \frac{\nu}{\nu+\beta} \quad \Leftrightarrow \quad q_{\rm c} = \sqrt{\frac{\nu + \beta}{\nu}}\,.
\end{equation}
Note that this does depend on $\beta$, and $q_{\rm c}$ diverges towards $\infty$ as $\nu \downarrow 0$, as expected.\\

Next, to find the nonzero critical overlap $\chi_{\rm c}$ for $\nu > 1$, we evaluate the derivative of the Cauchy transform at $z = 1$
\begin{align}
    G'(1) = -\int_0^1 \frac{W(D)}{(1-D)^2} \mathrm{d} D = -\frac{1}{B(\beta, \nu+1)} \int_0^1 D^{\beta-1} (1-D)^{\nu-2} \mathrm{d}D = -\frac{B(\beta, \nu-1)}{B(\beta, \nu+1)} = -\frac{(\nu+\beta)(\nu+\beta-1)}{\nu(\nu-1)}\,.
\end{align}
From Eq.~\eqref{eq:critical-projection}, we then find the order parameter jump across the transition:
\begin{align}
    \chi_{\rm c} = \frac{1}{\kappa_{\rm c}^2 |G'(1)|} = \frac{(\nu+\beta)(\nu-1)}{\nu(\nu+\beta-1)}\,,
\end{align}
which explicitly shows that the jump approaches $0$ as $\nu \downarrow 1$ (matching the continuous regime) and is strictly positive for $\nu > 1$ (first-order transition).\\

Finally, to find the phase boundaries for the rate function, we note that
\begin{align}
    y_+ &= \Psi'(q_{\rm c}^+)  = 2 q_{\rm c} = 2 \sqrt{\frac{\nu + \beta}{\nu}}\\
    y_- &= \Psi'(q_{\rm c}^-) =  2 q_{\rm c}  \bigg(1 -  \delta(\nu)\bigg) = 2 \sqrt{ \frac{\nu + \beta}{\nu}} \frac{\beta}{\beta + \nu - 1}
\end{align}
with relative gap size
\begin{align}
\delta(\nu) = \frac{\Psi'(q_{\rm c}^+) - \Psi'(q_{\rm c}^-)}{\Psi'(q_{\rm c}^+)} = \frac{\nu - 1}{\beta + \nu - 1}\,,
\end{align}
which tends to $0$ as $\nu \downarrow 1$, and to $1$ as $\nu \to \infty$. 

\section{Deriving the SD rate function from the RTP mapping}
\label{app:rto-to-sd}

In this appendix, we want to provide more details to derive from the results for $I_{\rm RTP}$ of~\cite{mori-le-doussal-majumdar-etal:2021} and Eq.~\eqref{eq:sd-rtp-rf} the rate function of the SD $I_{\rm SD}$ as obtained in Ref.~\cite{gueneau-majumdar-schehr:2025}. We set $D_{\max} = r = 1$ in the following for notational simplicity and suppress the explicit dependence on $r$ in the RTP notation, writing in particular $p_{1,{\rm RTP}}\equiv p_{\rm RTP}$.\\

 We first briefly outline how to obtain $I_{\rm RTP}$ via explicit saddle point analysis of the  Laplace transform of $p_{\rm RTP}$. Consider a configuration of an RTP trajectory of fixed duration $t$ consisting of $N-1$  completed runs and one final, uncompleted run, with $N-1\sim{\rm Poi}(t)$. We have $A(t) = \int_0^t D(s) \mathrm{d} s = \sum_{i=1}^{N-1} D_i \, \tau_i + D_N \Delta t_N$ and $t = \sum_{i=1}^{N-1} \tau_i + \Delta t_N$, where we denote the waiting times between switches by $\tau_i = T_{i} - T_{i-1}$. Since both the $D_i$'s and $\tau_i$'s are iid variables distributed respectively via $W(D)$ and $p(\tau) = e^{-\tau}$, we can express the position distribution $p_{\rm RTP}(A,t)$ as follows
\begin{align} 
p_{\rm RTP}(A,t) = \sum_{N =1}^\infty \int_0^1 \mathrm{d}D_{N} W(D_N) \prod_{i=1}^{N-1} \int_0^1 \mathrm{d}D_i \, W(D_i) \int_0^\infty \mathrm{d}\tau_i \, e^{-\tau_i} \delta\left(\sum_{k=1}^{N-1} D_k\,\tau_k + D_N \Delta t_N - A\right)\, 1_{\left\{ t \geq \sum_{k=1}^{N-1} \tau_k \right\} } e^{-\Delta t_N}\,,
 \label{PRTP_expl}
 \end{align}  
where the last term corresponds to the survival probability of the final run. Taking the double Laplace transform with respect to $A$ and $t$ decouples the product over $i$ and gives, by evaluating the geometric series:
\begin{eqnarray} \label{RTP_Laplace}
\tilde p_{\rm RTP}(\lambda,s) &=& \int_0^\infty \mathrm{d} A\, e^{-\lambda A} \int_0^\infty \mathrm{d}t\, e^{-st} p_{\rm RTP}(A,t) = \frac{\int_0^1 \mathrm{d}D\, \frac{W(D)}{1+s+\lambda D}}{1- \int_0^1 \mathrm{d}D\, \frac{W(D)}{1+s+\lambda D} } \;.
\end{eqnarray}
By inverting the double Laplace transform formally we get
\begin{eqnarray} \label{RTP_inv_Laplace}
p_{\rm RTP}(A,t) = \int_{\Gamma_1} \frac{\mathrm{d}\lambda}{2 \pi i} e^{\lambda A} \int_{\Gamma_2} \frac{\mathrm{d}s}{2 \pi i} e^{s t}  \frac{\int_0^1 \mathrm{d}D\, \frac{W(D)}{1+s+\lambda D} }{1- \int_0^1 \mathrm{d}D\, \frac{W(D)}{1+s+\lambda D} } \;,
\end{eqnarray}
where $\Gamma_1$ and $\Gamma_2$ are two vertical Bromwich contours respectively in the $\lambda$ and $s$ complex planes. Next we perform the change of variable
$
\lambda= -(1+s) \tilde \lambda
$,
which gives
\begin{eqnarray} \label{RTP_inv_Laplace_rescaled}
p_{\rm RTP}(A,t) = \int_{\tilde \Gamma_1} \frac{\mathrm{d} \tilde \lambda}{2 \pi i} e^{- \tilde \lambda A} \int_{\Gamma_2} \frac{\mathrm{d}s}{2 \pi i} (1+s)e^{s (t-\tilde \lambda A)}  \frac{\int_0^1 \mathrm{d}D\, \frac{W(D)}{1-\tilde \lambda D} }{s+1-\int_0^1 \mathrm{d}D\, \frac{W(D)}{1-\tilde \lambda D} } \;,
\end{eqnarray}
where $\tilde \Gamma_1$ runs vertically up and to the left of all singularities in the complex $\tilde \lambda$ plane. Evaluating the residue in the complex $s$-plane at the pole at
\begin{eqnarray}\label{pole}
s^* =F(\tilde \lambda) - 1 \quad {\rm where} \quad F(\tilde \lambda) = \int_0^1 \mathrm{d}D \, \frac{W(D)}{1-\tilde \lambda D} \;,
\end{eqnarray}
gives
\begin{eqnarray} \label{PXt}
p_{\rm RTP}(A,t) = \int_{\tilde \Gamma_1} \frac{\mathrm{d}\tilde \lambda}{2 \pi i}  \left[ F(\tilde \lambda)\right]^2 \,e^{-t S(\tilde \lambda,a = A/t)} \;,
\end{eqnarray}
where
\begin{eqnarray} \label{defS}
S(\tilde \lambda, a) = 1  -(1-\tilde \lambda  a)\,F(\tilde \lambda) \;.
\end{eqnarray}
This formula is exact for any $t$. We now take the large $t$ limit and evaluate the integral over $\tilde \lambda$ by the saddle point method. 
This gives, ignoring pre-exponential factors,
$
p_{\rm RTP}(A,t) \asymp e^{- t I_{\rm RTP}(a=A/t)}
$,
where
\begin{eqnarray} \label{Psi_RTP}
I_{\rm RTP}(a) = \max_{\tilde \lambda \in \mathbb{R}} \left[1 - (1-\tilde \lambda \,a) F(\tilde \lambda) \right] \;.
\end{eqnarray}
Note that for $F(\tilde \lambda)$ in Eq. (\ref{pole}) to exist, one must necessarily have $\tilde \lambda \leq 1$ on the real axis. Hence the maximisation in Eq. (\ref{Psi_RTP}) reads
 \begin{eqnarray} \label{Psi_RTP2}
I_{\rm RTP}(a) = \max_{\tilde \lambda \leq 1} \left[1 - (1-\tilde \lambda  a) F(\tilde \lambda) \right] = \max_{\tilde \lambda \leq 1} S(\tilde \lambda,a)\;.
\end{eqnarray}
The saddle point equation $\partial S/\partial \tilde \lambda = 0$ associated with the maximization in Eq.~(\ref{Psi_RTP2}) gives
\begin{eqnarray} \label{eq_SP1}
a = \frac{F'(\tilde \lambda)}{F(\tilde \lambda) + \tilde \lambda \, F'(\tilde \lambda)}  \;,
\end{eqnarray}
where $F(\tilde \lambda)$ is defined in Eq.~(\ref{pole}). We recall that $\tilde \lambda \leq 1$ and $a = A/t \in [0,1]$, since the maximal ``speed'' of the RTP is $D_{\max} = 1$. Defining the right hand side as a new function $h(\tilde \lambda)$ the saddle point equation (\ref{eq_SP1}) then reads
\begin{eqnarray} \label{defh}
a = h(\tilde \lambda) \equiv  \frac{\int_0^1 \mathrm{d}D \frac{D\,W(D)}{(1-\tilde \lambda D)^2}}{\int_0^1 \mathrm{d} D \frac{W(D)}{(1-\tilde \lambda D)} + \tilde \lambda \int_0^1 \mathrm{d}D \frac{D\, W(D)}{(1-\tilde \lambda D)^2}} \;.
\end{eqnarray}
For a given $0 \leq a \leq 1$, we need to invert this relation and find the saddle-point value 
$
\tilde{\lambda}^* = h^{-1}(a)
$.
We first note that as $\tilde \lambda$ approaches its maximal allowed value $\tilde \lambda = 1$, the function $h(\tilde \lambda)$ approaches a constant given by (assume $W(D) = (\nu + 1) (1 - D)^\nu$ for $\nu > 0$ and all $D \in [0,1]$ in the following for concreteness)
\begin{align}\label{h_1}
\lim_{\tilde \lambda \to 1}\, h(\tilde \lambda) =
\begin{cases}
&1 \quad, \quad 0<\nu\leq 1 \\
& \frac{1}{\nu} \quad, \quad \nu > 1 \;.
\end{cases}
\end{align}
Thus, for all $0<\nu\leq1$ and any given $a \in [0,1]$, there is a unique inverse $\tilde{\lambda}^*=h^{-1}(a)$, corresponding to the RTP fluid phase, in which all runs are of duration $O(1)$. In contrast, for $\nu > 1$, a solution $\tilde\lambda^* = h^{-1}(a)$ exists only as long as $a<a_{\rm c} = 1/\nu$. For $a>a_{\rm c}$, the saddle point breaks down and a single run with speed close to $D_{\max}=1$ and duration $\tau_{\rm cond}=(A-a_{\rm c}t)/(1-a_{\rm c})=O(t)$, the condensate, coexists with a fluid background which moves at the critical speed $a_{\rm c}$ during the remaining time and thus carries $a_{\rm c}(t-\tau_{\rm cond})$ (cf.~Ref.~\cite{mori-le-doussal-majumdar-etal:2021} for the analogous fixed-$N$ picture). This is the RTP counterpart of the lever rule~\eqref{eq:taustar} for the SD.\\

In order to connect this back to SD, we use Eq.~\eqref{eq:sd-rtp-rf} in the main text.
To find the optimal $a^* = a^*(y)$ for the minimization in Eq.~\eqref{eq:sd-rtp-rf}, we define $J_y(a) := I_{\rm RTP}(a) + y^2/(4a)$ and first try to find the minimum by setting $J_y'(a^*(y)) = 0$, or
\begin{eqnarray} \label{eq:rtp-to-sd-cond}
I_{\rm RTP}'(a^*(y)) = \frac{y^2}{4 (a^*(y))^2}\,.
\end{eqnarray}
Note that this is only valid if $a^*(y)$ as determined from this condition is in the interior of the interval $(0,1)$. In particular, we anticipate that at large $y$ and for any power law exponent $\nu > 0$ of $W$, we will find the optimal value at the right boundary of the interval, $a^* = 1$, which then yields the quadratic rate function of~\cite{gueneau-majumdar-schehr:2025} at large $y$, cf.~Eq.~\eqref{eq:microcanonical-summary} of this paper: 
\begin{eqnarray}
I_{\rm SD}(y) \sim I_{\rm RTP}(1) + \frac{y^2}{4} = 1 + \frac{y^2}{4}\,.
\end{eqnarray}
To show that this is true, we note that, given that $I_{\rm RTP}'$ is bounded from above on $(0,1]$ (by $F(1)$ for $\nu>0$, see the envelope property $I_{\rm RTP}'(a)=\tilde\lambda^*(a)F(\tilde\lambda^*(a))$ with $\tilde\lambda^*\leq1$ below; note that $I_{\rm RTP}'(a)\to-\infty$ as $a\to0$ if $W(0)>0$), we have at very large $ y $ that the derivative $J_y'(a) = I_{\rm RTP}'(a) - y^2/(4a^2) < 0$ is negative for all $a \in (0,1)$. Hence $J_y$ is strictly decreasing, and the minimum will indeed be achieved at the right boundary $a^* = 1$. Transitioning from $a^* < 1$ to $a^* = 1$ as $y$ is varied will signal one dynamical phase transition at $y = y_+$ for the SD problem.\\

Now, assuming the minimum $a^*(y)$ is achieved in the interior, such that Eq.~\eqref{eq:rtp-to-sd-cond} holds, and using the Legendre envelope property $I_{\rm RTP}'(a) = \tilde\lambda^*(a) F(\tilde\lambda^*(a))$ for the derivative of the RTP rate function, we get
\begin{eqnarray}
y = 2 a^*(y) \sqrt{\tilde \lambda^*(a^*(y)) F(\tilde \lambda^*(a^*(y)))}\,.
\end{eqnarray}
If $\nu > 1$, we have that for $a_{\rm c} = \tfrac{1}{\nu} < a \leq 1$ in the RTP condensate phase, by substituting $\tilde\lambda^*(a) = 1$, we get $I_{\rm RTP}(a) = F(1) a - (F(1) - 1)$. Note that this is a linear function in $a$, and that $F(1)$ is exactly the Cauchy transform of $W$ evaluated at the right edge. We then find
$
    a^*(y) = y / (2 \sqrt{F(1)})
$
in this case, and the condition $a^*(y_{-}) \overset{!}{=} a_{\rm c} = 1 / \nu$ determines the critical $y_-$ for this second phase transition at $\nu > 1$. Using this expression, the SD rate function for intermediate $y$, such that $a_{\rm c} < a^*(y) < 1$, becomes linear as well
\begin{eqnarray}
I_{\rm SD}(y) = \sqrt{F(1)}  y  - (F(1) - 1)\,.
\end{eqnarray}
In the notation of Ref.~\cite{gueneau-majumdar-schehr:2025}, we have $q_{\rm c} = \sqrt{F(1)}$, and the intercept $F(1) - 1 = q_{\rm c}^2 - 1$ (denoted $\gamma$ there), and hence perfectly reproduce the second line of Eq.~(15) from Ref.~\cite{gueneau-majumdar-schehr:2025}.\\

Lastly, for the fluid phase, we write
\begin{eqnarray}
F(\tilde \lambda) = \int_0^1 \mathrm{d}D \, \frac{W(D)}{1-\tilde \lambda D} = \frac{1}{\tilde \lambda} G \left(\frac{1}{\tilde \lambda} \right)
\end{eqnarray}
in terms of the Cauchy transform of $W$. Using the identity for the $R$-transform of the distribution $W$, $G\left(R(z) + \tfrac{1}{z} \right) = z$, we can then write $F(\tilde \lambda) = 1 + w R(w)$ if we define $G(1/\tilde \lambda) =: w$. With this change of variables, the RTP rate function is
\begin{eqnarray}
    I_{\rm RTP}(a) = \max_{\tilde \lambda} \left(1 - F(\tilde \lambda) + a \tilde \lambda F(\tilde \lambda) \right) = \max_w \left(aw - w R(w) \right)\,.
\end{eqnarray}
The transformation to the SD rate function, upon exchanging max and min (convex duality), then yields
\begin{eqnarray}
I_{\rm SD}(y) = \min_a \left( \max_w \left(aw - w R(w) \right) + \frac{y^2}{4a} \right) = \max_w \bigg(-wR(w) + \underbrace{\min_a \left(wa + \frac{y^2}{4 a} \right)}_{= \sqrt{w} y } \bigg)\,,
\end{eqnarray}
so that, if we define $q := \sqrt{w}$, we end up with
\begin{eqnarray}
I_{\rm SD}(y) = \max_q \left(y  q - q^2 R\left(q^2\right) \right)
\end{eqnarray}
as in Eqs.~(10) and (14) and (15) of Ref.~\cite{gueneau-majumdar-schehr:2025}.

\section{Analysis of the single-run distribution}\label{SingleRunAppendix}

In this appendix, we collect the technical calculations underlying the analysis of the single-run distribution $P_t(\xi\mid y)$ presented in Sec.~\ref{sec:single-run-main-text}. All of them start from the exact renewal equation~\eqref{eq:exactfirst}, with the first-run factor $Q(\xi,\tau)$ defined in Eq.~\eqref{eq:C-Qdef}.

\subsection{Precise large deviations of the endpoint distribution}
\label{sec:endpoint-ldt-precise}
\label{app:endpoint-ld}

Here, we derive the asymptotic forms of the endpoint PDF $Z_t(y)$ in the fluid and saturated phases, Eqs.~\eqref{eq:Z-regular-intro}, \eqref{eq:Breg-intro}, \eqref{eq:Z-sat-intro}, and~\eqref{eq:A-y-intro}, stated in Sec.~\ref{sec:renewal}. The time-Laplace transform of the moment generating function is
\begin{equation}
        \widetilde Z_q(p)=\int_0^\infty e^{-pt}\langle e^{q\,x(t)} \rangle\,\mathrm{d}t
        ={R(q,p)\over 1-rR(q,p)},
        \qquad
        R(q,p)=\int_0^{D_{\max}}{W(D) \,\mathrm{d}D\over p+r-q^2D} .
        \label{eq:Zq-resolvent-app}
\end{equation}
For $q<q_{\rm c}$, the dominant singularity is a simple pole at $p=\Psi(q)$, determined by
$1-rR(q,\Psi(q))=0$. Its residue is $R/(-r\partial_pR)=1/(r^2J_2(q))$, which coincides with the squared overlap $\chi(q)$ of Eq.~\eqref{eq:projection}, so that $\langle e^{qx(t)}\rangle\sim \chi(q)e^{t\Psi(q)}$. A saddle-point inversion in $q$ then gives
\begin{equation}
        Z(yt,t)\sim
        {\chi(q_*)\over \sqrt{2\pi\Psi''(q_*)t}}
        e^{-tI(y)},
        \qquad
        \chi(q)={1\over r^2J_2(q)},
        \label{eq:Z-regular-app}
\end{equation}
where
\begin{equation}
        J_2(q)=\int_0^{D_{\max}}{W(D) \,\mathrm{d}D \over [r+\Psi(q)-q^2D]^2}  = -\frac{1}{q^4} G'\left(\frac{r + \Psi(q)}{q^2} \right).
        \label{eq:bq-app}
\end{equation}
At $q=q_{\rm c}$ this regular construction reaches the endpoint $D_{\max}$ and the
rate function develops the linear branch stated in Eq.~\eqref{eq:I-linear-intro}. The corresponding prefactor $B_{\rm int}(y)$ in Eq.~\eqref{eq:Bint-intro} requires resolving the one-condensate structure of the conditioned trajectories, and is derived in Sec.~\ref{sec:cond-intermediate} and Appendix~\ref{app:critical-amplitudes}. For $q>q_{\rm c}$, the dominant singularity comes from $D=D_{\max}$.  Writing
$p=q^2D_{\max}-r+\delta p$ and $D=D_{\max}-\varepsilon$, the denominator in $R(q,p)$ is
$\delta p+q^2\varepsilon$. Hence $R=R_{\rm reg}+R_{\rm sing}$, where $R_{\rm reg}$ is analytic at $\delta p=0$ with $R_{\rm reg}(q,q^2D_{\max}-r)=q^{-2}\int_0^{D_{\max}}W(D)\,\mathrm{d}D/(D_{\max}-D)=q_{\rm c}^2/(rq^2)$ by Eq.~\eqref{eq:critical-q}, and $R_{\rm sing}$ is the non-analytic edge part, which is the Laplace transform of $\int_0^{D_{\max}}W(D)e^{-(r-q^2D)t}\,\mathrm{d}D\sim C\Gamma(\nu+1)q^{-2\nu-2}t^{-(\nu+1)}e^{t(q^2D_{\max}-r)}$. To first order in $R_{\rm sing}$, $\widetilde Z_q=R/(1-rR)$ has the singular part $R_{\rm sing}/(1-rR_{\rm reg})^2$, with $1-rR_{\rm reg}\to1-q_{\rm c}^2/q^2>0$. Inverting the Laplace transform therefore gives
\begin{equation}
        \langle e^{q\,x(t)} \rangle
        \sim
        {C\Gamma(\nu+1)\over
        q^{2\nu+2}\left(1-q_{\rm c}^2/q^2\right)^2}
        t^{-(\nu+1)}
        \exp\left(t(q^2D_{\max}-r)\right) .
        \label{eq:Zq-sat-app}
\end{equation}
The inverse transform in $q$ is then evaluated at $q_y=y/(2D_{\max})$, which
gives Eqs.~\eqref{eq:Z-sat-intro}--\eqref{eq:A-y-intro}.

\subsection{Edge asymptotics of the first-run factor}
\label{app:Q-edge}

All macroscopic contributions to $P_t(\xi\mid y)$ -- the condensate in the coexistence and saturated phases, as well as the interior sectors -- involve the first-run factor $Q(\xi,\tau)$ of Eq.~\eqref{eq:C-Qdef} for $\xi$ and $\tau$ of order $t$. Here, we derive its large-$t$ form, Eq.~\eqref{eq:C-Q-edge-global}. This shows that a macroscopic run is produced by diffusion constants in
a narrow layer near $D_{\max}$, and it also gives the algebraic prefactor which
must be compared with the prefactor of the denominator $Z_t(y)$.

Set
\begin{equation}
        \xi=\rho t,
        \qquad
        \tau=\theta t,
        \qquad
        0<\theta<1,
        \label{eq:C-rho-theta}
\end{equation}
and assume $\rho>0$.  The exponential part of the $D$-integrand in
Eq.~\eqref{eq:C-Qdef} is
\begin{equation}
        \exp\left(-r\theta t-{\rho^2t\over4D\theta}\right).
        \label{eq:C-D-exponential}
\end{equation}
For fixed $(\rho,\theta)$, the exponent is maximal at the endpoint
$D=D_{\max}$.  The $D$-integral is therefore localized near this endpoint.  To
resolve the corresponding boundary layer, we make the change of variable
\begin{equation}
        D=D_{\max}-{\zeta\over\tau},
        \qquad
        \mathrm{d}D=-{\mathrm{d}\zeta\over\tau}.
        \label{eq:C-D-edge-global}
\end{equation}
Since $\tau=\theta t$, keeping $\zeta=O(1)$ amounts to
$D_{\max}-D=O(t^{-1})$.  Defining
\begin{equation}
        q_{\rho,\theta}={\rho\over2D_{\max}\theta},
        \label{eq:C-q-rho-theta}
\end{equation}
we obtain
\begin{equation}
        {\xi^2\over4D\tau}
        ={\rho^2t\over4D_{\max}\theta}
        +q_{\rho,\theta}^2\zeta+o(1),
        \label{eq:C-D-expansion-global}
\end{equation}
while the edge behavior of $W$ gives
\begin{equation}
        W\left(D_{\max}-{\zeta\over\tau}\right)
        \sim C\left({\zeta\over\tau}\right)^\nu.
        \label{eq:C-W-edge-global}
\end{equation}
The upper limit of the resulting $\zeta$-integral may be sent to infinity, and we
obtain Eq.~\eqref{eq:C-Q-edge-global} of the main text.  This is the large-$t$ expansion of the first-run
factor entering the exact renewal equation~\eqref{eq:exactfirst}.  In particular,
it justifies both the restriction to $D$ close to $D_{\max}$ and the power of
$t$ carried by this factor.

\subsection{Derivation of the condensate peak in the coexistence phase}
\label{app:critical-profile}

We now derive the position, width, mass, and limiting shape of the condensate
peak in the coexistence phase discussed in Sec.~\ref{sec:cond-intermediate},
\begin{equation}
        y_-<y<y_+ .
        \label{eq:C-intermediate-domain}
\end{equation}
The starting point is the exact continuous contribution in
Eq.~\eqref{eq:exactfirst}.  The trajectory following the first completed run
is an independent copy of the process, with endpoint density $Z(x,t)$ defined in
Eq.~\eqref{eq:Z-def}.  We therefore analyze the numerator
\begin{equation}
\begin{split}
        \mathcal N_t(\xi;y)
        =\int_0^t \mathrm{d}\tau\,Q(\xi,\tau)
        Z(yt-\xi,t-\tau),
\end{split}
        \label{eq:C-exact-start}
\end{equation}
with $Q(\xi,\tau)$ defined in Eq.~\eqref{eq:C-Qdef}.
Thus $\mathcal N_t(\xi;y)$ is the first-run numerator associated with a fixed
first-run displacement $\xi$.  It should not be confused with $N_t$, used
below for the number of microscopic runs.  The conditional density is then
\begin{equation}
        P_t^{\rm cont}(\xi\mid y)
        ={\mathcal N_t(\xi;y)\over Z_t(y)}.
        \label{eq:C-P-from-N}
\end{equation}
The no-switching term of Eq.~\eqref{eq:exactfirst} is a Dirac mass at
$\xi=yt$ and does not contribute to the bump considered here.

The macroscopic saddle $(\rho_*,\theta_*)$, which fixes the position $\xi_*(t)$ and duration $\tau_*(y)$ of the bump, was determined in Sec.~\ref{sec:cond-intermediate}. For completeness, we note that an interior stationary point of $s_y(\theta,\rho)$ in Eq.~\eqref{eq:C-effective-action} with $z<y_-$ would require, writing $q=\phi_\nu'(z)$,
\begin{equation}
        {\rho\over\theta}=2D_{\max} q,
        \qquad
        r+\Psi(q)-q^2D_{\max}=0,
        \label{eq:C-edge-condition}
\end{equation}
which is excluded for $q<q_{\rm c}$ by Eq.~\eqref{eq:gap-positive}. On the boundary $z=y_-$, i.e., $y=\rho+y_-(1-\theta)$, using $I(y)=q_{\rm c}y-\Psi(q_{\rm c})$, $\phi_\nu(y_-)=q_{\rm c}y_- -\Psi(q_{\rm c})$ and $r+\Psi(q_{\rm c})=q_{\rm c}^2D_{\max}$ yields the perfect square~\eqref{eq:C-action-square}, and hence Eqs.~\eqref{eq:taustar} and~\eqref{eq:coexistenceidentity}.\\

The saddle determines the location of the bump only at the leading, extensive order. Its width, its limiting shape, and the position of its maximum on the subextensive scale around $\xi_*(t)$ are obtained in the remainder of this appendix by expanding the exact integral in Eq.~\eqref{eq:C-exact-start} around $(\xi_*(t),\tau_*(y))$. We shall find a scale $w_t=o(t)$ and a limiting profile whose mode $b_{\rm mode}$ gives
\begin{equation}
        \xi_{\rm mode}(t)
        =\xi_*(t)+b_{\rm mode}w_t+o(w_t).
        \label{eq:C-mode-shift}
\end{equation}

\subsubsection{Expansion of the exact integral around the saddle}

We now return to the exact $\tau$-integral in Eq.~\eqref{eq:C-exact-start}
and expand it in a neighborhood of the macroscopic saddle
$(\xi_*(t),\tau_*(y))$ determined above.  We fix $b=O(1)$ and evaluate the
numerator at
$
        \xi=\xi_*(t)+bw_t.
$
For this value of $\xi$, the saddle-point contribution to the $\tau$-integral
in Eq.~\eqref{eq:C-exact-start} comes from a subextensive neighborhood of
$\tau_*(y)$.  We therefore introduce the local variable $a$ by the change of
variable
\begin{equation}
        \tau=\tau_*(y)+{aw_t\over y_+},
        \qquad
        \mathrm{d}\tau={w_t\over y_+}\,\mathrm{d}a.
        \label{eq:C-local-tau}
\end{equation}
At this stage the scale $w_t$ is still unknown.  We only require
$w_t=o(t)$, so that Eq.~\eqref{eq:C-local-tau} indeed describes a
subextensive neighborhood of the macroscopic saddle.  The scale will be fixed
in Sec.~\ref{sec:crit-background-analysis} by requiring that the endpoint fluctuations of the remaining
trajectory have a non-degenerate limiting law. The original bounds $0\leq\tau\leq t$ become
\begin{equation}
        -{y_+\tau_*(y)\over w_t}\leq a\leq
        {y_+[t-\tau_*(y)]\over w_t}.
        \label{eq:C-a-bounds}
\end{equation}
Since both $\tau_*(y)$ and $t-\tau_*(y)$ are of order $t$, while
$w_t=o(t)$, the two bounds in Eq.~\eqref{eq:C-a-bounds} tend to
$-\infty$ and $+\infty$, respectively.  Moreover, the saddle-point expansion
localizes the integral to $a=O(1)$.  We may therefore extend the $a$-integration
to the whole real line, with an error which is negligible at the order retained
below. For each value of $a$ in the integral, let
\begin{equation}
        T_*=t-\tau_*(y),
        \qquad
        T=t-\tau=T_*-{aw_t\over y_+}
        \label{eq:C-T-def}
\end{equation}
be, respectively, the saddle-point duration and the actual duration of the
remaining trajectory.  Its displacement is
\begin{equation}
        x=yt-\xi
        =yt-\xi_*(t)-bw_t
        =y_-T_*-bw_t,
        \label{eq:C-x-def}
\end{equation}
where Eq.~\eqref{eq:coexistenceidentity} was used in the last equality.  We also set
\begin{equation}
        \beta={y_-\over y_+},
        \qquad
        \gamma_y=1-\beta={y_+-y_-\over y_+}.
        \label{eq:C-beta-gamma}
\end{equation}
The deviations of the two factors in the integrand of
Eq.~\eqref{eq:C-exact-start} from their saddle values are then
$
        \xi-y_+\tau=(b-a)w_t,
$
and
\begin{equation}
        x-y_-T=(\beta a-b)w_t.
        \label{eq:C-u-identity}
\end{equation}
We first specialize the edge expansion of $Q$ to this scaling regime.  Since
\begin{equation}
        {\xi\over2D_{\max}\tau}=q_{\rm c}+o(1),
        \label{eq:C-local-qc}
\end{equation}
we can rewrite the exponential cost exactly as
\begin{equation}
        r\tau+{\xi^2\over4D_{\max}\tau}
        =q_{\rm c}\xi-\Psi(q_{\rm c})\tau
        +{(\xi-y_+\tau)^2\over4D_{\max}\tau}.
        \label{eq:C-complete-square}
\end{equation}
Introduce the normalized Gaussian kernel
\begin{equation}
        \Gamma_t(v)
        ={w_t\over\sqrt{4\pi D_{\max}\tau_*(y)}}
        \exp\left(-{w_t^2v^2\over4D_{\max}\tau_*(y)}\right),
        \qquad
        \int_{-\infty}^\infty\Gamma_t(v)\,\mathrm{d}v=1.
        \label{eq:C-Gammat}
\end{equation}
Then Eq.~\eqref{eq:C-Q-edge-global} gives, uniformly for bounded $a$ and $b$,
\begin{equation}
\begin{split}
        Q(\xi,\tau)
        \sim {}&{rC\Gamma(\nu+1)\over
        q_{\rm c}^{2\nu+2}\tau_*(y)^{\nu+1}w_t}
        e^{-q_{\rm c}\xi+\Psi(q_{\rm c})\tau}\Gamma_t(b-a).
\end{split}
        \label{eq:C-first-factor}
\end{equation}

We next expand the second factor in the exact integral
\eqref{eq:C-exact-start}, namely the endpoint density $Z(x,t)$ defined in
Eq.~\eqref{eq:Z-def}.  For any tentative scale $w_t=o(t)$, we use the tilted and rescaled background law $F_T(u)$ defined in Eq.~\eqref{eq:C-FT-real-def}. Equivalently,
\begin{equation}
\begin{split}
        Z(y_-T+uw_t,T)
        ={1\over r\mu_{\rm c} w_t}
        e^{-q_{\rm c}(y_-T+uw_t)+T\Psi(q_{\rm c})}F_T(u).
\end{split}
        \label{eq:C-Z-through-FT}
\end{equation}
This is an exact rewriting.  The scale $w_t$ will be fixed in Sec.~\ref{sec:crit-background-analysis} so that
$F_T$ has a non-degenerate limit.

Substituting Eqs.~\eqref{eq:C-first-factor} and
\eqref{eq:C-Z-through-FT} into the exact integral
\eqref{eq:C-exact-start}, and using Eq.~\eqref{eq:C-u-identity}, the two
exponential factors cancel exactly:
\begin{equation}
\begin{split}
        &e^{-q_{\rm c}\xi+\Psi(q_{\rm c})\tau}
        e^{-q_{\rm c}(yt-\xi)+\Psi(q_{\rm c})(t-\tau)}
        =e^{-q_{\rm c}yt+\Psi(q_{\rm c})t}=e^{-tI(y)}.
\end{split}
        \label{eq:C-exp-cancel}
\end{equation}
We therefore obtain directly for the numerator $\mathcal N_t$ defined in
Eq.~\eqref{eq:C-exact-start}
\begin{equation}
\begin{split}
        \mathcal N_t\bigl(\xi_*(t)+bw_t;y\bigr)
        \sim {}&{C\Gamma(\nu+1)\over
        \mu_{\rm c} q_{\rm c}^{2\nu+2}y_+\tau_*(y)^{\nu+1}w_t}
        e^{-tI(y)}
       \int_{-\infty}^\infty\mathrm{d}a\,
        \Gamma_t(b-a)F_{T(a)}(\beta a-b),
\end{split}
        \label{eq:C-N-local-raw}
\end{equation}
where
$
        T(a)=T_*- aw_t /  y_+.
$
The extension of the $a$-integral to $\mathbb R$ follows from the bounds
\eqref{eq:C-a-bounds} and from $w_t=o(t)$.  The integral is localized to
$a=O(1)$, and the analysis in Sec.~\ref{sec:crit-background-analysis} gives, uniformly for $a=O(1)$,
\begin{equation}
        F_{T(a)}(u)=F_{T_*}(u)+o(1).
        \label{eq:C-FT-Ta-replacement}
\end{equation}
Setting $v=b-a$ in Eq.~\eqref{eq:C-N-local-raw} then yields the simpler form
\begin{equation}
\begin{split}
        \mathcal N_t\bigl(\xi_*(t)+bw_t;y\bigr)
        \sim {}&{C\Gamma(\nu+1)\over
        \mu_{\rm c} q_{\rm c}^{2\nu+2}y_+\tau_*(y)^{\nu+1}w_t}
        e^{-tI(y)}\int_{-\infty}^\infty\mathrm{d}v\,
        \Gamma_t(v)F_{T_*}(-\gamma_y b-\beta v).
\end{split}
        \label{eq:C-N-local-integral}
\end{equation}
This is the local form of the numerator entering the conditional density
\eqref{eq:C-P-from-N}. It is useful to normalize the last integral.  Since
$\widehat F_T(0)=\int_{\mathbb R}F_T(u)\,\mathrm{d}u$, define
\begin{equation}
\begin{split}
        \Phi_{\nu,y}^{(t)}(b)
        :={\gamma_y\over\widehat F_{T_*}(0)}
       \int_{-\infty}^\infty\mathrm{d}v\,
        \Gamma_t(v)F_{T_*}(-\gamma_y b-\beta v).
\end{split}
        \label{eq:C-Phi-direct}
\end{equation}
The change of variable $u=-\gamma_y b-\beta v$ shows immediately that
\begin{equation}
       \int_{-\infty}^\infty\Phi_{\nu,y}^{(t)}(b)\,db=1.
        \label{eq:C-Phi-normalization}
\end{equation}
Moreover, this is a genuine convolution after the single rescaling
$\tilde v=-\gamma_y b/\beta$: the normalized density in the variable $\tilde v$ is
\begin{equation}
        {\beta\over\gamma_y}
        \Phi_{\nu,y}^{(t)}\!\left(-{\beta \tilde v\over\gamma_y}\right)
        ={1\over\widehat F_{T_*}(0)}
       \int_{-\infty}^\infty\mathrm{d}v\,\Gamma_t(v)\,
        \beta F_{T_*}\bigl(\beta(\tilde v-v)\bigr).
        \label{eq:C-true-convolution}
\end{equation}
Thus the bump is the convolution of the fluctuations of the macroscopic run
with those of the critical background, written in compatible units. Using Eq.~\eqref{eq:C-Phi-direct}, Eq.~\eqref{eq:C-N-local-integral} becomes
\begin{equation}
\begin{split}
        \mathcal N_t\bigl(\xi_*(t)+bw_t;y\bigr)
        \sim {}&{C\Gamma(\nu+1)\over
        \mu_{\rm c} q_{\rm c}^{2\nu+2}(y_+-y_-)\tau_*(y)^{\nu+1}w_t}
        e^{-tI(y)}
        \,\widehat F_{T_*}(0)\Phi_{\nu,y}^{(t)}(b).
\end{split}
        \label{eq:C-N-local-profile}
\end{equation}

\subsubsection{Fourier representation of the local profile}

Fourier space is convenient, since it turns the convolution in
Eq.~\eqref{eq:C-Phi-direct} into a product.  We use
\begin{equation}
        \widehat f(k)=\int_{\mathbb R}e^{ikx}f(x)\,\mathrm{d}x.
        \label{eq:C-Fourier-convention}
\end{equation}
Taking the Fourier transform of Eq.~\eqref{eq:C-Phi-direct}, exchanging the
$b$ and $v$ integrations, and setting $u=-\gamma_y b-\beta v$, gives
\begin{equation}
\begin{split}
        \widehat\Phi_{\nu,y}^{(t)}(k)
        &={\widehat F_{T_*}(-k/\gamma_y)\over\widehat F_{T_*}(0)}
        \widehat\Gamma_t\left(-{\beta k\over\gamma_y}\right),
\end{split}
        \label{eq:C-Fourier-product}
\end{equation}
where
\begin{equation}
        \widehat\Gamma_t(\ell)
        =\exp\left(-{D_{\max}\tau_*(y)\over w_t^2}\ell^2\right).
        \label{eq:C-Gamma-Fourier}
\end{equation}
Hence all that remains for the shape of the bump is to determine the scale
$w_t$ and the characteristic function $\widehat F_T$.  This is done next.

\subsubsection{Critical background in Fourier space and determination of $w_t$}
\label{sec:crit-background-analysis}

We start from the bilateral Laplace transform of the endpoint density $Z(x,t)$
defined in Eq.~\eqref{eq:Z-def}:
\begin{equation}
        \widehat Z(q,T)
        =\int_{\mathbb R} e^{qx}Z(x,T)\,\mathrm{d}x.
        \label{eq:C-Zhat-def}
\end{equation}
Taking the Fourier transform of the definition \eqref{eq:C-FT-real-def} and
making the change of variable $x=y_-T+uw_t$ gives the exact identity
\begin{equation}
\begin{split}
        \widehat F_T(k)
        ={}&r\mu_{\rm c}\,
        e^{-T\Psi(q_{\rm c})-ik y_-T/w_t}
        \widehat Z\left(q_{\rm c}+{ik\over w_t},T\right).
\end{split}
        \label{eq:C-FT-def}
\end{equation}
This is the representation that will be used below.  It shows that the local
fluctuations near $x=y_-T$ are governed by the behavior of
$\widehat Z(q,T)$ when $q$ approaches $q_{\rm c}$ from the fluid side. On that branch, the large-$T$ behavior is controlled by the simple pole $p=\Psi(q)$ discussed in Appendix~\ref{sec:endpoint-ldt-precise}, with residue $\chi(q)$ given in Eq.~\eqref{eq:Z-regular-app}:
\begin{equation}
        \widehat Z(q,T)
        \sim \chi(q)e^{T\Psi(q)},
        \qquad T\to\infty.
        \label{eq:C-pole-form}
\end{equation}
Since $\chi(q)=1/(r\mu(q))$, cf.\ Sec.~\ref{sec:fluid-duration}, the residue has the finite limit
\begin{equation}
        \chi(q)\longrightarrow {1\over r\mu_{\rm c}},
        \qquad q\uparrow q_{\rm c},
        \label{eq:C-residue-muc}
\end{equation}
with $\mu_{\rm c}$ from Eq.~\eqref{eq:muc}.
Substituting Eq.~\eqref{eq:C-pole-form} into Eq.~\eqref{eq:C-FT-def} yields Eq.~\eqref{eq:C-FT-master} of the main text.
The scale $w_t$ is fixed by requiring the exponent in
Eq.~\eqref{eq:C-FT-master} to have a finite and non-zero limit. This requires the expansion of the fluid branch $\Psi(q)$ as $q\uparrow q_{\rm c}$ beyond linear order, which we now derive from the exact pole equation for all three edge regimes at once.\\

\noindent\textbf{Pole equation near $q_{\rm c}$.} Set
\begin{equation}
        h=q_{\rm c}-q,
        \qquad
        S(q)={r+\Psi(q)\over q^2}-D_{\max},
        \qquad
        \varepsilon=D_{\max}-D,
        \label{eq:C-marginal-defs}
\end{equation}
so that $r+\Psi(q)-q^2D=q^2[S(q)+\varepsilon]$ and $S(q)\downarrow0$ as $q\uparrow q_{\rm c}$. The pole equation~\eqref{eq:secular} reads $q^2/r=\int_0^{D_{\max}}W(D_{\max}-\varepsilon)\,\mathrm{d}\varepsilon/(\varepsilon+S)$, and the critical condition~\eqref{eq:critical-q} reads $q_{\rm c}^2/r=\int_0^{D_{\max}}W(D_{\max}-\varepsilon)\,\mathrm{d}\varepsilon/\varepsilon$. Subtracting the two, and using $1/[\varepsilon(\varepsilon+S)]=1/\varepsilon^2-S/[\varepsilon^2(\varepsilon+S)]$, gives the exact identity
\begin{equation}
        {q_{\rm c}^2-q^2\over r}
        =S(q)\int_0^{D_{\max}}
        {W(D_{\max}-\varepsilon)\over \varepsilon\,[\varepsilon+S(q)]}\,\mathrm{d}\varepsilon
        =S(q)J_{2,c}-S(q)^2\int_0^{D_{\max}}
        {W(D_{\max}-\varepsilon)\over \varepsilon^2\,[\varepsilon+S(q)]}\,\mathrm{d}\varepsilon,
        \label{eq:C-marginal-pole}
\end{equation}
with $J_{2,c}$ from Eq.~\eqref{eq:C-J2c-def}. This is the same identity as Eq.~(133) of the Supplemental Material of Ref.~\cite{gueneau-majumdar-schehr:2025}. Since $q_{\rm c}^2-q^2=2q_{\rm c}h-h^2$, to leading order
\begin{equation}
        S(q)={2\Delta\over q_{\rm c}}\,h+o(h),
        \qquad
        \Delta:={q_{\rm c}^2\over rJ_{2,c}},
        \label{eq:C-S-linear}
\end{equation}
and inserting this into $\Psi(q)=q^2[D_{\max}+S(q)]-r$ gives the slope $\Psi'(q_{\rm c}^-)=2q_{\rm c}(D_{\max}-\Delta)$. Comparing with $y_-=2q_{\rm c}D_{\rm eff}$, we identify $\Delta=D_{\max}-D_{\rm eff}$, in agreement with $y_+-y_-=2r\chi_{\rm c}/q_{\rm c}$ of Sec.~\ref{overlapsection}. The three regimes differ in the behavior of the last integral in Eq.~\eqref{eq:C-marginal-pole} as $S\downarrow0$, which is controlled by the edge law $W(D_{\max}-\varepsilon)\sim C\varepsilon^\nu$.

\paragraph{Case $1<\nu<2$.}
Here the last integral in Eq.~\eqref{eq:C-marginal-pole} diverges as $S\downarrow0$ and is dominated by $\varepsilon=O(S)$,
\begin{equation}
        \int_0^{D_{\max}}
        {W(D_{\max}-\varepsilon)\over \varepsilon^2(\varepsilon+S)}\,\mathrm{d}\varepsilon
        \simeq C\int_0^\infty{\varepsilon^{\nu-2}\,\mathrm{d}\varepsilon\over\varepsilon+S}
        ={\pi C\over|\sin(\pi\nu)|}S^{\nu-2}.
        \label{eq:C-stable-integral}
\end{equation}
Hence $(q_{\rm c}^2-q^2)/r=SJ_{2,c}-\pi CS^\nu/|\sin(\pi\nu)|+o(S^\nu)$. Solving iteratively,
\begin{equation}
        S(q)={q_{\rm c}^2-q^2\over rJ_{2,c}}
        +{\pi C\over|\sin(\pi\nu)|J_{2,c}}\left({2q_{\rm c}h\over rJ_{2,c}}\right)^\nu+o(h^\nu),
        \label{eq:C-S-stable}
\end{equation}
where the first term is analytic in $h$. Inserting Eq.~\eqref{eq:C-S-stable} into $\Psi(q)=q^2[D_{\max}+S(q)]-r$, all analytic contributions beyond the linear term are $O(h^2)=o(h^\nu)$, and the non-analytic term, multiplied by $q^2=q_{\rm c}^2+O(h)$, yields precisely the expansion~\eqref{eq:C-Psi-stable} with the coefficient $K_\nu$ given there.
Taking the Legendre transform of Eq.~(\ref{eq:C-Psi-stable}) gives the fluid-side expansion
\begin{equation}
\begin{split}
        \phi_\nu(y_--\delta)-I(y_-)+q_{\rm c}\delta
        \sim {\nu-1\over\nu}
        {\delta^{\nu/(\nu-1)}\over
        (\nu K_\nu)^{1/(\nu-1)}},
\end{split}
        \label{eq:C-I-stable-known}
\end{equation}
which corrects a typo in Eq.~(148) of the Supplemental Material of Ref.~\cite{gueneau-majumdar-schehr:2025}.
Equation~\eqref{eq:C-I-stable-known} shows that fluctuations around $y_-T$
occur on the scale $T^{1/\nu}$.  The multiplicative constant is obtained from
Eq.~\eqref{eq:C-FT-master}. Indeed, with $q=q_{\rm c}+ik/w_t$, i.e., $h=-ik/w_t$, the exponent in Eq.~\eqref{eq:C-FT-master} equals $TK_\nu(-ik/w_t)^\nu+o(1)$. Using the principal branch, $(-ik)^\nu=|k|^\nu\cos(\pi\nu/2)[1-i\,\mathrm{sgn}(k)\tan(\pi\nu/2)]$ with $\cos(\pi\nu/2)<0$, and choosing $w_t$ as in
Eq.~\eqref{eq:C-wt-stable},
we find, with $T=T_*[1+o(1)]$,
\begin{equation}
        \widehat F_T(k)\longrightarrow
        \widehat{V}_\nu(k)
        :=\exp\left(-|k|^\nu
        \left(1-i\,\mathrm{sgn}(k)\tan{\pi\nu\over2}\right)\right).
        \label{eq:C-stable-char}
\end{equation}
This equation defines the completely asymmetric stable scaling function $V_\nu$ of Eq.~\eqref{eq:stablelaw}, following the notation commonly used in the condensation literature. It is totally skewed to the right: $V_\nu(u)$ has the power-law tail $A_\nu u^{-(1+\nu)}$ for $u\to+\infty$ and a stretched-exponential tail for $u\to-\infty$, cf.\ Eq.~\eqref{eq:B-tails-intro}. This corresponds to a L{\'e}vy central limit theorem, since the critical fluid is heavy-tailed with infinite variance in the present case~\cite{nolan:2020}.\\

\paragraph{Case $\nu=2$.}
The marginal case was not worked out explicitly in Ref.~\cite{gueneau-majumdar-schehr:2025}:
Eq.~(137) of the Supplemental Material of that reference treats $1<\nu<2$ and $\nu>2$, but does not include
$\nu=2$.  For
$\nu=2$, the edge law reads
\begin{equation}
        W(D_{\max}-\varepsilon)=C\varepsilon^2+o(\varepsilon^2),
        \qquad \varepsilon\downarrow0,
        \label{eq:C-edge-nu2}
\end{equation}
and the last integral in Eq.~\eqref{eq:C-marginal-pole} diverges only logarithmically,
\begin{equation}
\begin{split}
        \int_0^{D_{\max}}{W(D_{\max}-\varepsilon)\over \varepsilon^2(\varepsilon+S)}\,\mathrm{d}\varepsilon
        =C\log{1\over S}+O(1).
\end{split}
        \label{eq:C-marginal-integral}
\end{equation}
Using $q_{\rm c}^2-q^2=2q_{\rm c}h-h^2$ and $\Delta=D_{\max}-D_{\rm eff}$, Eqs.~\eqref{eq:C-marginal-pole} and
\eqref{eq:C-marginal-integral} give directly
\begin{equation}
\begin{split}
        S(q)={2(D_{\max}-D_{\rm eff})\over q_{\rm c}}h
        +{4C(D_{\max}-D_{\rm eff})^2\over q_{\rm c}^2J_{2,c}}
        h^2\log{1\over h}+O(h^2).
\end{split}
        \label{eq:C-S-marginal}
\end{equation}
Since $\Psi(q)=q^2[D_{\max}+S(q)]-r$, Eq.~\eqref{eq:C-S-marginal} yields the expansion~\eqref{eq:C-Psi-marginal},
where $A_2$ is the regular, non-universal quadratic coefficient, and $B_{\log}$ is given in Eq.~\eqref{eq:C-Psi-marginal}.
Equivalently,
\begin{equation}
        \Psi''(q_{\rm c}-h)=2B_{\log}\log{1\over h}+O(1).
        \label{eq:C-Psipp-marginal}
\end{equation}
Thus Eq.~\eqref{eq:C-Psi-marginal} is precisely the marginal
$h^2\log(1/h)$ singularity that is missing from the non-marginal discussion in
the cited reference.  Its Legendre transform gives
\begin{equation}
        \phi_\nu(y_--\delta)-I(y_-)+q_{\rm c}\delta
        \sim {\delta^2\over4B_{\log}\log(1/\delta)}.
        \label{eq:C-I-marginal}
\end{equation}

We next insert Eq.~\eqref{eq:C-Psi-marginal} into the Fourier formula
\eqref{eq:C-FT-master}.  On the principal branch, for real $k$,
\begin{equation}
        \log{1\over -ik/w_t}
        =\log{w_t\over |k|}+{i\pi\over2}\mathrm{sgn}(k).
        \label{eq:C-marginal-log-branch}
\end{equation}
Hence the non-analytic part of the exponent in
Eq.~\eqref{eq:C-FT-master} is
\begin{equation}
        -{B_{\log}Tk^2\over w_t^2}
        \left[\log{w_t\over|k|}
        +{i\pi\over2}\mathrm{sgn}(k)\right].
        \label{eq:C-marginal-Fourier-exponent}
\end{equation}
Choosing, to leading order, $w_t$ as in Eq.~\eqref{eq:C-wt-marginal},
and using $T=T_*[1+o(1)]$, we recover
\begin{equation}
        \widehat F_T(k)\longrightarrow e^{-k^2/2}.
        \label{eq:C-FT-marginal-limit}
\end{equation}
The logarithmic singularity therefore changes the width from $T^{1/2}$ to
$(T\log T)^{1/2}$, but it does not change the leading scaling function at fixed
rescaled argument: the latter is Gaussian. The $k^2\log|k|$ structure has not disappeared, however.  It survives in the
first correction to scaling.  With the leading scale
$w_T^2=B_{\log}T\log T$, Eq.~\eqref{eq:C-marginal-Fourier-exponent} and the
regular term $A_2h^2$ in Eq.~\eqref{eq:C-Psi-marginal} give, for fixed $k$,
\begin{equation}
\begin{split}
        \log\widehat F_T(k)
        =&-{k^2\over2}
        -{k^2\over2\log T}
        \left[\log\log T+\log B_{\log}
        +{2A_2\over B_{\log}}\right]
        +{k^2\over\log T}
        \left[\log|k|-{i\pi\over2}\mathrm{sgn}(k)\right]
        +o\left({1\over\log T}\right).
\end{split}
        \label{eq:C-marginal-correction-raw}
\end{equation}
The first correction in the first line of
Eq.~\eqref{eq:C-marginal-correction-raw} is still purely quadratic in $k$ and
only renormalizes the width.  In fact it is of relative order
$\log\log T/\log T$, and is therefore larger than the genuinely non-Gaussian
correction.  To isolate the latter, define a refined width $w_T^{\rm ref}$ by
\begin{equation}
        (w_T^{\rm ref})^2
        =2T\left[B_{\log}\log w_T^{\rm ref}+A_2\right].
        \label{eq:C-refined-width-implicit}
\end{equation}
It has the expansion
\begin{equation}
        (w_T^{\rm ref})^2
        =B_{\log}T\left[
        \log T+\log\log T+\log B_{\log}
        +{2A_2\over B_{\log}}+o(1)\right].
        \label{eq:C-refined-width-expansion}
\end{equation}
If the characteristic function is defined with this refined scale, then
\begin{equation}
        \log\widehat F_T^{\rm ref}(k)
        =-{k^2\over2}
        +{k^2\over\log T}
        \left[\log|k|-{i\pi\over2}\mathrm{sgn}(k)\right]
        +o\left({1\over\log T}\right).
        \label{eq:C-marginal-universal-correction}
\end{equation}
This formula provides a direct way of observing the marginal singularity.  For
example, after using the refined width, the rescaled quantity
\begin{equation}
        \log T\left[
        \log\widehat F_T^{\rm ref}(k)+{k^2\over2}\right]
        \longrightarrow
        k^2\left[\log|k|-{i\pi\over2}\mathrm{sgn}(k)\right]
        \label{eq:C-marginal-Fourier-diagnostic}
\end{equation}
retains the full $k^2\log|k|$ dependence.  In real space, the inverse Fourier transform of the correction in
Eq.~\eqref{eq:C-marginal-universal-correction} is real and asymmetric, with an
algebraic $u^{-3}$ tail on the heavy-tail side.  Its amplitude is only
$O(1/\log T)$ for fixed $u$, but it overtakes the Gaussian core when
$u=O(\sqrt{\log\log T})$.\\

\paragraph{Case $\nu>2$.}
In this case the edge integral $J_{3,c}:=\int_0^{D_{\max}}W(D)\,\mathrm{d}D/(D_{\max}-D)^3$ is finite, and the last integral in Eq.~\eqref{eq:C-marginal-pole} tends to $J_{3,c}$ as $S\downarrow0$. Hence $(q_{\rm c}^2-q^2)/r=SJ_{2,c}-S^2J_{3,c}+o(S^2)$. Writing $S=a_1h+a_2h^2+o(h^2)$ and matching powers of $h$ gives
\begin{equation}
        a_1={2q_{\rm c}\over rJ_{2,c}},
        \qquad
        a_2={a_1^2J_{3,c}-1/r\over J_{2,c}}.
        \label{eq:C-S-Gaussian}
\end{equation}
Inserting this into $\Psi(q)=q^2[D_{\max}+S(q)]-r$, the coefficient of $h^2$ is $\sigma_{\rm c}^2/2=D_{\max}-2q_{\rm c}a_1+q_{\rm c}^2a_2$, i.e.,
\begin{equation}
        \sigma_{\rm c}^2=2D_{\max}-10\Delta+8\Delta^2{J_{3,c}\over J_{2,c}},
        \label{eq:C-sigma-compact}
\end{equation}
with $\Delta=D_{\max}-D_{\rm eff}$ from Eq.~\eqref{eq:C-S-linear}. Expanding $(D_{\rm eff}-D)^2=(\varepsilon-\Delta)^2$ and using $\int_0^{D_{\max}}W(D)\,\mathrm{d}D/(D_{\max}-D)=q_{\rm c}^2/r=\Delta J_{2,c}$, one checks that Eq.~\eqref{eq:C-sigma-compact} is identical to the expression for $\sigma_{\rm c}^2$ in Eq.~\eqref{eq:C-Psi-Gaussian}. Thus $\Psi(q)$ is twice differentiable on the fluid side of $q_{\rm c}$, with the expansion given in Eq.~\eqref{eq:C-Psi-Gaussian} and $\sigma_{\rm c}^2=\Psi''(q_{\rm c}^-)$.
This is equivalent to the quadratic rate-function expansion given in Eq.~(149)
of the Supplemental Material of Ref.~\cite{gueneau-majumdar-schehr:2025},
\begin{equation}
        \phi_\nu(y_--\delta)-I(y_-)+q_{\rm c}\delta
        \sim {\delta^2\over2\sigma_{\rm c}^2}.
        \label{eq:C-I-Gaussian-known}
\end{equation}
Choosing $w_t$ as in Eq.~\eqref{eq:C-wt-Gaussian},
Eq.~\eqref{eq:C-FT-master} gives
\begin{equation}
        \widehat F_T(k)\longrightarrow e^{-k^2/2}.
        \label{eq:C-FT-Gaussian-limit}
\end{equation}

In all three regimes, $\widehat F_T(0)\to1$.  This fact, together with the
normalized form \eqref{eq:C-N-local-profile}, will now allow us to calculate the
pre-exponential amplitude of $\mathcal N_t$ directly and compare it with that
of $Z_t(y)$.

\subsubsection{Direct calculation of the numerator and denominator amplitudes}
\label{app:critical-amplitudes}

We first integrate the local expression for the actual numerator
$\mathcal N_t$ in Eq.~\eqref{eq:C-N-local-profile}.  Since
$\Phi_{\nu,y}^{(t)}(b)$ is normalized by Eq.~\eqref{eq:C-Phi-normalization} and
$\widehat F_{T_*}(0)\to1$, the contribution of the condensate scaling regime is
\begin{equation}
\begin{split}
        \int_{\rm bump}\mathrm{d}\xi\,\mathcal N_t(\xi;y)
        \sim {}&{C\Gamma(\nu+1)\over
        \mu_{\rm c} q_{\rm c}^{2\nu+2}(y_+-y_-)}
        {e^{-tI(y)}\over\tau_*(y)^{\nu+1}}.
\end{split}
        \label{eq:C-numerator-full-prefactor}
\end{equation}
Here the integral means $\xi=\xi_*(t)+O(w_t)$; after the change
$\xi=\xi_*(t)+bw_t$, its bounds may be extended to $b\in\mathbb R$ at the
order retained.  Equation~\eqref{eq:C-numerator-full-prefactor} is therefore
the complete pre-exponential amplitude of the numerator defined in
Eq.~\eqref{eq:C-exact-start}.\\

We now compute the denominator with the same accuracy.  Recall the exact
resolvent
\begin{equation}
        R(q,p)=\int_0^{D_{\max}}{W(D) \,\mathrm{d}D\over p+r-q^2D},
        \qquad
        \widetilde Z_q(p)={R(q,p)\over1-rR(q,p)}.
        \label{eq:C-resolvent-reminder}
\end{equation}
Near $(q_{\rm c},\Psi(q_{\rm c}))$, write $R=R_{\rm fl}+R_{\rm edge}$, where
$R_{\rm edge}$ is the non-analytic contribution generated by the layer
$D \uparrow D_{\max}$, while $R_{\rm fl}$ is the regular critical-background part.
Keeping one power of $R_{\rm edge}$ selects the one-condensate sector; higher
powers correspond to several macroscopic runs. That the latter are negligible
follows from power counting. The term of order $R_{\rm edge}^k$ in the expansion
of Eq.~\eqref{eq:C-resolvent-reminder} is $\simeq R_{\rm edge}^k/(1-rR_{\rm fl})^{k+1}$.
In the time domain, each factor $R_{\rm edge}$ represents one macroscopic run of
duration $\tau_j=O(t)$ which, after integration over its displacement,
contributes a factor $\propto\tau_j^{-(\nu+1)}$ by Eq.~\eqref{eq:C-Q-edge-global}.
The critical pole of order $k+1$ contributes a factor $\propto T^k$ from the
$k+1$ fluid stretches surrounding the macroscopic runs (cf.\
Eq.~\eqref{eq:C-pole-integrals} below). Finally, along the zero-cost line
$\rho=y_+\theta$ of Eq.~\eqref{eq:C-action-square}, the total duration
$\tau_*(y)$ of the macroscopic runs is fixed, but it may be split freely among
the $k$ runs, which yields $k-1$ additional integrations over durations of order
$t$. Altogether, the $k$-condensate contribution to $Z_t(y)$ scales as
$t^k\cdot t^{k-1}\cdot t^{-k(\nu+1)}=t^{k-1-k\nu}$, i.e., it is suppressed by a
factor $O(t^{(k-1)(1-\nu)})$ relative to the single condensate ($k=1$), for
which one recovers the power $t^{-\nu}$ of Eq.~\eqref{eq:I-linear-intro}.\\

The regular denominator has the linear expansion
\begin{equation}
\begin{split}
        1-rR_{\rm fl}(q,p)
        =\mu_{\rm c}\bigl[p-\Psi(q_{\rm c})-y_-(q-q_{\rm c})\bigr]
        +o\bigl(|p-\Psi(q_{\rm c})|+|q-q_{\rm c}|\bigr).
\end{split}
        \label{eq:C-critical-denominator}
\end{equation}
Indeed,
\begin{equation}
        \left.\partial_p(1-rR)\right|_{(q_{\rm c},\Psi(q_{\rm c}))}=\mu_{\rm c},
        \label{eq:C-denominator-p-derivative}
\end{equation}
and differentiating the pole equation
$1-rR(q,\Psi(q))=0$ gives the tangent slope
$\Psi'(q_{\rm c}^-)=y_-$, which proves Eq.~\eqref{eq:C-critical-denominator}.\\

The double transform of the integrated first-run numerator in
Eq.~\eqref{eq:C-numerator-full-prefactor} contains the edge contribution in
the first-run factor and the critical fluid resolvent for the remainder.  Its singular part is
\begin{equation}
\begin{split}
        rR_{\rm edge}(q,p)
        {R_{\rm fl}(q,p)\over1-rR_{\rm fl}(q,p)}
        \sim
        {R_{\rm edge}(q,p)\over
        \mu_{\rm c}\,[p-\Psi(q_{\rm c})-y_-(q-q_{\rm c})]},
\end{split}
        \label{eq:C-numerator-transform-edge}
\end{equation}
where $rR_{\rm fl}\to1$ at the critical point.  For the denominator, expanding the exact resolvent \eqref{eq:C-resolvent-reminder} to first order in the same edge contribution gives
\begin{equation}
\begin{split}
        {R_{\rm fl}+R_{\rm edge}\over
        1-r(R_{\rm fl}+R_{\rm edge})}
        -{R_{\rm fl}\over1-rR_{\rm fl}}
        &={R_{\rm edge}\over(1-rR_{\rm fl})^2}
        +O(R_{\rm edge}^2)
        \sim {R_{\rm edge}(q,p)\over
        \mu_{\rm c}^2[p-\Psi(q_{\rm c})-y_-(q-q_{\rm c})]^2}.
\end{split}
        \label{eq:C-denominator-transform-edge}
\end{equation}
Thus numerator and denominator contain exactly the same edge amplitude.  Their only difference is one additional critical pole in the denominator.\\

To invert this extra pole, set
$s=p-\Psi(q_{\rm c})-y_-(q-q_{\rm c})$.  Then
\begin{equation}
        {1\over s}=\int_0^\infty e^{-sT}\,\mathrm{d}T,
        \qquad
        {1\over s^2}=\int_0^\infty T e^{-sT}\,\mathrm{d}T.
        \label{eq:C-pole-integrals}
\end{equation}
The variable $T$ is precisely the duration of the critical background in
Eq.~\eqref{eq:C-T-def}.  In the saddle region,
\begin{equation}
        T=T_*-{aw_t\over y_+}=T_*+O(w_t),
        \qquad w_t=o(t),
        \label{eq:C-Tstar-local}
\end{equation}
so that $T/T_*=1+o(1)$ throughout the region $a=O(1)$ contributing to the
integral.  Comparing Eqs.~\eqref{eq:C-numerator-transform-edge} and
\eqref{eq:C-denominator-transform-edge} therefore gives directly
\begin{equation}
        Z_t(y)
        \sim {T_*\over\mu_{\rm c}}
        \int_{\rm bump}\mathrm{d}\xi\,\mathcal N_t(\xi;y).
        \label{eq:C-ratio-Nt}
\end{equation}
Substituting Eq.~\eqref{eq:C-numerator-full-prefactor} yields
\begin{equation}
\begin{split}
        Z_t(y)
        \sim {}&{C\Gamma(\nu+1)\over
        \mu_{\rm c}^2q_{\rm c}^{2\nu+2}(y_+-y_-)}
        {T_*\over\tau_*(y)^{\nu+1}}e^{-tI(y)}.
\end{split}
        \label{eq:C-denominator-full}
\end{equation}
Using $T_*=t(y_+-y)/(y_+-y_-)$ and $\tau_*(y)=t(y-y_-)/(y_+-y_-)$, this is precisely Eq.~\eqref{eq:I-linear-intro} with the prefactor $B_{\rm int}(y)$ given in Eq.~\eqref{eq:Bint-intro}. The numerator is of order $t^{-(\nu+1)}e^{-tI(y)}$, whereas the denominator is
of order $t^{-\nu}e^{-tI(y)}$.  The missing factor of $t$ is exactly the factor
$T_*$ produced by the inverse transform of the double pole.\\

Recalling Eq.~\eqref{eq:Nt}, i.e.,
$
        N_t\sim T_* / \mu_{\rm c},
$
Eqs.~\eqref{eq:C-N-local-profile}, \eqref{eq:C-ratio-Nt} and
$\widehat F_{T_*}(0)\to1$ give the conditional peak directly:
\begin{equation}
        P_t^{\rm peak}\bigl(\xi_*(t)+bw_t\mid y\bigr)
        ={\mathcal N_t(\xi_*(t)+bw_t;y)\over Z_t(y)}
        \sim {1\over N_tw_t}\Phi_{\nu,y}^{(t)}(b).
        \label{eq:C-B-profile}
\end{equation}
Consequently,
\begin{equation}
        \int_{\rm bump}P_t^{\rm peak}(\xi\mid y)\,\mathrm{d}\xi
        \sim {1\over N_t}.
        \label{eq:C-bump-mass-Nt}
\end{equation}
The factor $1/N_t$ has therefore been obtained from the explicit ratio of the
pre-exponential amplitudes of $\mathcal N_t$ and $Z_t(y)$. Its interpretation as the inverse number of background runs, Eq.~\eqref{eq:Nt}, reflects the fact that the single condensate may be located at any of the $\simeq T_*/\mu_{\rm c}$ positions within the critical background.

\subsubsection{Limiting profile and position of the mode}

We now take the limit of Eq.~\eqref{eq:C-Fourier-product} in the three regimes.
This last step is immediate once the scale $w_t$ has been fixed.\\

\paragraph{Case $1<\nu<2$.}
Here $w_t\asymp t^{1/\nu}$, and therefore
\begin{equation}
        {D_{\max}\tau_*(y)\over w_t^2}\longrightarrow0.
        \label{eq:C-Gamma-stable-vanishes}
\end{equation}
Thus $\widehat\Gamma_t(\ell)\to1$, while
Eq.~\eqref{eq:C-stable-char} gives
\begin{equation}
        \widehat\Phi_{\nu,y}(k)
        =\widehat{V}_\nu\left(-{k\over\gamma_y}\right).
        \label{eq:C-profile-stable-Fourier}
\end{equation}
Inverting the Fourier transform yields
\begin{equation}
        \Phi_{\nu,y}(b)
        =\gamma_yV_\nu(-\gamma_y b).
        \label{eq:C-profile-stable}
\end{equation}
If $m_\nu$ denotes the mode of $V_\nu$, then
\begin{equation}
        b_{\rm mode}=-{m_\nu\over\gamma_y},
        \qquad
        \xi_{\rm mode}(t)
        =\xi_*(t)-{m_\nu\over\gamma_y}w_t+o(w_t).
        \label{eq:C-bmode-stable}
\end{equation}

\paragraph{Case $\nu=2$.}
Now $w_t^2=B_{\log}T_*\log T_*$, so that the Gaussian factor associated with
the first run again disappears,
$
        \widehat\Gamma_t(\ell)\longrightarrow1.
$
Using Eq.~\eqref{eq:C-FT-marginal-limit} in
Eq.~\eqref{eq:C-Fourier-product}, we obtain
\begin{equation}
        \widehat\Phi_{2,y}(k)
        =\exp\left(-{k^2\over2\gamma_y^2}\right).
        \label{eq:C-profile-marginal-Fourier}
\end{equation}
Since a centered Gaussian ${\cal G}_\Sigma(b)$ has Fourier transform
$\exp(-\Sigma k^2/2)$, Eq.~\eqref{eq:C-profile-marginal-Fourier} gives directly
\begin{equation}
        \Phi_{2,y}(b)={\cal G}_{1/\gamma_y^2}(b),
        \qquad
        b_{\rm mode}=0.
        \label{eq:C-profile-marginal}
\end{equation}
This is the second line of the central profile formula in Eq.~\eqref{eq:Phi-profile-main} as stated in the main text. The correction derived in Eq.~\eqref{eq:C-marginal-universal-correction} is
also present in the finite-time bump itself.  This is most transparent in the
reflected and standardized bump coordinate
$
        \tilde u=-\gamma_y b,
$
which is measured in the same units as the argument $u$ of $F_{T_*}(u)$.
By Eq.~\eqref{eq:C-Fourier-product}, its characteristic function is
\begin{equation}
        \widehat\Phi_{2,y}^{(\tilde u,t)}(k)
        ={\widehat F_{T_*}(k)\over\widehat F_{T_*}(0)}
        \widehat\Gamma_t(\beta k)+o\left({1\over\log T_*}\right).
        \label{eq:C-marginal-z-Fourier}
\end{equation}
Here the error term collects the corrections neglected in the derivation of Eq.~\eqref{eq:C-Fourier-product}: the replacement~\eqref{eq:C-FT-Ta-replacement}, which is of relative order $w_t/t$, the corrections to the edge asymptotics of $Q$, which are $O(1/t)$, and the deviation of $r\mu_{\rm c}\chi(q_{\rm c}+ik/w_t)$ from $1$, which is $O(w_t^{-1}\log w_t)$ for $\nu=2$. All of them are $o(1/\log T_*)$, so that Eq.~\eqref{eq:C-marginal-z-Fourier} may be expanded to this order.
Using the refined background width of
Eq.~\eqref{eq:C-refined-width-implicit}, we obtain
\begin{equation}
\begin{split}
        \log\widehat\Phi_{2,y}^{(\tilde u,t)}(k)
        ={}&-{k^2\over2}
        +{k^2\over\log T_*}
        \left[\log|k|-{i\pi\over2}\mathrm{sgn}(k)\right]-{D_{\max}\beta^2\tau_*(y)\over
        B_{\log}T_*\log T_*}\,k^2
        +o\left({1\over\log T_*}\right).
\end{split}
        \label{eq:C-marginal-bump-correction}
\end{equation}
The second term on the RHS of
Eq.~\eqref{eq:C-marginal-bump-correction} is the genuinely non-Gaussian
marginal correction.  The last term, which stems from $\Gamma_t$, is
purely quadratic in $k$ and merely produces an additional finite-time
renormalization of the variance.  Thus, after fitting the finite-time variance,
the $k^2\log|k|$ diagnostic in Eq.~\eqref{eq:C-marginal-Fourier-diagnostic}
can be applied directly to the condensate bump.\\

\paragraph{Case $\nu>2$.}
In this regime $w_t^2=\sigma_{\rm c}^2T_*$.  The critical-background factor tends to
$\exp[-k^2/(2\gamma_y^2)]$, while the first-run factor has the non-trivial
limit
\begin{equation}
        \widehat\Gamma_t\left(-{\beta k\over\gamma_y}\right)
        \longrightarrow
        \exp\left(-{D_{\max}\beta^2\tau_*(y)\over
        \gamma_y^2\sigma_{\rm c}^2T_*}k^2\right).
        \label{eq:C-Gamma-Gaussian-limit}
\end{equation}
Their product is Gaussian.  Hence
$
        \Phi_{>,y}(b)={\cal G}_{\Sigma_{>,y}}(b),
$
with the variance $\Sigma_{>,y}$ stated in Eq.~\eqref{eq:C-Sigma-Gaussian}.
The profile is centered, and therefore
$
        b_{\rm mode}=0
$.

\subsection{Derivation of the saturated peak}
\label{app:saturated}

This appendix derives the continuous part $\mathcal{P}_y^{\rm cont}(u)$ of the $O(1)$ law of $u=yt-\xi$ in the saturated phase, its mass, and its tails, as stated in Sec.~\ref{sec:cond-saturated}.  The starting point is
the exact first-run marginal in Eq.~\eqref{eq:exactfirst}. Throughout this appendix,
\begin{equation}
        y>y_+,
        \qquad
        m_y={y_+^2\over y^2},
        \qquad
        w_y=(1-m_y)^2.
        \label{eq:E-saturated-parameters}
\end{equation}
The endpoint density has the asymptotic form
\begin{equation}
        Z_t(y)\sim
        {C\Gamma(\nu+1)\over
        \sqrt{4\pi D_{\max}}\,w_y}
        \left({2D_{\max}\over y}\right)^{2\nu+2}
        t^{-(\nu+3/2)}e^{-tJ(y)},
        \qquad
        J(y)=r+{y^2\over4D_{\max}},
        \label{eq:E-Z-saturated}
\end{equation}
which is Eq.~\eqref{eq:Z-sat-intro} with the prefactor
\eqref{eq:A-y-intro}.  The saturated peak contains two distinct pieces: a
Dirac mass produced by trajectories with no switch, whose weight $w_y$ was derived in Sec.~\ref{sec:no-switch}, and a continuous density
produced by trajectories whose first run ends shortly before time $t$.

\subsubsection{Continuous contribution}

We now return to the continuous term in the exact first-run formula
\eqref{eq:exactfirst}.  The trajectory following the first completed run is an
independent copy of the original process, and hence has the endpoint density
$Z(x,t)$ defined in Eq.~\eqref{eq:Z-def}.  Evaluated in the saturated scaling regime
$\xi=yt-u$, the continuous contribution reads
\begin{equation}
\begin{split}
        P_t^{\rm cont}(yt-u\mid y)
        ={1\over Z_t(y)}\int_0^t \mathrm{d}\tau\int_0^{D_{\max}}\mathrm{d}D\,
        r e^{-r\tau}W(D)G_{D,\tau}(yt-u)
        Z(u,t-\tau).
\end{split}
        \label{eq:E-exact-continuous-start}
\end{equation}
Equation~\eqref{eq:E-exact-continuous-start} is the exact starting point of the
calculation.  Here $\tau$ is the duration of the first run, while the
independent trajectory that follows it has duration $t-\tau$, displacement
$u$, and endpoint density $Z(u,t-\tau)$, with $Z(x,t)$ defined in
Eq.~\eqref{eq:Z-def}.\\

For fixed $u=O(1)$, a first run carrying a displacement $yt-u$ must occupy
almost the whole observation time.  We therefore introduce the local variables~\eqref{eq:E-local-variables}, in terms of which the exponential cost of the long first run has the expansion~\eqref{eq:E-exponent-expansion}.\\

Using again $W(D_{\max}-\zeta/t)\sim C(\zeta/t)^\nu$, integrating over $\zeta$, and then
extending the $s$-integral to $[0,\infty)$, while keeping the exact endpoint
density $Z(x,t)$ defined in Eq.~\eqref{eq:Z-def}, we find
\begin{equation}
\begin{split}
        &\int_0^t \mathrm{d}\tau\int_0^{D_{\max}}\mathrm{d}D\,
        r e^{-r\tau}W(D)G_{D,\tau}(yt-u)
        Z(u,t-\tau)
        \\
        &\quad\sim
        {rC\Gamma(\nu+1)\over\sqrt{4\pi D_{\max}}}
        \left({2D_{\max}\over y}\right)^{2\nu+2}
        t^{-(\nu+3/2)}e^{-tJ(y)}
        \exp\!\left({yu\over2D_{\max}}\right)
        \int_0^\infty
        \exp\!\left(-s\left({y^2\over4D_{\max}}-r\right)\right)
        Z(u,s)\,\mathrm{d}s .
\end{split}
        \label{eq:E-continuous-numerator}
\end{equation}
The numerator in Eq.~\eqref{eq:E-continuous-numerator} and the denominator in
Eq.~\eqref{eq:E-Z-saturated} have the same power of $t$, the same exponential
factor, and the same edge amplitude.  Their ratio therefore has a finite
limit.  Substituting them into Eq.~\eqref{eq:E-exact-continuous-start} yields
directly Eq.~\eqref{eq:Pcont-main} of the main text, in terms of the endpoint density $Z(x,t)$ defined in
Eq.~\eqref{eq:Z-def}.

\subsubsection{Mass of the continuous peak}

To check the normalization, note that the endpoint density of the residual trajectory obeys the renewal equation $Z(u,s)=e^{-rs}\int_0^{D_{\max}}W(D)G_{D,s}(u)\,\mathrm{d}D+\int_0^s\mathrm{d}\tau\int_{\mathbb R}\mathrm{d}\xi\,Q(\xi,\tau)Z(u-\xi,s-\tau)$, where the first term accounts for a residual trajectory without a switch, i.e., consisting of a single unfinished run. Applying the tilt $e^{q_yu-\Psi(q_y)s}$ and integrating over $s$, the completed-run kernel $Q$ becomes $\mathcal F_y=p_{q_y}$ of Eq.~\eqref{eq:fluid-microscopic-integral}, while the unfinished run, which lacks the factor $r$ of $Q$, becomes $\mathcal F_y/r$. Hence $\int_0^\infty e^{q_yu-\Psi(q_y)s}Z(u,s)\,\mathrm{d}s=r^{-1}\sum_{n\geq1}\mathcal F_y^{*n}(u)$, and Eq.~\eqref{eq:Pcont-main} becomes the
convolution representation~\eqref{eq:E-Pcont-renewal-series}, where the term with index $n$ represents a residual cloud containing $n$
microscopic runs, the last of which is unfinished.  Since
\begin{equation}
       \int_{-\infty}^\infty\mathcal{F}_y(\xi)\,\mathrm{d}\xi=m_y,
        \qquad
        w_y=(1-m_y)^2,
        \label{eq:E-fluid-mass-reminder}
\end{equation}
we obtain
\begin{equation}
\begin{split}
       \int_{-\infty}^\infty\mathcal{P}_y^{\rm cont}(u)\,\mathrm{d}u
        &=w_y\sum_{n\geq1}m_y^n
        =m_y(1-m_y).
\end{split}
        \label{eq:E-Pcont-mass}
\end{equation}
Adding the atom gives
\begin{equation}
       \int_{-\infty}^\infty\mathcal{P}_y^{\rm sat}(u)\,\mathrm{d}u
        =w_y+m_y(1-m_y)
        =1-m_y,
        \label{eq:E-Psat-mass}
\end{equation}
which is precisely the mass missing from the fluid component.

\subsubsection{Tails}

The renewal series \eqref{eq:E-Pcont-renewal-series} also gives the tails with no
further calculation.  For $u\to+\infty$,
\begin{equation}
        \mathcal{F}_y(u)\sim A_y^{(+)}u^{-(\nu+1)}.
        \label{eq:E-F-positive-tail}
\end{equation}
Because this tail is subexponential and $\mathcal{F}_y(\xi)$ has mass $m_y$,
\begin{equation}
        \mathcal{F}_y^{*n}(u)
        \sim n m_y^{n-1}A_y^{(+)}u^{-(\nu+1)}.
        \label{eq:E-convolution-tail}
\end{equation}
Substitution in Eq.~\eqref{eq:E-Pcont-renewal-series} gives
\begin{equation}
\begin{split}
        \mathcal{P}_y^{\rm cont}(u)
        \sim w_y A_y^{(+)}u^{-(\nu+1)}
        \sum_{n\geq1}n m_y^{n-1} =A_y^{(+)}u^{-(\nu+1)},
        \qquad u\to+\infty,
\end{split}
        \label{eq:E-Pcont-positive-tail}
\end{equation}
because $\sum_{n\geq1}n m_y^{n-1}=(1-m_y)^{-2}=w_y^{-1}$. Finally, the explicit expression \eqref{eq:pyfluid} implies the exact symmetry
\begin{equation}
        \mathcal{F}_y(-u)=\exp\!\left(-{yu\over D_{\max}}\right)\mathcal{F}_y(u),
        \qquad u>0.
        \label{eq:E-fluid-tilt-symmetry}
\end{equation}
This symmetry is preserved by convolution and hence by the sum in
Eq.~\eqref{eq:E-Pcont-renewal-series}.  Therefore
\begin{equation}
        \mathcal{P}_y^{\rm cont}(u)\sim
        A_y^{(+)}|u|^{-(\nu+1)}
        \exp\!\left(-{y|u|\over D_{\max}}\right),
        \qquad u\to-\infty.
        \label{eq:E-C-negative-tail}
\end{equation}
Equations~\eqref{eq:E-Pcont-positive-tail} and
\eqref{eq:E-C-negative-tail} reproduce the two tails stated in
Eq.~\eqref{eq:Pcont-tails-intro}.

\subsection{Derivation of the interior asymptotics}
\label{app:interior}

Here, we derive the asymptotics of the conditional first-run distribution in the macroscopic interior sector, $\xi=O(t)$, stated in Sec.~\ref{sec:interior}.

\subsubsection{Fluid interior sector $y<y_-$}

We first consider the fluid phase, $0<y<y_-$. In this regime, a typical
conditioned trajectory contains only microscopic runs, so that no
macroscopic contribution is present. We therefore probe the interior
macroscopic sector by imposing a first-run displacement of order $t$,
and show that such a fluctuation is exponentially suppressed. Starting from the exact renewal representation, we have
\begin{equation}
P_t^{\mathrm{cont}}(\xi\mid y)
=
\frac{1}{Z_t(y)}
\int_0^t \mathrm{d}\tau\,
Q(\xi,\tau)\,
Z(yt-\xi,t-\tau),
\label{eq:interior-renewal}
\end{equation}
with $Q(\xi,\tau)$ defined in Eq.~\eqref{eq:C-Qdef}.
We consider the macroscopic scaling
\begin{equation}
\xi=\rho t,
\qquad
0<\rho<y<y_-,
\qquad
\tau=\theta t\, .
\end{equation}
After the first switch, the remaining trajectory has a duration $(1-\theta)t$, a displacement $(y-\rho)t$, and hence an effective velocity
$
z=(y-\rho) / (1-\theta)\, .
$
We first assume that the saddle-point value of $z$ remains on the
regular fluid branch, $z<y_-$, and verify this assumption below. Using
the fluid endpoint asymptotics in Eq.~(\ref{eq:Z-regular-intro}), we have
\begin{equation}
Z\!\left((y-\rho)t,(1-\theta)t\right)
\sim
B_{\mathrm{reg}}(z)\,[(1-\theta)t]^{-1/2}
e^{-t(1-\theta)I(z)},
\label{eq:fluid-int-Zrem}
\end{equation}
whereas the normalization behaves as
\begin{equation}
Z_t(y)
\sim
B_{\mathrm{reg}}(y)\,t^{-1/2}
e^{-tI(y)}.
\label{eq:fluid-int-Znorm}
\end{equation}
On the regular branch,
\begin{equation}
I(z)=q_*(z)z-\Psi(q_*(z)),
\qquad
z=\Psi'(q_*(z)).
\end{equation}
For a macroscopic first run, $\xi=\rho t$ and $\tau=\theta t$, the first-run factor is given by the edge
asymptotics~\eqref{eq:C-Q-edge-global}, with $q_{\rho,\theta}=\rho/(2D_{\max}\theta)$. Combining these expressions and changing variables from $\tau$ to
$\theta=\tau/t$, we obtain
\begin{equation}
P_t^{\rm cont}(\rho t\mid y)
\sim
\frac{rC\Gamma(\nu+1)}{\sqrt{4\pi D_{\max}}}\,
t^{-(\nu+1/2)}
\int_0^1 \mathrm{d}\theta\,
\frac{B_{\rm reg}(z)}{B_{\rm reg}(y)}
\frac{
e^{-t[s_y(\theta,\rho)-I(y)]}
}{
q_{\rho,\theta}^{\,2\nu+2}
\theta^{\nu+3/2}(1-\theta)^{1/2}
},
\label{eq:fluid-interior-integral}
\end{equation}
where
\begin{equation}
s_y(\theta,\rho)
=
r\theta+
\frac{\rho^2}{4D_{\max}\theta}
+
(1-\theta)I(z),
\qquad
z=\frac{y-\rho}{1-\theta}.
\label{eq:fluid-interior-action}
\end{equation}

We now determine the saddle in $\theta$. Writing
$q_*(z)=I'(z)$ and using
$I(z)=q_*(z)z-\Psi(q_*(z))$, differentiation gives
\begin{equation}
\partial_\theta s_y(\theta,\rho)
=
r-\frac{\rho^2}{4D_{\max}\theta^2}
+\Psi(q_*(z)).
\label{eq:fluid-saddle-derivative}
\end{equation}
The saddle $\theta=\theta_\rho$ therefore satisfies
\begin{equation}
D_{\max}q_{\rho,\theta_\rho}^2
=
r+\Psi( q_*(z_\rho)),
\qquad
z_\rho=\frac{y-\rho}{1-\theta_\rho}.
\label{eq:fluid-saddle}
\end{equation}
Equivalently,
\begin{equation}
\theta_\rho
=
\frac{\rho}
{2\sqrt{D_{\max}[r+\Psi( q_*(z_\rho))]}},
\qquad
z_\rho=\Psi'( q_*(z_\rho)).
\label{eq:fluid-saddle-param}
\end{equation}
It remains to verify that the residual trajectory indeed stays in the
fluid regime, i.e. $z_\rho<y_-$. Since $
z(\theta)=(y-\rho)/(1-\theta)$ 
is increasing in $\theta$, the boundary $z=y_-$ corresponds to
$
\theta_- = 1-(y-\rho) / y_-
$.
The saddle satisfies $\partial_\theta s_y(\theta_\rho,\rho)=0$. Moreover,
$\partial_\theta s_y\to-\infty$ as $\theta\to0$, whereas at the boundary
\begin{equation}
\partial_\theta s_y(\theta_-,\rho)
=
D_{\max}q_{\rm c}^2
-
\frac{\rho^2}{4D_{\max}\theta_-^2}
>0,
\end{equation}
where we used $\theta_->\rho/y_+$ and
$y_+=2D_{\max}q_{\rm c}$. Hence a saddle occurs for
$\theta_\rho<\theta_-$, and therefore
$
z_\rho<y_-.
$
This saddle is the global minimum of $s_y(\theta,\rho)$ over $0<\theta<1$. Indeed, $s_y$ is convex in $\theta$, since $r\theta+\rho^2/(4D_{\max}\theta)$ is convex and $(1-\theta)I\bigl((y-\rho)/(1-\theta)\bigr)$ is the perspective of the convex function $I$, so that the saddle is unique. This can also be checked directly for $\theta>\theta_-$: on the linear branch, $y_-<z<y_+$, one has $\partial_\theta s_y=D_{\max}(q_{\rm c}^2-q_{\rho,\theta}^2)>0$, since $\rho/\theta<\rho/\theta_-<y_+$, while on the saturated branch, $z>y_+$, one has $\partial_\theta s_y=(z^2-\rho^2/\theta^2)/(4D_{\max})>0$, since $\rho/\theta<y_+<z$.
Thus the use of the regular fluid asymptotics is self-consistent. Expanding around the saddle gives
\begin{equation}
\int \mathrm{d}\theta\,
e^{-t[s_y(\theta,\rho)-I(y)]}
\sim
\sqrt{
\frac{2\pi}
{t\,s_y''(\theta_\rho,\rho)}
}
e^{-tJ_y^{(1)}(\rho)},
\end{equation}
where
\begin{equation}
J_y^{(1)}(\rho)
=
s_y(\theta_\rho,\rho)-I(y)=
q_*(z_\rho)(y-\rho)
-
\Psi\!\left(q_*(z_\rho)\right)
+
\rho
\sqrt{
\frac{r+\Psi\!\left(q_*(z_\rho)\right)}
{D_{\max}}
}
-
I(y)\, .
\label{eq:fluid-interior-rate}
\end{equation}
Consequently,
\begin{equation}
P_t^{\rm cont}(\rho t\mid y)
\sim
\mathcal{H}_y(\rho)\,
t^{-(\nu+1)}
e^{-tJ_y^{(1)}(\rho)},
\qquad
0<\rho<y,
\label{eq:fluid-interior-final}
\end{equation}
with
\begin{equation}
\mathcal{H}_y(\rho)
=
\frac{rC\Gamma(\nu+1)}
{\sqrt{4\pi D_{\max}}}
\frac{B_{\rm reg}(z_\rho)}
{B_{\rm reg}(y)}
\frac{1}
{q_{\rho,\theta_\rho}^{\,2\nu+2}
\theta_\rho^{\nu+3/2}
(1-\theta_\rho)^{1/2}}
\sqrt{
\frac{2\pi}
{s_y''(\theta_\rho,\rho)}
}.
\label{eq:fluid-interior-amplitude}
\end{equation}

The function $J_y^{(1)}(\rho)$ is strictly positive for every $\rho>0$. Differentiation gives
\begin{equation}
\frac{\mathrm d J_y^{(1)}}{\mathrm d\rho}
=
\partial_\rho s_y(\theta_\rho,\rho)
+
\partial_\theta s_y(\theta_\rho,\rho)
\frac{\mathrm d\theta_\rho}{\mathrm d\rho}.
\end{equation}
At the saddle, $\partial_\theta s_y(\theta_\rho,\rho)=0$, hence
\begin{equation}
\frac{\mathrm d J_y^{(1)}}{\mathrm d\rho}
=
\partial_\rho s_y(\theta_\rho,\rho)=
\frac{\rho}{2D_{\max}\theta}
-
I'(z)
=
q_{\rho,\theta}-q_*(z)\, .
\end{equation}
Using Eq.~(\ref{eq:fluid-saddle}) together with the regular fluid branch
inequality
\begin{eqnarray}
    r+\Psi(q_*(z_\rho))
>
D_{\max}q_*(z_\rho)^2\, ,
\end{eqnarray}
we obtain
\begin{equation}
q_{\rho,\theta_\rho}>q_*(z_\rho),
\qquad
\frac{\mathrm{d}J_y^{(1)}}{\mathrm{d}\rho}>0.
\end{equation}
Since $J_y^{(1)}(0)=0$, it follows that
$
J_y^{(1)}(\rho)>0$,
$
0<\rho<y$.
Hence, a macroscopic first run in the fluid phase is exponentially rare.\\

Finally, the limit $\rho\to0$ matches the positive tail of the
microscopic fluid law. Setting $q_y^*=q_*(y)$, one has
\begin{equation}
J_y^{(1)}(\rho)
=
\lambda_+(q_y^*)\,\rho+o(\rho),
\qquad
\lambda_+(q)
=
\sqrt{\frac{r+\Psi(q)}{D_{\max}}}-q,
\end{equation}
while
\begin{equation}
\mathcal{H}_y(\rho)
\sim
B(q_y^*)\,\rho^{-(\nu+1)}.
\end{equation}
Therefore, for $\xi=\rho t$ with $1\ll\xi\ll t$,
\begin{equation}
P_t^{\rm cont}(\xi\mid y)
\sim
B(q_y^*)\,\xi^{-(\nu+1)}
e^{-\lambda_+(q_y^*)\xi},
\end{equation}
which precisely recovers the positive tail of the fluid distribution
in Eq.~(\ref{fluidtailsgeneral}).\\

Thus, the fluid interior describes an exponentially rare fluctuation in which a single run is forced to carry an $O(t)$ displacement, while the remaining
trajectory stays entirely on the regular fluid branch.

\subsubsection{Coexistence interior sector $y_-<y<y_+$}

In this section, we investigate the interior macroscopic sector of the
single-run distribution, corresponding to first-run displacements
$
    0 < \xi < \xi_*(t),
$
i.e., between the microscopic fluid contribution around $\xi=O(1)$ and
the condensate bump centered at $\xi_*(t)$. Since this regime remains
away from the no-switching contribution at $\xi=yt$, only the continuous
part of the conditional first-run distribution contributes. We shall show that, in the coexistence phase, the leading contribution
to Eq.~\eqref{eq:interior-renewal} for $0<\xi<\xi_*(t)$ does not arise
from a single macroscopic run followed by a purely fluid background.
Instead, the dominant trajectories contain two macroscopic contributions:
the first run carries the prescribed displacement $\xi$, while the
remaining trajectory is itself in the coexistence phase and contains a
second condensate carrying the missing macroscopic fraction. In this
sense, the condensate is split between two distinct runs. More precisely, writing
$\xi=\rho t$, $\tau=\theta t$,
with $0<\rho<\rho_*$, the first run carries the macroscopic
displacement $\rho t$, while the remaining trajectory, of duration
$(1-\theta)t$, must produce the residual displacement $(y-\rho)t$.
Its effective velocity is therefore
$
    z= (y-\rho) / (1-\theta).
$
As shown below, the saddle-point value of $z$ remains inside the
coexistence interval $(y_-,y_+)$ for $0<\rho<\rho_*$. The remaining
trajectory is therefore itself condensed, which is the origin of the
split-condensate contribution.\\

For a macroscopic first run, $\xi=\rho t$ and $\tau=\theta t$, the edge
asymptotics is given again by Eq.~(\ref{eq:C-Q-edge-global}).
We first assume that $y_-<z<y_+$ lies within the coexistence interval. Under this assumption, the endpoint density of the remaining trajectory
is governed by the intermediate large-deviation asymptotics Eq.~(\ref{eq:I-linear-intro})
\begin{equation}
        Z(yt-\xi,(1-\theta )t)\overset{t \to \infty}{\sim} B_{\rm int}(z)((1-\theta)t)^{-\nu}e^{-t(1-\theta)I(z)},
        \qquad
        I(z)=q_{\rm c}z-\Psi(q_{\rm c}),
        \label{eq:I-linear-intderiv}
\end{equation}
where $B_{\rm int}(z)$ is the amplitude~\eqref{eq:Bint-intro}, evaluated at velocity $z$.
We shall verify below that the saddle-point value of $z$ indeed belongs
to this interval, so that the use of the intermediate asymptotics is
self-consistent. Since the conditioning velocity itself lies in the coexistence phase,
$y_-<y<y_+$, the normalization is given by
\begin{equation}
    Z_t(y)
    \underset{t\to\infty}{\sim}
    B_{\rm int}(y)t^{-\nu}e^{-tI(y)},
    \qquad
    I(y)=q_{\rm c}y-\Psi(q_{\rm c}) .
    \label{eq:Z-int-conditioning}
\end{equation}
Combining Eqs.~(\ref{eq:I-linear-intderiv}) and
(\ref{eq:Z-int-conditioning}) with the edge asymptotics of the first-run
factor, and changing variables from $\tau$ to $\theta=\tau/t$, we obtain
\begin{align}
P_t^{\rm cont}(\rho t\mid y)
\sim
\frac{rC\Gamma(\nu+1)}
{\sqrt{4\pi D_{\max}}}\,
t^{-(\nu+1/2)}
\int_0^1 \mathrm{d}\theta\, \frac{B_{\rm int}(z)}{B_{\rm int}(y)}
\frac{
\exp\!\left\{-t\left[s_y(\theta,\rho)-I(y)\right]\right\}
}{
q_{\rho,\theta}^{\,2\nu+2}
\theta^{\nu+3/2}
(1-\theta)^\nu
},
\label{eq:interior-theta-integral}
\end{align}
where
\begin{equation}
    s_y(\theta,\rho)
    =
    r\theta
    +
    \frac{\rho^2}{4D_{\max}\theta}
    +
    (1-\theta)I(z).
    \label{eq:interior-action}
\end{equation}
Since both $y$ and, by assumption, $z$ belong to the coexistence
interval, the rate function is linear for both arguments, we thus
obtain
\begin{align}
s_y(\theta,\rho)-I(y)
&=
r\theta
+\frac{\rho^2}{4D_{\max}\theta}
+(1-\theta)\big[q_{\rm c}z-\Psi(q_{\rm c})\big]
-\big[q_{\rm c}y-\Psi(q_{\rm c})\big]
\nonumber\\
&=
\frac{\rho^2}{4D_{\max}\theta}
-q_{\rm c}\rho
+\theta\big[r+\Psi(q_{\rm c})\big].
\end{align}
At the critical point,
$
    r+\Psi(q_{\rm c})=D_{\max}q_{\rm c}^2
$
and
$y_+=2D_{\max}q_{\rm c}
$,
and therefore the excess action can be written as the perfect square
\begin{equation}
    s_y(\theta,\rho)-I(y)
    =
    \frac{(\rho-y_+\theta)^2}
         {4D_{\max}\theta}.
    \label{eq:split-action-square}
\end{equation}
For fixed $\rho$, the exponent in Eq.~(\ref{eq:interior-theta-integral})
is therefore minimized at
$
    \theta_\rho=\rho / y_+,
$
for which
\begin{equation}
    s_y(\theta_\rho,\rho)=I(y),
    \qquad
    q_{\rho,\theta_\rho}=q_{\rm c}.
    \label{eq:split-zero-cost}
\end{equation}
Hence, there is no residual
exponential suppression in $t$. It remains to verify that the saddle indeed lies within the assumed
coexistence sector for the remaining trajectory. At
$\theta=\theta_\rho$, its scaled endpoint displacement is
\begin{equation}
    z_\rho
    =
    \frac{y-\rho}{1-\rho/y_+}.
    \label{eq:split-z-saddle}
\end{equation}
Using the condensate position
\begin{equation}
    \rho_*
    =
    y_+\frac{y-y_-}{y_+-y_-},
    \label{eq:split-rhostar}
\end{equation}
one finds
\begin{equation}
    z_\rho-y_-
    =
    \frac{y-y_-}
         {1-\rho/y_+}
    \left(1-\frac{\rho}{\rho_*}\right)>0.
    \label{eq:split-zminus}
\end{equation}
Moreover,
\begin{equation}
    y_+-z_\rho
    =
    \frac{y_+-y}{1-\rho/y_+}>0.
\end{equation}
It follows that
\begin{equation}
    0<\rho<\rho_*
    \qquad\Longleftrightarrow\qquad
    y_-<z_\rho<y_+,
\end{equation}
which confirms a posteriori that the intermediate asymptotics used in
Eq.~(\ref{eq:I-linear-intderiv}) is self-consistent.\\

We now evaluate Eq.~(\ref{eq:interior-theta-integral}) beyond
exponential order. Expanding the action around the saddle gives
\begin{equation}
    s_y(\theta,\rho)-I(y)
    =
    \frac{1}{2}
    s_y''(\theta_\rho,\rho)
    (\theta-\theta_\rho)^2
    +o\!\left((\theta-\theta_\rho)^2\right),
\end{equation}
with
\begin{equation}
    s_y''(\theta_\rho,\rho)
    =
    \frac{2D_{\max}q_{\rm c}^2}{\theta_\rho}.
    \label{eq:split-curvature}
\end{equation}
Consequently,
\begin{equation}
    \int \mathrm{d}\theta\,
    e^{-t[s_y(\theta,\rho)-I(y)]}
    \sim
    \sqrt{
        \frac{2\pi}
             {t\,s_y''(\theta_\rho,\rho)}
    }
    =
    \sqrt{
        \frac{\pi\theta_\rho}
             {D_{\max}q_{\rm c}^2t}
    }.
    \label{eq:split-gaussian}
\end{equation}
Evaluating the smooth prefactors in
Eq.~(\ref{eq:interior-theta-integral}) at the saddle,
$q_{\rho,\theta_\rho}=q_{\rm c}$, and using
Eq.~(\ref{eq:split-gaussian}), we obtain
\begin{equation}
P_t^{\rm cont}(\rho t\mid y)
\sim
\frac{rC\Gamma(\nu+1)}
     {2D_{\max}q_{\rm c}^{2\nu+3}}
\,t^{-(\nu+1)}
\frac{B_{\rm int}(z_\rho)}
     {B_{\rm int}(y)}
\frac{1}
     {\theta_\rho^{\nu+1}
      (1-\theta_\rho)^\nu}.
\label{eq:split-after-saddle}
\end{equation}
The remaining ratio of intermediate prefactors can be simplified
explicitly to
\begin{equation}
\frac{B_{\rm int}(z_\rho)}
     {B_{\rm int}(y)}
=
\frac{y_+-z_\rho}{y_+-y}
\left(
    \frac{y-y_-}{z_\rho-y_-}
\right)^{\nu+1}.
\end{equation}
Using Eqs.~(\ref{eq:split-z-saddle}) and
(\ref{eq:split-zminus}), this becomes
\begin{equation}
    \frac{B_{\rm int}(z_\rho)}
         {B_{\rm int}(y)}
    =
    \frac{(1-\theta_\rho)^\nu}
         {\left(1-\rho/\rho_*\right)^{\nu+1}}.
    \label{eq:split-B-ratio}
\end{equation}
Substituting Eq.~(\ref{eq:split-B-ratio}) into
Eq.~(\ref{eq:split-after-saddle}), the factors
$(1-\theta_\rho)^\nu$ cancel. Since
$\theta_\rho=\rho/y_+$ and $y_+=2D_{\max}q_{\rm c}$, we finally find
\begin{equation}
    P_t^{\rm cont}(\rho t\mid y)
    \sim
    \frac{A_{\rm c}}
         {(\rho t)^{\nu+1}}
    \frac{1}
         {\left(1-\rho/\rho_*\right)^{\nu+1}},
    \qquad
    0<\rho<\rho_*,
    \label{eq:split-final}
\end{equation}
where
\begin{equation}
    A_{\rm c}
    =
    \frac{
        rC\Gamma(\nu+1)
        2^\nu D_{\max}^{\nu}
    }{
        q_{\rm c}^{\nu+2}
    }.
    \label{eq:split-Ac}
\end{equation}
We now introduce $\omega=\frac{\rho}{\rho_*}
=\frac{\xi}{\xi_*(t)}$, $0<\omega<1$. Since $\rho=\omega\rho_*$, Eq.~(\ref{eq:split-final}) can be written as
\begin{eqnarray}
    P_t^{\rm cont}\!\left(\omega\rho_* t\mid y\right)
\sim
\frac{A_{\rm c}}
{(\rho_* t)^{\nu+1}[\omega(1-\omega)]^{\nu+1}},
\qquad 0<\omega<1.
\label{coexist_final_interior}
\end{eqnarray}
The two mesoscopic limits of Eq.~(\ref{coexist_final_interior}) recover the neighboring asymptotic regimes given in the main text. For $\omega\to0$, with $\xi=\omega\rho_* t$ and $1\ll\xi\ll t$, one finds
\begin{equation}
P_t^{\rm cont}(\xi\mid y)
\sim
A_{\rm c}\,\xi^{-(\nu+1)},
\end{equation}
which matches the positive tail of the critical fluid law in Eq.~(\ref{eq:edge-tails}). Conversely, for $\omega\to1$, i.e., $\xi_*(t)-\xi=(1-\omega)\rho_* t$ with $1\ll \xi_*(t)-\xi\ll t$, one obtains
\begin{equation}
P_t^{\rm cont}(\xi\mid y)
\sim
A_{\rm c}\,[\xi_*(t)-\xi]^{-(\nu+1)}.
\label{eq:split-left-limit}
\end{equation}
For $1<\nu<2$, this matches the algebraic left tail of the condensate bump, including its amplitude. Indeed, inserting the left tail~\eqref{eq:B-tails-intro} of $\Phi_{\nu,y}$ into Eq.~\eqref{eq:inter-main-result} with $b=-[\xi_*(t)-\xi]/w_t$ gives
\begin{equation}
P_t(\xi\mid y)\sim{\mu_{\rm c}\over T_*}\,A_\nu\gamma_y^{-\nu}\,w_t^\nu\,[\xi_*(t)-\xi]^{-(\nu+1)}
=\mu_{\rm c}A_\nu\gamma_y^{-\nu}K_\nu\left|\cos{\pi\nu\over2}\right|[\xi_*(t)-\xi]^{-(\nu+1)},
\end{equation}
where we used Eq.~\eqref{eq:C-wt-stable}. With
\begin{equation}
\mu_{\rm c}={rJ_{2,c}\over q_{\rm c}^4},
\qquad
\gamma_y={y_+-y_-\over y_+}={q_{\rm c}^2\over rD_{\max}J_{2,c}},
\qquad
|\sin(\pi\nu)|=2\sin{\pi\nu\over2}\left|\cos{\pi\nu\over2}\right|,
\end{equation} and $K_\nu$, $A_\nu$ from Eqs.~\eqref{eq:C-Psi-stable} and~\eqref{eq:stable-tail-constants}, the prefactor reduces identically to $rC\Gamma(\nu+1)2^\nu D_{\max}^\nu/q_{\rm c}^{\nu+2}=A_{\rm c}$, in agreement with Eq.~\eqref{eq:split-left-limit}. This provides a non-trivial check connecting the microscopic edge amplitude $C$, the stable-law constants, and the split-condensate sector.\\

We have therefore shown that the saddle describes a genuine split condensate. Indeed, the first run carries the macroscopic displacement $\rho t$, while the remaining trajectory has velocity $z_\rho\in(y_-,y_+)$. Since the fluid background in the coexistence phase is pinned at the critical velocity $y_-$, the excess displacement $(z_\rho-y_-)(1-\theta_\rho)t=O(t)$ cannot be carried by the fluid background and must therefore form a second macroscopic run of size $(\rho_*-\rho)t$, which yields the decomposition of the total displacement stated in Sec.~\ref{sec:interior}.\\

Integrating Eq.~\eqref{coexist_final_interior}
over $\xi=\omega\rho_*t$, with $\omega$ in any fixed compact subinterval of $(0,1)$,
gives a total weight
\begin{equation}
        \int P_t^{\rm cont}(\xi\mid y)\,\mathrm{d}\xi
        \sim\frac{A_{\rm c}}{(\rho_*t)^{\nu}}
        \int\frac{\mathrm{d}\omega}{[\omega(1-\omega)]^{\nu+1}}
        =O\!\left(t^{-\nu}\right),
        \label{eq:C-split-weight}
\end{equation}
where the non-integrable behavior at $\omega\to0$ and $\omega\to1$ is not part of the
split-condensate sector: by the matching discussed above, these limits connect to the tail of the
critical fluid and to the left tail of the condensate bump, respectively.
The weight~\eqref{eq:C-split-weight} should be compared with the weight
$1/N_t=O(t^{-1})$ of the single condensate, Eq.~\eqref{eq:Nt}. Their ratio is
$O(t^{1-\nu})\to0$ for $\nu>1$, which confirms the power counting of Appendix~\ref{app:critical-amplitudes} for $k=2$. Equivalently, summing
Eq.~\eqref{eq:C-split-weight} over the $O(t)$ runs of a trajectory shows that
the probability that a conditioned trajectory contains two macroscopic runs
vanishes as $O(t^{1-\nu})$, whereas a single condensate is present with
probability one. Note, however, that the convergence is slow, as
$t^{1-\nu}$, when $\nu$ is close to $1$.\\

Finally, macroscopic first runs beyond the condensate, $\rho_*<\rho<y$, are exponentially suppressed. Indeed, for general $z$, the excess cost can be written as
\begin{equation}
        s_y(\theta,\rho)-I(y)={(\rho-y_+\theta)^2\over4D_{\max}\theta}+(1-\theta)\bigl[I(z)-q_{\rm c}z+\Psi(q_{\rm c})\bigr],
        \label{eq:split-general-excess}
\end{equation}
where both terms are non-negative, since $I(z)\geq q_{\rm c}z-\Psi(q_{\rm c})$ by Eq.~\eqref{eq:legendre-summary}. The first term vanishes only at $\theta=\rho/y_+$, where Eq.~\eqref{eq:split-zminus} gives $z_\rho<y_-$ for $\rho>\rho_*$; there the second term is strictly positive, because $I(z)$ lies strictly above its linear continuation for $z<y_-$. Hence $s_y(\theta,\rho)-I(y)>0$ for all $\theta$. For $\rho\downarrow\rho_*$, $y_--z_\rho\propto\rho-\rho_*$, and the excess cost vanishes as $(\rho-\rho_*)^{\nu/(\nu-1)}$ for $1<\nu<2$ by Eq.~\eqref{eq:C-I-stable-known}, which connects to the stretched-exponential right tail of the condensate bump in Eq.~\eqref{eq:B-tails-intro}.

\subsubsection{Saturated interior sector $y>y_+$}

We now consider the saturated phase, $y>y_+$, in the macroscopic
interior regime
$
    \xi=\rho t
$,
$0<\rho<y$,
and again write $\tau=\theta t$. The remaining trajectory has duration
$(1-\theta)t$, displacement $(y-\rho)t$, and scaled endpoint displacement
\begin{equation}
    z=\frac{y-\rho}{1-\theta}.
    \label{eq:sat-z}
\end{equation}
We first assume $z>y_+$ and verify below that this assumption is
self-consistent at the saddle. Using the saturated endpoint asymptotics
\begin{equation}
    Z\big((y-\rho)t,(1-\theta)t\big)
    \underset{t\to\infty}{\sim}
    {\cal A}(z)\big[(1-\theta)t\big]^{-(\nu+3/2)}
    e^{-t(1-\theta)J(z)},
    \label{eq:sat-Z-rem}
\end{equation}
where
\begin{equation}
    J(z)=r+\frac{z^2}{4D_{\max}},
\end{equation}
and ${\cal A}(z)$ is the saturated amplitude~\eqref{eq:A-y-intro}, evaluated at velocity $z$,
while the normalization is
\begin{equation}
    Z_t(y)
    \underset{t\to\infty}{\sim}
    {\cal A}(y)t^{-(\nu+3/2)}e^{-tJ(y)}.
    \label{eq:sat-Z-norm}
\end{equation}

Combining Eqs.~(\ref{eq:sat-Z-rem}) and
(\ref{eq:sat-Z-norm}) with the edge asymptotics of the first-run
factor in Eq.~(\ref{eq:interior-renewal}) gives
\begin{align}
P_t^{\rm cont}(\rho t\mid y)
\sim{}&
\frac{rC\Gamma(\nu+1)}
{\sqrt{4\pi D_{\max}}}
t^{-(\nu+1/2)}
\int_0^1 \mathrm{d}\theta\,
\frac{{\cal A}(z)}{{\cal A}(y)}
\frac{
e^{-t[s_y(\theta,\rho)-J(y)]}
}{
q_{\rho,\theta}^{2\nu+2}
\theta^{\nu+3/2}
(1-\theta)^{\nu+3/2}
},
\label{eq:sat-theta-int}
\end{align}
where
\begin{equation}
    q_{\rho,\theta}
    =
    \frac{\rho}{2D_{\max}\theta} \, , \qquad  s_y(\theta,\rho)
    =
    r\theta
    +
    \frac{\rho^2}{4D_{\max}\theta}
    +(1-\theta)J(z)\, .
    \label{eq:sat-action}
\end{equation}
Using Eq.~(\ref{eq:sat-z}) and the explicit form of $J$, the excess
action becomes
\begin{align}
s_y(\theta,\rho)-J(y)
&=
\frac{1}{4D_{\max}}
\left[
\frac{\rho^2}{\theta}
+
\frac{(y-\rho)^2}{1-\theta}
-y^2
\right]=
\frac{(\rho-y\theta)^2}
{4D_{\max}\theta(1-\theta)}.
\label{eq:sat-square}
\end{align}
It is therefore minimized at
$
    \theta_\rho= \rho / y,
$
for which
$
    s_y(\theta_\rho,\rho)=J(y).
$
Thus, the exponential factor of the numerator cancels that of the
conditioning normalization. At the saddle,
\begin{equation}
    z_\rho
    =
    \frac{y-\rho}{1-\rho/y}
    =y>y_+,
\end{equation}
which confirms the saturated asymptotics assumed above. Moreover,
\begin{equation}
    q_{\rho,\theta_\rho}
    =
    \frac{y}{2D_{\max}}
    \equiv q_y.
\end{equation}

Introducing
$
    \omega= \rho / y
    = \xi / yt,
$
the curvature at the saddle is
\begin{equation}
    s_y''(\theta_\rho,\rho)
    =
    \frac{y^2}
    {2D_{\max}\omega(1-\omega)}.
\end{equation}
Hence
\begin{equation}
    \int \mathrm{d}\theta\,
    e^{-t[s_y(\theta,\rho)-J(y)]}
    \sim
    \frac{\sqrt{4\pi D_{\max}}}{y}
    \sqrt{\frac{\omega(1-\omega)}{t}}.
\end{equation}

Since $z_\rho=y$, the ratio ${\cal A}(z_\rho)/{\cal A}(y)$ is equal to unity.
Evaluating the remaining prefactors at the saddle therefore gives
\begin{equation}
    P_t^{\rm cont}(\omega yt\mid y)
    \sim
    \frac{rC\Gamma(\nu+1)}
         {yq_y^{2\nu+2}}
    t^{-(\nu+1)}
    \frac{1}
         {[\omega(1-\omega)]^{\nu+1}},
    \qquad
    0<\omega<1.
    \label{eq:sat-bridge-raw}
\end{equation}
Using (see Eq.~(\ref{eq:Ac-intro}))
\begin{equation}
    A_y^{(+)}
    =
    A_{\rm c}
    \left(\frac{y_+}{y}\right)^{\nu+2},
\end{equation}
this can be written as
\begin{equation}
    P_t^{\rm cont}(\omega yt\mid y)
    \sim
    \frac{A_y^{(+)}}
         {(yt)^{\nu+1}[\omega(1-\omega)]^{\nu+1}},
    \qquad 0<\omega<1 .
    \label{eq:sat-bridge}
\end{equation}
The two matching limits of Eq.~(\ref{eq:sat-bridge}) recover the
neighboring asymptotic regimes. For $\omega\to0$, with
$1\ll \xi=\omega yt\ll t$,
\begin{equation}
    P_t^{\rm cont}(\xi\mid y)
    \sim A_y^{(+)}\xi^{-(\nu+1)},
\end{equation}
which matches the positive tail of the microscopic fluid law in
Eq.~(\ref{eq:edge-tails}). Conversely, for $\omega\to1$, setting
$u=yt-\xi=(1-\omega)yt$ as in Sec.~\ref{sec:cond-saturated}, with $1\ll u\ll t$,
\begin{equation}
    P_t^{\rm cont}(yt-u\mid y)
    \sim A_y^{(+)}u^{-(\nu+1)},
\end{equation}
which matches the $u\to+\infty$ tail of the continuous saturated
macroscopic density in Eq.~(\ref{eq:Pcont-tails-intro}).\\

Thus, the saturated interior sector also describes a split condensate: at the saddle, the first run carries the macroscopic displacement $\omega yt$, while the remaining trajectory has $z_\rho=y>y_+$ and is therefore itself saturated, carrying the complementary macroscopic displacement $(1-\omega)yt$. Hence, to leading order,
$
    yt=\omega yt+(1-\omega)yt
$,
with only subextensive contributions from the fluid background. Integrating Eq.~\eqref{eq:sat-bridge} over $\xi=\omega yt$, with $\omega$ in a fixed compact subinterval of $(0,1)$, gives a weight $O(t^{-\nu})$ for such a split condensate, which is negligible compared with the $O(1)$ weight of the saturated condensate.

\section{Details of the rare event simulation}
\label{app:numerical-algorithm}

In this appendix, we detail the numerical scheme used to sample the conditioned SD trajectories discussed in Sec.~\ref{sec:numerics}. The algorithm proceeds directly in the configuration space of the paths (cf.\ Ref.~\cite{mori-le-doussal-majumdar-etal:2021}), which are fully characterized by the waiting times $\tau_1, \tau_2, \dots$, diffusion coefficients $D_1, D_2, \dots$ and displacements $\Delta x_1, \Delta x_2, \dots$, so that no error from time discretization is incurred. In this parameterization, we use MCMC with a combination of standard Metropolis--Hastings updates and Gibbs sampling to update the values of $\tau, D, \Delta x$, as well as RJMCMC to split or merge constant $D$-segments, in order to sample from the conditional distribution
\begin{align}
    (\tau_1, \tau_2, \dots, D_1, D_2, \dots, \Delta x_1, \Delta x_2, \dots) \mid x(t) = yt\,.
\end{align}

\subsection{Unconditional simulation of SD paths}
We first remind the reader that the \textit{unconditional} distributions of the configuration variables, without prescribed end point or time, are given by
\begin{align}
\begin{cases}
    \tau_i &\overset{\rm iid}{\sim} p(\tau) = r e^{-r \tau}\,,\\
    D_i &\overset{\rm iid}{\sim} W(D)\,,\\
    \Delta x_i \mid \tau_i, D_i &\overset{\rm iid}{\sim} {\cal N}(0, 2 D_i \tau_i)\,.
    \end{cases}
\end{align}
For an unconditional ``forward simulation'' of the SD on the time interval $[0,t]$, we can hence for instance\footnote{Alternatively, one could first generate the number of segments $N-1 \sim {\rm Poi}(rt)$ and then sample $T_1, \dots, T_{N-1}$ iid from ${\cal U}([0,t])$.} first generate a sufficiently large number of the $\tau_i$'s, clip them to the interval $[0,t]$ as illustrated in Fig.~\ref{fig:sketch-setup}, and then evaluate the final time displacement of the path as
\begin{align}
    x(t) = \sum_{i=1}^N \Delta x_i
\end{align}
with $\Delta x_i \sim {\cal N}(0, 2 D_i \tau_i)$ for $i = 1, \dots, N-1$, and $\Delta x_N \sim {\cal N}(0, 2 D_N \Delta t_N)$ with $\Delta t_N = t - \sum_{i=1}^{N-1} \tau_i > 0 $. Of course, since we are interested in large deviations where the event $x(t) \approx yt$ is exponentially unlikely, it is hopeless to attempt to estimate, e.g., the conditional single run distribution $P_t(\xi\mid y)$ from direct sampling. Importantly, note that the set of ``configuration variables''
\begin{align}
    {\cal C} := (\tau_1, \dots, \tau_{N-1}, \Delta t_{N}, D_1, \dots, D_N, \Delta x_1, \dots, \Delta x_N) \in \mathbb{R}^{3N}
\end{align}
is sufficient to fully characterize the path for our purposes; in particular, we may take any (or all, when reweighting by the number of runs per trajectory) of the $\Delta x_i$'s to measure the single run marginal PDF. Here, $N -1 \sim {\rm Poi}(rt)$ is a random variable. Given the set of configuration variables~${\cal C}$, the process $x(s)$ for $s$ in the time interval from $T_{i-1} = \sum_{j = 1}^{i-1} \tau_j$ to $T_{i} = T_{i-1} + \tau_{i}$ follows a Brownian bridge with end points prescribed through $x(T_{i-1}) = \sum_{j=1}^{i-1} \Delta x_j$ and $x(T_i) = x(T_{i-1}) + \Delta x_i$, and diffusivity $D_i$. This will not be necessary for our MCMC simulations, but is useful for visualizing sample paths given their configuration variables ${\cal C}$.

\subsection{MCMC for the conditional distribution}

To sample from the conditional distribution of paths satisfying the rare constraint $x(t) = yt$, the idea is now to first initialize a Markov chain $({\cal C}_i)_{i \geq 0}$ with an initial configuration ${\cal C}_0 = {\cal C}$ from a path that already satisfies the constraint $x(t) = yt$, and then propose only moves for the chain that preserve the constraint exactly. Recall that the SD path is fully characterized by the configuration variables ${\cal C} \in \mathbb{R}^{3N}$ with $N$ random. In each step of the chain ${\cal C}_i \to {\cal C}_{i+1}$, we select among four possible moves which we term A,B,C,D with equal probability. The proposed new configuration is denoted by ${\cal C}'$, and ${\cal C}_{i+1}$ will either be equal to ${\cal C}'$ or ${\cal C}_i$, depending on the acceptance rule. Moves A,B,C preserve the number $N$ of constant-$D$ segments of the configuration, such that ${\cal C}' \in \mathbb{R}^{3N}$ for ${\cal C}_i \in \mathbb{R}^{3N}$, whereas move D implements a RJMCMC step~\cite{green:1995} that either splits one or joins two adjacent segment(s), such that ${\cal C}' \in \mathbb{R}^{3(N+1)}$ or ${\cal C}' \in \mathbb{R}^{3(N-1)}$, respectively. We now discuss the initialization and possible moves in more detail. All numerical parameters, as introduced below, are also summarized in Table~\ref{tab:numerical-params}. In practice, we run a large number of independent chains in parallel for improved statistics, vectorized per core via Numpy arrays holding \texttt{N\_VECTORIZED} many chains, and further parallelized over \texttt{N\_PROC} many cores. Convergence of the chains, after an initial burn-in phase, is monitored via the Gelman--Rubin statistic, across the  \texttt{N\_VECTORIZED} many chains on each core. For this, we compute the within-chain and between-chain variances for the observable $\sum_i D_i \Delta t_i / t$ to calculate the $\hat{R}$ statistic, and consider the simulation well-thermalized when $\hat{R}$ is close to $1$. We only collect a single sample of~${\cal C}$ per chain for the histograms shown in the main text.\\

We remark that in addition to the sampling strategy presented here, we experimented with a rare event sampling algorithm based on subset simulation~\cite{au-beck:2001}. Subset simulation is a standard sequential Monte Carlo method for simulating rare events~\cite{cerou-del-moral-furon-etal:2012}, which we combined with preconditioned Crank--Nicolson proposals~\cite{cotter-roberts-stuart-etal:2013} for the configurations after mapping them to standard normal space. However, due to poor mixing at large $t$ in the coexistence phase, we abandoned that approach in favor of direct MCMC sampling, with moves and initialization designed to provide better mixing in the presence of a condensate.

\begin{table}
\caption{Summary of the default numerical parameters used in the MCMC simulations for the conditional SD ensemble.}
\label{tab:numerical-params}
\begin{ruledtabular}
\begin{tabular}{llc}
\textbf{parameter} & \textbf{description} & \textbf{value} \\
\colrule
\texttt{N\_VECTORIZED}  & number of trajectories per core & $4000$ \\
\texttt{N\_PROC}  & number of cores & $30$ \\
$n_{\rm steps}$ & maximum number of MCMC steps & $10^4 \cdot [rt]$  \\
$n_{\rm burn-in}$ & number of burn-in steps where $\sigma_D, \sigma_\tau$ are adjusted & $[n_{\rm steps} / 3]$ \\
$M$ & fixed maximum number of random variables to represent configurations & $[2 r t + 10]$\\
$\sigma_D$ & initial proposal width for move B & $0.05$ \\
$\sigma_\tau$ & initial proposal width for move C & $0.5$ \\
$\sigma_{\rm RJ}$ & fixed proposal width for move D & $1$
\end{tabular}
\end{ruledtabular}
\end{table}

\subsubsection{Initialization with an ``approximate condensate''}

To ensure that the Markov chain strictly satisfies the rare constraint $\sum_{i=1}^N \Delta x_i = y t$ and the fixed time constraint $\sum_{i=1}^N \tau_i = t$
(where we denote $\tau_{N} := \Delta t_N$ in the following for notational simplicity)
from the first step, we start by drawing an unconditioned path of random length $N$ by simulating $\tau_i$, $D_i$, and $\Delta x_i$ from their unconditional distributions as described above. Let $c = \mathrm{argmax}_i \tau_i$ denote the index of the segment with the longest duration in this unconditional trajectory, which serves as our initial candidate for the macroscopic condensate. To inject the necessary displacement while respecting the time constraint, we artificially inflate this segment to macroscopic proportions. We draw a uniform random variable $u \sim \mathcal{U}(0.5, 0.9)$ and assign the ``condensate'' a duration $\tau_c' = u t$. The remaining segment times $\tau_j$, $j \neq c$, are then uniformly scaled down to preserve the total time $t$:
\begin{align}
    \tau_j' = \tau_j \frac{t - \tau_c'}{t - \tau_c}\,.
\end{align}
To satisfy the rare spatial constraint exactly, the displacement of the condensate is deterministically set to absorb the difference between the target state and the background fluctuations, yielding $\Delta x_c' = y t - \sum_{j \neq c} \Delta x_j$. Finally, we set the ``condensate'' diffusion coefficient to $D_c' = 0.999$. Overall, this constructs a strictly valid initial configuration $\mathcal{C}_0$ containing an artificial condensate. Note that we do not use the analytical expression for the condensate here, and further start from this initialization in all phases. It can hence be seen as a device to accelerate thermalization in the challenging coexistence and saturated phases. We have observed in our experiments that -- unsurprisingly -- proper initialization is important to decrease the number of necessary steps for the MCMC simulation.

\subsubsection{Move A: exchange displacement between two segments via Gibbs}

To thermalize the spatial displacements without violating the exact constraint $\sum_{i=1}^N \Delta x_i = yt$, we perform a block Gibbs update on two uniformly selected distinct segments, $i$ and $j$, such that the sum of the two displacements $S_x = \Delta x_i + \Delta x_j$ is preserved.  The variances of the independent increments are $\sigma_i^2 = 2 D_i \tau_i$ and $\sigma_j^2 = 2 D_j \tau_j$. Conditioned on their sum $S_x$, the displacement $\Delta x_i$ is exactly Gaussian distributed with mean $\mu_{\rm cond}$ and variance $\sigma_{\rm cond}^2$ given by
  \begin{align}
      \mu_{\rm cond} = S_x \frac{\sigma_i^2}{\sigma_i^2 + \sigma_j^2}\,, \qquad \sigma_{\rm cond}^2 = \frac{\sigma_i^2 \sigma_j^2}{\sigma_i^2 + \sigma_j^2}\,.
  \end{align}
We propose a new displacement $\Delta x_i' \sim \mathcal{N}(\mu_{\rm cond}, \sigma_{\rm cond}^2)$ and set $\Delta x_j' = S_x - \Delta x_i'$. As this is an exact Gibbs step, the acceptance probability is strictly $1$, such that ${\cal C}_{i+1} = {\cal C}'$.

\subsubsection{Move B: update one diffusion coefficient via Metropolis}

We update a single diffusion coefficient $D_i$ for a randomly selected segment $i$ using a standard random-walk Metropolis--Hastings step. A proposal $D_i' = D_i + \delta_D$ is generated with $\delta_D \sim \mathcal{N}(0, \sigma_D^2)$, subject to reflecting boundary conditions at $D=0$ and $D=D_{\max}$.  The acceptance probability $\min(1, \alpha)$ is given by
  \begin{align}
  \log \alpha = \log \left( \frac{W(D_i')}{W(D_i)} \right) - \frac{1}{2} \log \frac{D_i'}{D_i} - \frac{\Delta x_i^2}{4 \tau_i} \left( \frac{1}{D_i'} - \frac{1}{D_i} \right)\,.
  \end{align}
The proposal width $\sigma_D$ is tuned adaptively during a burn-in phase to achieve an acceptance rate near $30\%$.  

\subsubsection{Move C: update one or exchange two segment length(s) via Metropolis}

To alter the temporal durations while exactly preserving the total time $t$, we randomly select two distinct segments $i$ and $j$, and propose a pairwise exchange keeping the sum $S_t = \tau_i + \tau_j$ constant. The proposal is drawn from a mixture distribution: with probability $0.2$, we execute a global jump proposing $\tau_i' = u S_t$ with $u \sim \mathcal{U}(0,1)$, and with probability $0.8$, we employ a local random walk $\tau_i' = \tau_i + \delta_\tau$ with $\delta_\tau \sim \mathcal{N}(0, \sigma_\tau^2)$ (this mixture proposal is symmetric, which is why there is no corresponding Hastings proposal ratio term appended to the acceptance probability $\alpha$ below). Proposals yielding negative times are rejected. The acceptance probability $\min(1, \alpha)$ for these proposals is determined by:
  \begin{align}
  \log \alpha = - \frac{1}{2} \log \left( \frac{D_i \tau_i' + D_j \tau_j'}{D_i \tau_i + D_j \tau_j} \right) - \frac{S_x^2}{4} \left( \frac{1}{D_i \tau_i' + D_j \tau_j'} - \frac{1}{D_i \tau_i + D_j \tau_j} \right)\,.
  \end{align}
Following this accept/reject step for the times, we immediately apply the exact Gibbs block update (move A) to resample $\Delta x_i$ and $\Delta x_j$ conditioned on the (possibly updated) values of $\tau_i$ and $\tau_j$. Again, the proposal width $\sigma_\tau$ is tuned adaptively during a burn-in phase to achieve an acceptance rate near $30\%$ for the update to the $\tau$'s.  

\subsubsection{Move D: birth/death of a segment via RJMCMC}

To allow the number of constant-$D$ segments $N$ to fluctuate, we implement a reversible jump MCMC (RJMCMC) update~\cite{green:1995}. We choose between a ``birth'' (splitting a segment into two) and a ``death'' (merging two adjacent segments) with equal probability.\\

\textbf{Birth step:} A segment $i$ is chosen uniformly. We propose to split it into two new adjacent segments, $1$ and $2$. We draw $u \sim \mathcal{U}(0,1)$ and partition the time as $\tau_1' = u \tau_i$ and $\tau_2'= (1-u) \tau_i$. To partition the displacement, we introduce a random perturbation $\delta_x \sim \mathcal{N}(0, \sigma_{\rm RJ}^2)$ such that $\Delta x_1' = u \Delta x_i + \delta_x$ and $\Delta x_2' = (1-u)\Delta x_i - \delta_x$. We discard the original coefficient $D_i$ and draw two new, independent diffusion coefficients $D_1', D_2'$ from $W(D)$. The log-acceptance ratio then takes the form
  \begin{align}
  \log \alpha_{\rm birth} = \log r + \log P(\Delta x_1'\mid D_1', \tau_1') + \log P(\Delta x_2' \mid D_2', \tau_2') - \log P(\Delta x_i \mid D_i, \tau_i) + \frac{1}{2} \log(2\pi \sigma_{\rm RJ}^2) + \frac{\delta_x^2}{2\sigma_{\rm RJ}^2} + \log \tau_i\,,
  \end{align}
where the first term $\log r$ comes from the ratio $p(\tau_1')p(\tau_2')/p(\tau_i)$, and the conditionals are 
\begin{align}
   P(\Delta x \mid D, \tau) = (4\pi D \tau)^{-1/2} \exp\left(-\Delta x^2 / (4D\tau) \right) 
\end{align}
for short. The final term $\log \tau_i$ originates from the Jacobian determinant of the transformation between the different model spaces~\cite{green:1995}.\\

\textbf{Death step:} A random pair of adjacent segments $i$ and $i+1$ is chosen. We propose merging them into a single segment with $\tau' = \tau_i + \tau_{i+1}$ and $\Delta x' = \Delta x_i + \Delta x_{i+1}$. A new diffusion coefficient $D' \sim W(D)$ is drawn. The auxiliary variables that would have yielded the original state from a birth move are $u = \tau_i / \tau'$ and $\delta_x = \Delta x_i - u \Delta x'$. The acceptance probability $\alpha_{\rm death}$ is then exactly one over the $\alpha_{\rm birth}$ corresponding to this implied reverse split.

\subsection{Further numerical results: sample paths}
\label{app:further-num-results}

\begin{figure}
    \centering
    \includegraphics[width=\linewidth]{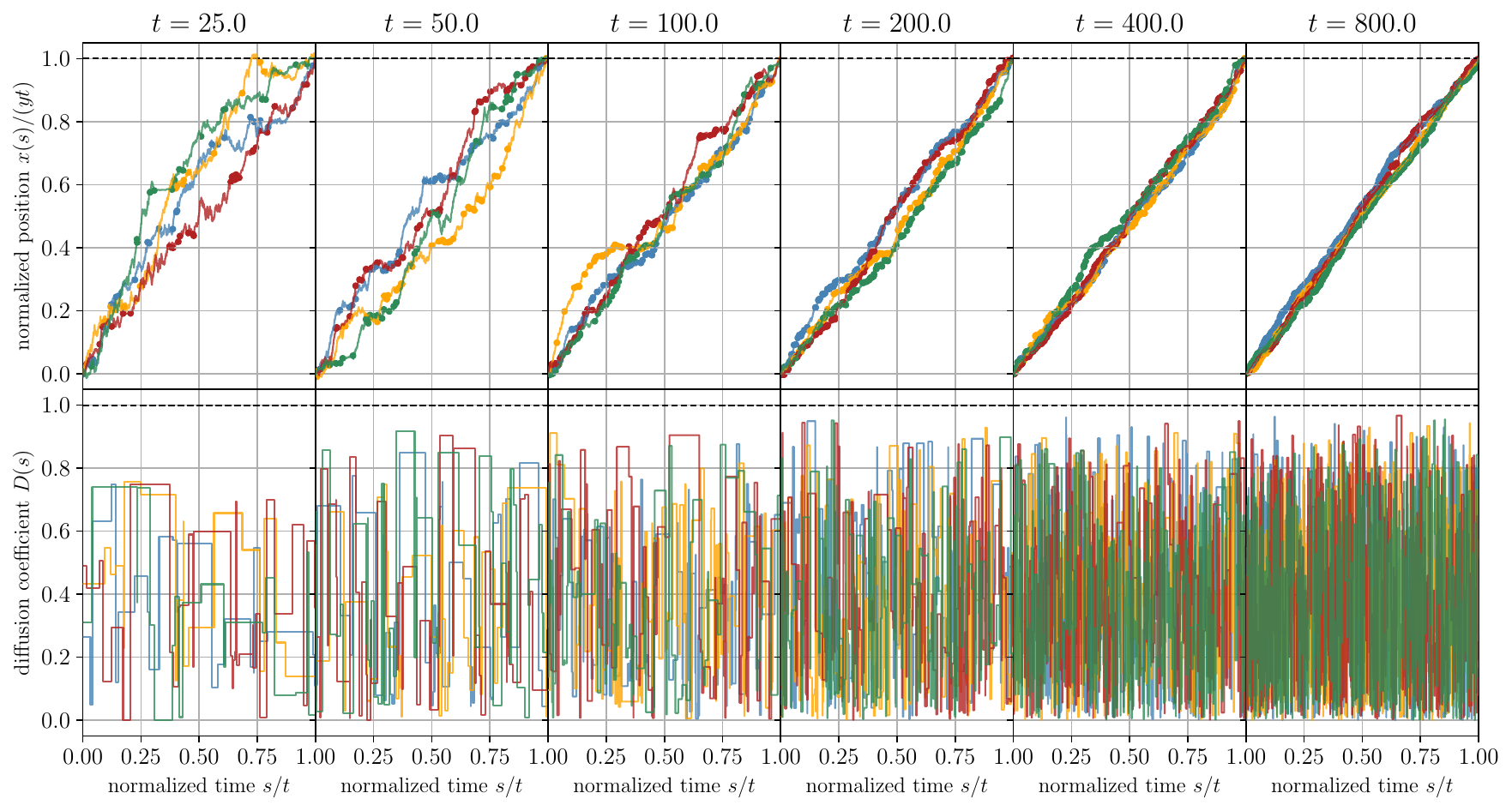}
    \caption{\textbf{Fluid phase.} Four sample paths of $x(s)$ (top row) and $D(s)$ (bottom row) from MCMC, conditioned on $x(t) = y t$ for $y = 1$ at $\nu = 1.8$ (fluid phase), at different $t$ (columns). Dots in the top row indicate switching times. Parameters $W(D) = (\nu + 1) (1 - D)^\nu$, and $r = D_{\max} = 1$. The fluid phase shows many short switching intervals, with no single segment dominating the displacement.}
    \label{fig:path-fluid}
\end{figure}

\begin{figure}
    \centering
    \includegraphics[width=\linewidth]{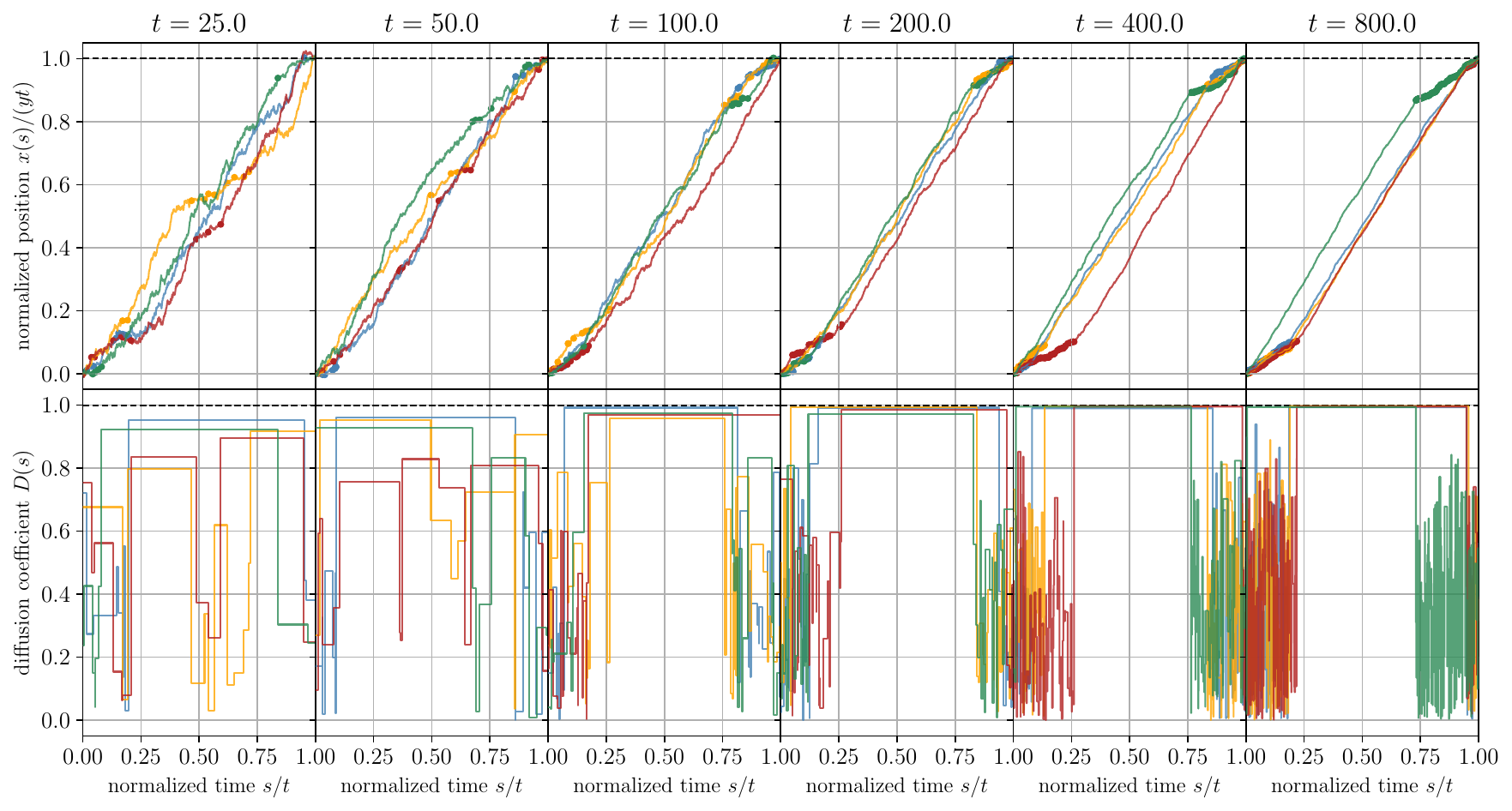}
    \caption{\textbf{Coexistence phase, normal condensate.} Four sample paths of $x(s)$ (top row) and $D(s)$ (bottom row) from MCMC, conditioned on $x(t) = y t$ for $y = 2$ at $\nu = 2.5$ (coexistence phase, normal condensate), at different $t$ (columns). Dots in the top row indicate switching times. Parameters $W(D) = (\nu + 1) (1 - D)^\nu$, and $r = D_{\max} = 1$. In the coexistence phase, one macroscopic high-diffusivity segment coexists with a fluctuating background of shorter runs.}
    \label{fig:path-norm-cond}
\end{figure}

\begin{figure}
    \centering
    \includegraphics[width=\linewidth]{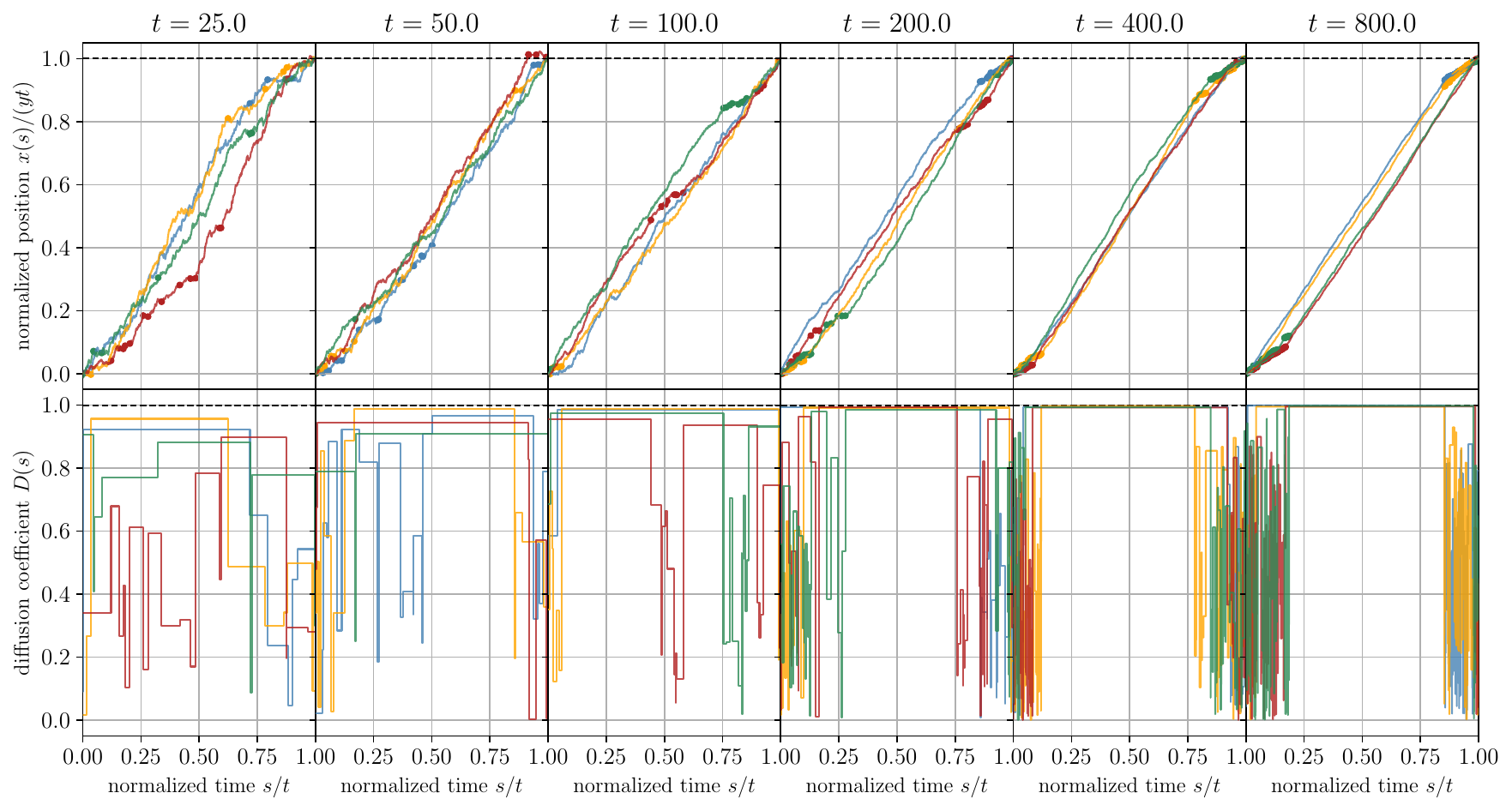}
    \caption{\textbf{Coexistence phase, anomalous condensate.} Four sample paths of $x(s)$ (top row) and $D(s)$ (bottom row) from MCMC, conditioned on $x(t) = y t$ for $y = 2.25$ at $\nu = 1.8$ (coexistence phase, anomalous condensate), at different $t$ (columns). Dots in the top row indicate switching times. Parameters $W(D) = (\nu + 1) (1 - D)^\nu$, and $r = D_{\max} = 1$. In the coexistence phase, one macroscopic high-diffusivity segment coexists with a fluctuating background of shorter runs.}
    \label{fig:path-anom-cond}
\end{figure}

\begin{figure}
    \centering
    \includegraphics[width=\linewidth]{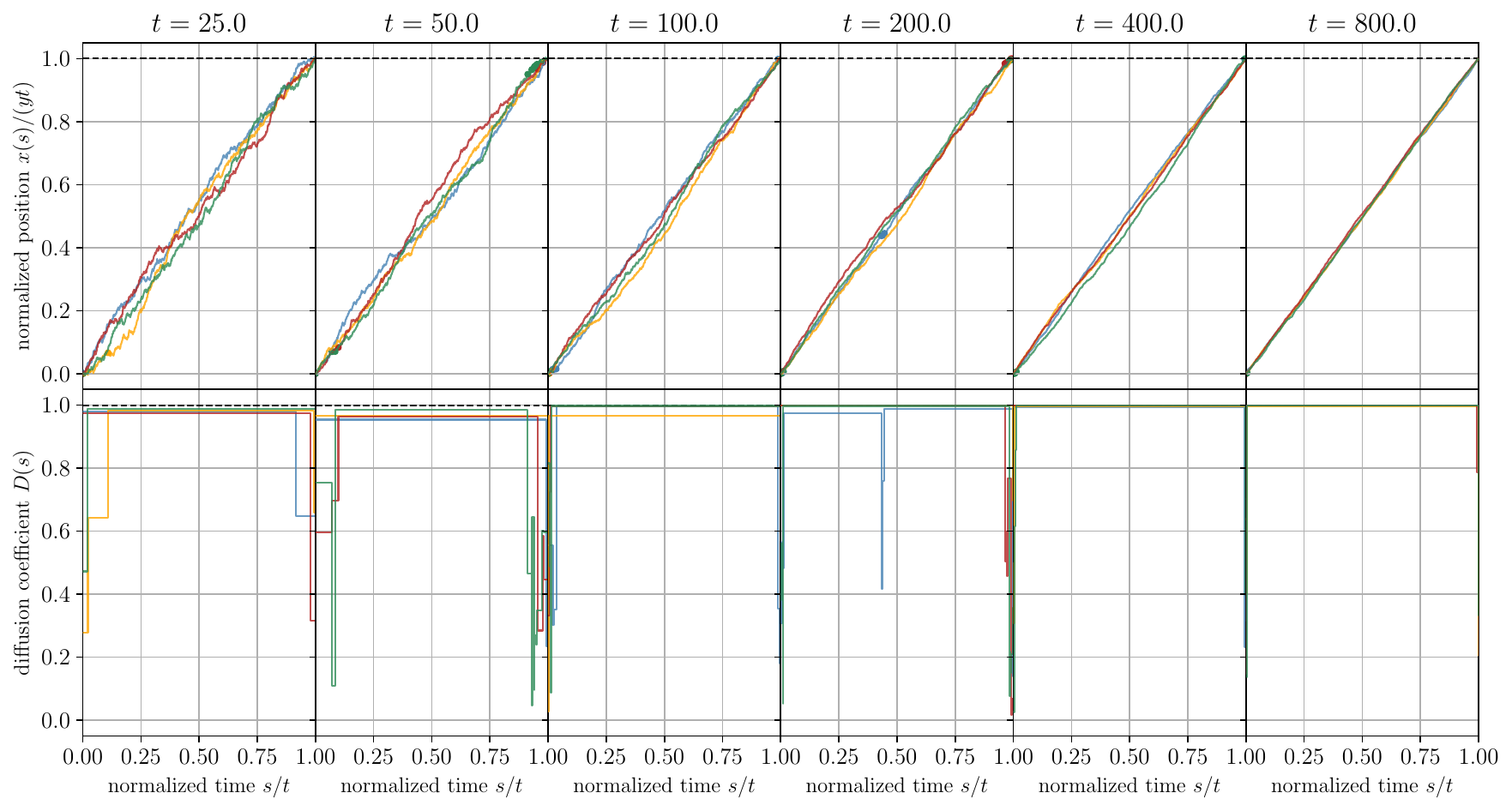}
    \caption{\textbf{Saturated phase.} Four sample paths of $x(s)$ (top row) and $D(s)$ (bottom row) from MCMC, conditioned on $x(t) = y t$ for $y = 3$ at $\nu = 1.8$ (saturated phase), at different $t$ (columns). Dots in the top row indicate switching times. Parameters $W(D) = (\nu + 1) (1 - D)^\nu$, and $r = D_{\max} = 1$. In the saturated phase, one long high-diffusivity segment increasingly dominates the entire trajectory.
    }
    \label{fig:path-saturated}
\end{figure}

In this appendix, we show additional numerical results to expand upon Sec.~\ref{sec:numerics} in the main text. Figs.~\ref{fig:path-fluid}, \ref{fig:path-norm-cond}, \ref{fig:path-anom-cond} and \ref{fig:path-saturated} show a few sample paths from the MCMC simulations in all phases as $t$ increases. They display the run structures sketched in Fig.~\ref{fig:runs-dist} (top), which become increasingly pronounced as $t$ increases.

\section{Numerical solution of the single-run marginal PDF renewal equation}
\label{app:renewal-numerics}

In this appendix, we describe how the solid lines in Fig.~\ref{fig:cond-mcmc-vs-theory} are obtained by evaluating the exact renewal equation~\eqref{eq:exactfirst} numerically at finite $t$, without any asymptotic approximation. Throughout our numerical results, $W(D)=(\nu+1)(1-D)^\nu$ and $r=D_{\max}=1$ as in Sec.~\ref{sec:numerics}, but we keep them in the formulas here for clarity.

\subsection{Form of the renewal equation}

We write
\begin{equation}
        K(x,t):=e^{-rt}\int_0^{D_{\max}}W(D)\,G_{D,t}(x)\,\mathrm{d}D
        \label{eq:F-K-def}
\end{equation}
for the density of trajectories without any switch up to time $t$ that end at $x$, so that $Q(\xi,\tau)=rK(\xi,\tau)$ in Eq.~\eqref{eq:C-Qdef}. Equations~\eqref{eq:exactfirst} and~\eqref{eq:at-def} then read
\begin{equation}
        P_t(\xi\mid y)=a_t(y)\,\delta(\xi-yt)+\frac{r}{Z(yt,t)}\int_0^t K(\xi,\tau)\,Z(yt-\xi,t-\tau)\,\mathrm{d}\tau,
        \qquad
        a_t(y)=\frac{K(yt,t)}{Z(yt,t)},
        \label{eq:F-renewal-numerics}
\end{equation}
and the endpoint density $Z$ obeys the renewal equation obtained by decomposing with respect to the first switch, as in Eq.~\eqref{eq:jointfirst},
\begin{equation}
        Z(x,t)=K(x,t)+r\int_0^t\mathrm{d}\tau\int_{\mathbb R}\mathrm{d}\xi\,K(\xi,\tau)\,Z(x-\xi,t-\tau).
        \label{eq:F-Z-renewal}
\end{equation}
This form singles out the two ingredients we need: the explicit no-switch kernel $K$, and the endpoint density $Z$, which enters both the residual factor $Z(yt-\xi,t-\tau)$ and the normalization $Z(yt,t)$. We split $Z=K+Z^{\rm sw}$, where $Z^{\rm sw}$ is the contribution of trajectories with at least one switch, and treat the two parts differently. The kernel $K$ is singular in two ways that are hard to resolve on a grid: it approaches a Dirac delta as $t\to0$, and for $|x|=O(t)$ its $D$-integral is dominated by a boundary layer $D_{\max}-D=O(1/t)$, cf.\ Sec.~\ref{sec:no-switch}. For the present $W(D)$, however, it is known in closed form. Substituting $D=D_{\max}/(1+s)$ in Eq.~\eqref{eq:F-K-def} gives, for $D_{\max}=1$,
\begin{equation}
        K(x,t)=\frac{\Gamma(\nu+2)}{\sqrt{4\pi t}}\,e^{-rt-x^2/(4t)}\,U\!\left(\nu+1,\tfrac12,\tfrac{x^2}{4t}\right),
        \label{eq:F-K-tricomi}
\end{equation}
with $U$ the Tricomi confluent hypergeometric function. In contrast, $Z^{\rm sw}$ is a time-integrated convolution of at least two kernels, and is smooth enough to be computed spectrally, as described next.

\subsection{Exponential tilt}

Because $Z(yt,t)\asymp e^{-tI(y)}$, the densities in Eq.~\eqref{eq:F-renewal-numerics} are exponentially small and cannot be computed directly at large~$t$. We therefore work with tilted densities $e^{qx-\Psi(q)t}K(x,t)$ and $e^{qx-\Psi(q)t}Z(x,t)$, choosing $q=q(y)$ from Eq.~\eqref{eq:q-of-y-summary}, i.e., $q_*(y)$ obtained from the secular equation~\eqref{eq:fluid-branch-summary} in the fluid phase, $q_{\rm c}$ in the coexistence phase, and $q_y=y/(2D_{\max})$ in the saturated phase. The tilt factors cancel exactly in Eq.~\eqref{eq:F-renewal-numerics}, since $e^{q\xi-\Psi\tau}\,e^{q(yt-\xi)-\Psi(t-\tau)}=e^{qyt-\Psi t}$. After tilting, the densities near the conditioning point $x\simeq yt$ are no longer exponentially small in $t$, and they can be evaluated in double precision without underflow. The tilted kernel follows from Eq.~\eqref{eq:F-K-tricomi} after completing the square in the exponent, $qx-(r+\Psi)t-x^2/(4t)=-(x-2qt)^2/(4t)+(q^2-r-\Psi)t$. We evaluate $U$ through a spline of $\log U$ for moderate arguments, and through its asymptotic series for large arguments.

\subsection{Switched propagator in Fourier space}

The Fourier transform of the tilted endpoint density is a MGF at the complex argument $z=q+ik$,
\begin{align}
   \int_{-\infty}^\infty e^{ikx}e^{qx-\Psi t}Z(x,t)\,\mathrm{d}x=e^{-\Psi t}\langle e^{zx(t)}\rangle.
\end{align}
To evaluate it, we discretize $W$ by a Gauss--Jacobi quadrature with nodes $D_1,\dots,D_M$ and normalized weights $w_1,\dots,w_M$ ($\sum_jw_j=1$), which integrates the factor $(1-D)^\nu$ exactly. As in Sec.~\ref{sec:rmt}, the diffusivity-resolved generating functions then evolve linearly, and symmetrizing with $\bm v=(\sqrt{w_1},\dots,\sqrt{w_M})^\top$ gives
\begin{equation}
        e^{-\Psi t}\langle e^{zx(t)}\rangle\simeq\bm v^\top e^{t\bm H(z)}\bm v,
        \qquad
        \bm H(z)=\mathrm{diag}\big(z^2D_j-r-\Psi\big)_{j=1}^M+r\,\bm v\bm v^\top .
        \label{eq:F-Hz}
\end{equation}
This is the rank-one perturbed matrix of Eq.~\eqref{eq:auxiliary-matrix}, with the uniform weights $1/M$ replaced by quadrature weights and with a complex argument $z$. The matrix exponential sums the renewal series~\eqref{eq:F-Z-renewal} to all orders, i.e., it includes all numbers of switches. Since $\bm H(z)$ is complex symmetric, we diagonalize $\bm H(z)=\bm V\bm\Lambda\bm V^{-1}$ with eigenvectors normalized with respect to the bilinear form $\bm V_n^\top\bm V_n$ (not the Hermitian one), so that
\begin{equation}
        \bm v^\top e^{t\bm H(z)}\bm v=\sum_{n=1}^M e^{t\lambda_n(z)}\frac{(\bm v^\top\bm V_n)^2}{\bm V_n^\top\bm V_n}.
        \label{eq:F-spectral}
\end{equation}
A single eigendecomposition per frequency $k$ thus yields the propagator at all times. We subtract the no-switch contribution $\sum_jw_je^{t(z^2D_j-r-\Psi)}$, which is accounted for exactly through Eq.~\eqref{eq:F-K-tricomi}, and invert the Fourier transform of the remaining smooth part $Z^{\rm sw}$ by FFT. The FFT grid has length $L=2(2qt_{\rm max}+12\sqrt{t_{\rm max}}+12)$, which contains the tilted density (centered at $\Psi'(q)t\le2qD_{\max}t$, with width $O(\sqrt t)$), and a maximal wavenumber of order $14/\sqrt{t_{\rm min}}$, which resolves its smallest spatial scale $O(\sqrt{t_{\min}})$. Here $t_{\rm min}$, $t_{\max}$ are the smallest and largest times computed on the same grid; to keep the grids efficient, the times entering the integral below are grouped into blocks of geometrically increasing duration (by factors of four), each with its own grid. The values at the required displacements are obtained by cubic spline interpolation. Gauss--Jacobi nodes cluster near $D_{\max}$ with spacing $O(M^{-2})$, so that the boundary layer $D_{\max}-D=O(1/t)$, which carries the macroscopic runs, is resolved for $M\gtrsim\sqrt t$; accordingly, we use $M\propto\sqrt t$, with $M\simeq 210$ at $t=800$.

\subsection{Time integral and assembly}

The $\tau$-integral in Eq.~\eqref{eq:F-renewal-numerics} has integrable singularities $\propto\tau^{-1/2}$ and $\propto(t-\tau)^{-1/2}$, from the first run when $\xi\to0$ and from the residual trajectory when $\xi\to yt$, respectively. Both are removed by the substitution $\tau=t\cos^2\theta$, $t-\tau=t\sin^2\theta$, with $\theta\in[0,\pi/2]$ and $\mathrm{d}\tau=t\sin(2\theta)\,\mathrm{d}\theta$. In $\theta$, we use a composite Gauss--Legendre rule with $48$ nodes per subinterval, with break points at residual durations $t-\tau\in\{0.02,0.08,0.25,1,4,16,\dots\}<t/2$, mirrored at $t/2$. This resolves microscopic first runs and microscopic residual trajectories as well as macroscopic ones. At each node, the integrand is $K(\xi,\tau)$ from Eq.~\eqref{eq:F-K-tricomi}, multiplied by $Z(yt-\xi,t-\tau)=K(yt-\xi,t-\tau)+Z^{\rm sw}(yt-\xi,t-\tau)$ (all tilted). The normalization $Z(yt,t)$ and the atom $a_t(y)$ are obtained in the same way. We evaluate the density on a non-uniform grid in $\xi$, which is refined geometrically near $\xi=0$ and $\xi=yt$ to resolve the microscopic peaks there.\\

Two further details control round-off errors. First, the FFT has an absolute error of the order of machine precision relative to the maximum of the tilted density. For $|x|>\sqrt{2D_{\max}t}$, the tilted $Z^{\rm sw}(x,t)$ is bounded by $(1-e^{-rt})$ times the tilted Gaussian $G_{D_{\max},t}(x)$, since the integrated variance is at most $2D_{\max}t$. Where this bound falls below $10^{-12}$, we set $Z^{\rm sw}$ to zero instead of using the noisy FFT values, and record the omitted bound. Second, for $\xi\geq yt$ the residual displacement $u=yt-\xi$ is negative, and the tilt $e^{qu}$ suppresses the tilted propagator there far below the FFT noise floor. In this region, we compute the untilted propagator ($q=\Psi=0$, for which $\bm H$ is real symmetric) and multiply by $e^{qu-\Psi(t-\tau)}\leq1$ analytically. This is an exact change of representation, so that the small right tail of $P_t(\xi\mid y)$ beyond $yt$ is resolved accurately.\\

Finally, we check accuracy independently. We do not renormalize the densities, so that $\int P_t^{\rm cont}(\xi\mid y)\,\mathrm{d}\xi+a_t(y)-1$, evaluated by Simpson's rule, measures the total discretization error; it is below $10^{-6}$ for all curves in Fig.~\ref{fig:cond-mcmc-vs-theory}. Convergence can further be tested by increasing $M$, the number of time nodes, and the Fourier resolution. As a further check, the atom weights in the saturated phase approach the predicted limit $w_y=(1-m_y)^2$ of Eq.~\eqref{eq:my-w-intro} as $t$ increases. For times $t\gg800$, as used for the convergence check mentioned in Sec.~\ref{sec:numerics}, we replace the Gauss--Jacobi rule by a composite rule in $D_{\max}-D$ with geometrically graded intervals down to $1/(4t)$, reuse the eigendecompositions~\eqref{eq:F-spectral} across all times of a block, and drop wavenumbers for which a bound on $e^{t\lambda_n(z)}$ shows that their contribution is negligible.

\bibliography{bib}

\end{document}